\documentclass[useAMS,usenatbib]{mn2e}
\usepackage{epsfig,rotate,graphicx}
\usepackage[fleqn]{amsmath}
\usepackage{subfigure}
\usepackage{lscape}
\usepackage{bm}

\newcommand{\p}{\partial}
\newcommand{\mnras}{MNRAS}
\newcommand{\apj}{ApJ}
\newcommand{\aap}{A\&A}
\newcommand{\apjl}{ApJL}
\newcommand{\araa}{ARAA}

\newcommand{\be}{\begin{equation}}
\newcommand{\ee}{\end{equation}}
\newcommand{\gtrsim}{\;\raisebox{-.8ex}{$\buildrel{\textstyle>}\over\sim$}\;}
\newcommand{\lesssim}{\; \raisebox{-.8ex}{$\buildrel{\textstyle<}\over\sim$}\;}

\newcommand{\apjs}{{\it ApJS, }}

\newcommand{\prd}{{\it Phys. Rev. D }}

\newcommand{\avg}[1]{\langle #1 \rangle}
\newcommand{\bu}{\bm{u}}
\newcommand{\lmax}{l_\mathrm{max}}
\newcommand{\mmax}{m_\mathrm{max}}

\newcommand{\zeus}{{\tt ZEUS-MP }}
\newcommand{\ii}{\mathrm{i}}
\DeclareMathOperator{\erf}{erf}

\title[Gap stability in 3D]{Vortex and spiral instabilities at gap edges 
in three-dimensional self-gravitating disc-satellite simulations}

\author[Lin]{ Min-Kai Lin
  \thanks{E-mail: mklin924@cita.utoronto.ca} \\ 
Canadian Institute for Theoretical Astrophysics,  
60 St. George Street, Toronto, ON, M5S 3H8, Canada \\
}

\begin{document}

\maketitle
\begin{abstract}
Numerical simulations of global three-dimensional (3D), self-gravitating
discs with a gap opened by an embedded planet are presented. The 
simulations are customised to examine planetary gap stability.  
Previous results, obtained by  \citeauthor{lin11a} 
from two-dimensional (2D) disc models, are reproduced in 3D. These
include (i) the development of vortices associated with
local vortensity minima at gap edges and their merging on 
dynamical timescales in weakly self-gravitating discs, (ii) the increased
number of vortices as the strength of self-gravity is increased and their
resisted merging, and (iii) suppression of the vortex instability and
development of global spiral arms associated with local vortensity
maxima in massive discs.     
The vertical structure of these disturbances are examined. 
In terms of the relative density perturbation, the vortex disturbance
has weak vertical dependence when self-gravity is neglected. Vortices
become more vertically stratified with increasing self-gravity. This effect is
seen even when the unperturbed region around the planet's orbital
radius has a Toomre stability parameter $\sim 10$. The spiral modes
display significant vertical structure at the gap edge, with the
midplane density enhancement being several times larger than that near
the upper disc boundary. However, for both  
instabilities the vertical Mach number is typically a few per cent,
and on average vertical motions near the gap edge do not dominate
horizontal motions.  
\end{abstract}

\begin{keywords}
planetary systems: formation --- planetary systems:
protoplanetary discs
\end{keywords}

\section{Introduction}\label{intro}
Astrophysical discs may develop radial structure for several
reasons \citep{armitage11}. It has been suggested that protoplanetary
discs contain `dead zones' in which the magneto-rotational instability
is inefficient, leading to a reduced accretion rate in this region 
\citep{gammie96}. Matter then accumulates at the radial boundary of a  
dead zone and the actively accreting region, leading to a local
density bump. 

Structure can also be induced by an external
potential such as  an embedded satellite. A sufficiently massive
planet opens a gap in the disc \citep{lin86}, and the gap edges
involve radial structure with characteristic length-scales of the  
local disc scale-height.   

It is well established that localised radial structure in thin 
discs can be dynamically unstable \citep{lovelace99,li00,li01}. The
evolution of radially structured discs may then be affected by such
instabilities. More 
specifically, a necessary condition for instability is the existence
of an extremum in the ratio of vorticity to surface density, or 
vortensity\footnote{This is modified by a factor involving the entropy
  for non-barotropic discs.}. 
Indeed,  
instabilities have been demonstrated explicitly for dead zone
boundaries \citep{varniere06,lyra08,lyra09,crespe11} as well as
planetary gap edges
\citep{koller03,li05,valborro07,lyra09b,lin10}.

These studies consider non-self-gravitating or weakly
self-gravitating discs. In fact, instabilities in   
structured astrophysical discs can be traced back to 
\cite{lovelace78}, who considered self-gravitating particle 
discs. \cite{sellwood91} studied a similar system, while
self-gravitating gaseous discs were examined by \cite{papaloizou89},
\cite{papaloizou91} and \citet[who adopted disc
profiles to mimic planetary gaps ]{meschiari08}.  
  
\cite{lin11a,lin11b} explored in more detail the role of
self-gravity on the stability of gaps self-consistently opened by a
planet. They performed a series of linear and 
nonlinear calculations for a range of disc masses. 
They found vortex formation in weakly self-gravitating discs and global
spiral arms in massive discs, but both are associated with the gap
edge.

The above studies have employed the razor-thin or two-dimensional (2D)
disc approximation. It is natural to extend these models to
three-dimensions (3D). However, note that  
non-axisymmetric instabilities in pressure-supported thick discs,
i.e. the Papaloizou-Pringle instability (PPI), was originally  
studied in 3D 
\citep{papaloizou84,papaloizou85,papaloizou87,goldreich86}. The  
vortex-forming instability mentioned above is essentially the
PPI operating in a thin disc with the density bump being analogous to  
a torus. Some early studies of slender tori also included self-gravity
\citep[e.g.][]{goodman88,christodoulou92}.

Recently, 3D non-self-gravitating, rotationally-supported global discs
have been simulated with a local density bump, either set as an
initial condition \citep{meheut10,meheut12a,meheut12b}, or     
self-consistently generated by a resistivity jump in magnetic discs 
\citep{lyra12}. The latter models the dead zone
scenario. These simulations display vortex formation similar to 2D
discs. On the other hand, instabilities at planetary gap edges in 3D
self-gravitating discs have not yet been simulated.    


In this work we extend the 2D self-gravitating disc models of
\cite{lin11a,lin11b} to 3D. Because the instabilities are associated
with radial structure with comparable size to the disc thickness, it
is not obvious at first that 2D is a good approximation. Thus,  
our priority in this first study is to verify results obtained in 
\cite{lin11a,lin11b} by simulating equivalent systems in 3D. We also 
identify some three-dimensional effects that sets the direction
for future investigations.     

This paper is organised as follows. After reviewing the main 2D
results in the next subsection, we describe our 3D disc-planet models
in \S\ref{model}. Numerical methods are stated in \S\ref{method}. We
go through our simulations in \S\ref{results} with additional result
analyses presented in \S\ref{additional}. We summarise and
conclude in \S\ref{summary} with a discussion of several limitations
of our simulations. 

\subsection{Gap stability in 2D}
For discussion purposes here we consider a barotropic disc 
\footnote{Our locally isothermal numerical models are not strictly barotropic. However,
the instabilities of interest are associated with localised structure and we adopt sound-speed profiles that vary
slowly in these regions. Hence we can consider it to be isothermal and barotropic. 
}. For a radially structured disc the quantity governing    
stability is the vortensity profile   
\begin{align}
  \eta \equiv \frac{\kappa^2}{2\Omega\Sigma},  
\end{align}
where $\kappa^2\equiv R^{-3}d(R^4\Omega^2)/dR$ is the square of the
epicycle frequency, $\Omega$ is the disc angular velocity and $\Sigma$
its surface density. $R$ is the cylindrical radius. If the disc is
self-gravitating, the Toomre parameter $Q$ is also important, 
\begin{align}\label{toomreq_vortensity}
  Q \equiv \frac{c_s\kappa}{\pi G \Sigma} = \frac{c_s}{\pi
    G}\left(\frac{2\Omega\eta}{\Sigma}\right)^{1/2},  
\end{align}
where $c_s$ is the sound-speed. $Q < 1$ signifies local
axisymmetric gravitational instability \citep{toomre64}. 
We remark that some studies of instabilities in structured
discs employ values of $\kappa^2$ marginally above zero or even
negative \citep[e.g.][]{li00,li01}, implying the classic Toomre
instability may operate. Of course, if the disc is strictly
non-self-gravitating and $\kappa^2>0$, then $\Sigma$ can be rescaled so that
$Q\gg1$, giving a self-consistent model.    

\begin{figure}
  \centering
  \includegraphics[width=\linewidth,clip=true,trim=0cm .2cm 0cm
    0cm]{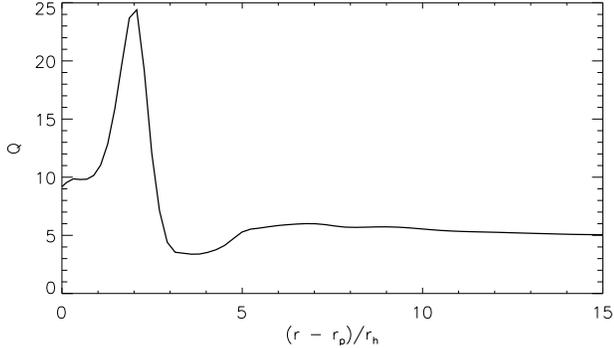} 
  \caption{Typical structure of the outer edge of a planetary gap in
    terms of the Toomre parameter $Q$. The horizontal axis is the
    displacement away from the planet in units of its Hill radius
    (defined in 
    \S\ref{planet_config}). 
    \label{vortex2_gap}} 
\end{figure}

The connection between $Q$ and $\eta$ results in 
the vortensity profile of a planetary gap to resemble its Toomre
parameter profile because of vortensity generation and
destruction across planet-induced shocks 
\citep{lin10}. 
Extrema in $Q$ and $\eta$ nearly
coincide at the same radius. Fig. \ref{vortex2_gap} shows a typical
Toomre profile. Here we focus on the outer disc where we find 
instabilities strongest in the numerical simulations. 

As mentioned in \S\ref{intro}, disc profiles with stationary points in
$\eta$ (and therefore in $Q$ for planetary gaps) can be dynamically 
unstable. The gap profile fulfils this requirement and   
has the following stability properties \citep[taken
from][]{lin11a,lin11b}:

\begin{enumerate} 
\item In weakly or non-self-gravitating discs, instability is
  associated with the vortensity minimum or $\mathrm{min}(Q)$,
  leading to local vortex formation. 
\item As the strength of self-gravity is increased, the vortex mode
  shifts to higher azimuthal wavenumber $m$. This is partly due to the
  stabilisation effect of self-gravity on low $m$ vortex modes. \label{higher_m}
\item The timescale for vortex-merging increases with the strength of
  self-gravity. In non-self-gravitating discs, vortices merge on
  dynamical timescales and the result is a single,
  azimuthally extended vortex. Multi-vortex configurations can last
  much longer with increased disc gravity. Merging eventually takes
  place but the resulting vortex is azimuthally localised. In the
  moderately self-gravitating case discussed in \cite{lin11a}, a
  vortex-pair persists until the end of the
  simulation.  \label{post_merge} 
\item The vortex mode is
  suppressed with sufficiently strong self-gravity and replaced by 
  a global spiral instability 
  associated with the local vortensity maxima or
  $\mathrm{max}(Q)$. The instability can be physically understood as
  gravitational coupling between the gap edge and the wider
  disc exterior to it. 
\end{enumerate}
We shall confirm the above in 3D disc models, except   
for the azimuthally localised, post-merger vortices in
\ref{post_merge}. This requires very long simulations with self-gravity \citep[$\sim
  200$ orbits in][]{lin11a}, which are currently
impractical in 3D. 

These instabilities also affect planetary migration, leading to
vortex-planet and spiral-planet interactions
\citep{lin10,lin11a,lin11b,lin12b}. In order to focus on gap stability
we will not consider migration (and because of resolution limits in a
3D simulation), but we can still measure disc-planet torques. In
particular, we will confirm that spiral modes make the
disc-on-planet torques more positive with increasing instability
strength.



\section{Disc-planet models}\label{model}
We consider a three-dimensional gas disc of mass $M_d$ with an  
embedded planet of mass $M_p$, both rotating about a central 
star of mass $M_*$. To describe the system, we use both spherical
polar co-ordinates $\bm{r}=(r,\theta,\phi)$ and cylindrical 
co-ordinates $\bm{r}=(R,\phi,z)$ centred on the star. The frame is 
non-rotating. The disc is governed by the standard fluid
equations 
\begin{align}\label{3d_gov_eq}
  &\frac{\p\rho}{\p t}+\nabla\cdot(\rho \bu)=0,\\
  & \frac{\p\bu}{\p t}+\bu\cdot\nabla\bu= -\frac{1}{\rho}\nabla p 
   - \nabla{\Phi_\mathrm{eff}} \\
  & p = c_s^2\rho. 
\end{align}
where $\rho$ is the mass density, $\bu$
is the velocity field, $p$ is the pressure and $\Phi_\mathrm{eff}$ is
an effective potential. Physical viscosity is not included in this
study (but artificial viscosity is employed in the numerical
  simulations to treat shocks). The effect of viscosity on vortex modes and spiral modes have 
been investigated previously \citep{valborro07,lin11b}. We do
not expect its effect to differ in 3D. 

\subsection{Equation of state}
We adopt a modified isothermal equation of state (EOS) where the
sound-speed $c_s$ depends on $R$ and the planet position if present.  
Without a planet, we set  
\begin{align}
   \begin{array}{lr}
     c_s = c_\mathrm{iso} \equiv H\Omega_k & \quad
     \quad\quad\quad \text{No planet}, 
     \\   
   \end{array}
\end{align}
where $H=hR$ is the disc scale-height with constant
aspect-ratio $h$ and $\Omega_k=\sqrt{GM_*/R^3}$ is the Keplerian
frequency. When a planet is present, we set 
\begin{align}
  \begin{array}{lr}
    c_s=\frac{H H_p \sqrt{\Omega_k^2 +
        \Omega_\mathrm{kp}^2}  }{\left(H^{7/2} + H_p^{7/2}\right)^{2/7}} &
    \quad \text{With planet}, \\ 
  \end{array}
\end{align}
where $H_p=h_pd_p$, $\Omega_\mathrm{kp}^2 = GM_p/d_p^3$ and $d_p =
\sqrt{|\bm{r}-\bm{r}_p|^2 + \epsilon_p^2}$ is the softened distance to
the planet at position $\bm{r}_p$ and $\epsilon_p $ is the softening 
length defined later. This EOS is taken from \cite{peplinski08a} and
is used here to increase the temperature near to planet in order to
reduce mass accumulation in this region. The dimensionless parameter
$h_p$  controls the temperature increase at $\bm{r}_p$ relative to
that given by $c_\mathrm{iso}$.  

\subsection{Effective potential}
The effective potential is:
\begin{align}\label{pot_eff}
  \Phi_\mathrm{eff} = \Phi_* + \Phi_p + \Phi + \Phi_i,
\end{align}
where
\begin{align}
  \Phi_* = -\frac{GM_*}{r} 
\end{align}
is the stellar potential and
\begin{align}
  \Phi_p = -\frac{GM_p}{d_p}
\end{align}
is the softened planet potential. $\Phi$ is the gravitational potential 
due to the disc material and is given via the Poisson equation 
\begin{align}\label{3d_poisson}
  \nabla^2 \Phi = 4\pi G \rho.
\end{align}
In Eq. \ref{pot_eff}, $\Phi_i$ is the indirect potential due to the
disc and the planet, 
\begin{align}\label{indirect_potential}
  \Phi_i(\bm{r}) = \int \frac{G\rho(\bm{r}')}{r'^3}\bm{r}\cdot\bm{r}'
  d^3\bm{r}' + \frac{GM_p}{|\bm{r}_p|^3}\bm{r}\cdot\bm{r}_p. 
\end{align}
The indirect potential accounts for the acceleration of the
co-ordinate origin relative to the inertial frame. This term is
included for consistency but is unimportant for the instabilities of
interest.

\subsection{Initial disc}
The physical disc occupies $r\in[r_i,r_o]$,
$\theta\in[\theta_\mathrm{min}, \pi - \theta_\mathrm{min}]$ and
$\phi\in[0,2\pi]$.  The vertical domain is such that $\tan{(\pi/2 -
  \theta_\mathrm{min})}/h=n_H$ scale-heights.  The density field is
initialised to 
\begin{align}
  \rho(t=0) = \beta\rho_0,  
\end{align}
where $\rho_0$ is the density profile corresponding to a
non-self-gravitating disc,
\begin{align}\label{initial_density}
  &\rho_0(R,z) = \frac{\Sigma_0}{\sqrt{2\pi}H}
  \left(\frac{R}{r_i}\right)^{-\sigma}
   \left[1 - \sqrt{\frac{r_i}{R + hr_i}}\,\right]\notag\\
   &     \phantom{\rho_0(R,z)=} \times \exp{\left(-\frac{\Phi_*}{c_\mathrm{iso}^2}
     - \frac{1}{h^2}  
    \right)},
\end{align}
with fixed power-law index $\sigma=3/2$. $\beta$ is a function to 
account for vertical self-gravity, such that the surface 
density of the initial disc is the same as that corresponding to
$\rho_0$. We calculate $\beta$ in Appendix
\ref{vertsg_mod} with some approximations. Because of this, we always
first evolve the disc without a planet.  

The constant $\Sigma_0$ is chosen via the Keplerian Toomre parameter
$Q_o$ at the outer boundary: 
\begin{align}\label{3d_q0}
&Q_o \equiv \left. \frac{c_\mathrm{iso}\Omega_k}{\pi G \Sigma}\right|_{R=r_o}, \\
&\Sigma \equiv \int_{z_\mathrm{min}}^{z_\mathrm{max}}
\rho_0 dz.
\end{align}
Note that the integration for surface density $\Sigma$ is taken 
over the finite vertical domain being considered. 

The disc is initialised with zero meridional velocity
($u_r=u_\theta=0$). The azimuthal velocity is set by centrifugal
balance with stellar gravity, pressure and self-gravity, but for the
disc models being considered, which are thin and not very massive, the
initial azimuthal velocity is essentially Keplerian.

Our disc models are labelled by $Q_0\propto M_d^{-1}$. This gives an
indication of the strength of self-gravity. Specifically it measures
gravitational stability against local axisymmetric perturbations at
the outer disc boundary. All of our discs satisfy the Toomre
criterion for stability.

\subsection{Planet configuration}\label{planet_config}
In this work the planet is treated as a fixed external potential. 
Its purpose is to create and maintain a
structured disc. The planet is held on a circular orbit, 
$\bm{r}_p=(r_p, \pi/2 , \phi_p)$ with $\phi_p(t)=\Omega_k(r_p)t$ in
spherical co-ordinates. The
softening length  of the planet potential is fixed to
$\epsilon_p=0.1r_h$ where $r_h=(q/3)^{1/3}r_p$ is the Hill radius and
$q\equiv M_p/M_*$. The EOS parameter is set to $h_p=0.5$. 

\section{Numerical method}\label{method}
 We evolve the disc-planet system using the \zeus
  finite-difference code  in 
spherical coordinates \citep{hayes06}. The computational domain is
divided into $(N_r, N_\theta, N_\phi)$ zones, logarithmically spaced
in radius and uniformly spaced in the angular coordinates. We assume
symmetry about the midplane, so the computational domain only covers
the upper plane  ($z>0$). Hydrodynamic boundary conditions are outflow
at $r_i,\,r_o$, reflecting at $\theta_\mathrm{min}$ and periodic in
$\phi$.  

\zeus was chosen for its ability to treat self-gravity
on a spherical grid with parallelisation. It solves the discretised
Poisson equation using a conjugate gradient method \citep[for details,
see][]{hayes06}. To supply boundary conditions to the solver, we
approximate the boundary potential using spherical harmonic expansion
as described in \cite{boss80}. The expansion in spherical
harmonics $Y_{lm}$ is truncated at $\lmax,\,\mmax$. We assume negligible 
contributions to the disc potential beyond the 
physical disc boundaries. 


\subsection{Simulation setup}
Computational units are such that $G=M_*=1$. The radial range
of the disc is $(r_i,r_o)=(1,25)$. The vertical extent is $n_H=2$
scale-heights. The grid resolution is $(N_r, N_\theta,
N_\phi)=(256,32,512)$. We quote time in units of
$P_0=2\pi/\Omega_k(r_p)$. Between $0\leq t < 10P_0$ the disc is
evolved without a planet and $(\lmax, \mmax)=(48,0)$. The planet is
introduced at $t=10P_0$ and its mass smoothly increased from zero to
its full value between $10P_0\leq t\leq 20P_0$. For $t>10P_0$ we set
$(\lmax, \mmax)=(16,10)$.


The unstable modes of interest are associated with vortensity 
extrema at gap edges, so these features need to be resolved. Numerical
diffusivity, e.g. due to low resolution or grid choice, may inhibit such
instabilities. For example, in their 2D studies of vortex generation
at gap edges, \cite{valborro07} did not find instability in 
Cartesian simulations.   

Test runs with half the resolution in each co-ordinate 
also produce vortex and spiral modes, but with reduced 
growth rate. While the resolution adopted here can
confirm these instabilities operate in 3D, it may  
be inadequate to probe secondary instabilities on longer
timescales (see \S\ref{caveats}).    

We set $\mmax=10$ for the boundary potential
as a compromise between accounting for the non-axisymmetry 
of unstable modes and computation time.
We typically find vortex modes with azimuthal wave-number
$m= 3$---6. We tested a run (case 3 below) with $\mmax=12$ 
and obtained similar results. However with $\mmax=0$, 
vortex merging proceeds soon after their formation, whereas it is
resisted with $\mmax=10$ (consistent with high-resolution 2D
results). The global spiral modes 
have $m=2$, so large $\mmax$ is not crucial. Indeed, 
tests with $\mmax=4$  also yield the spiral instability.

\section{Results}\label{results}
Our simulations are summarised in Table
\ref{experiments}. If the stellar mass is taken to be $M_*=M_{\sun}$
then $q=10^{-3}$ corresponds to a Jupiter-mass planet. The planetary
masses considered here are larger than our previous 2D investigations
\citep{lin11a,lin11b} in order to achieve higher instability growth
rates and shorten the computation time. We will examine gap stability
as a function of $Q_0$.       

\begin{table}
  \centering
    \caption{Simulation parameters. $Q_p$ is the Keplerian Toomre
      parameter evaluated at $r_p$. Case 0 was ran without
      self-gravity.}    
  \begin{tabular}{cllllc}
    \hline
    Case & $h$ & $10^{3}q$ & $Q_0, \,Q_p$ & $M_d/M_*$ & mode \\ 
    \hline\hline
    0  & 0.07  &  2  &  $\infty$ & 0.021 & vortex \\
    1  & 0.07  &  2  &  8.0, 14.8 & 0.021 & vortex \\

    2  & 0.07  &  2  &  4.0, 7.40 & 0.042 & vortex \\
    3  & 0.07  &  2  &  3.0, 5.54 & 0.056 & vortex \\

    4  & 0.05  &  1  &  4.0, 7.39 & 0.030  & vortex \\
    5  & 0.05  &  1  &  3.0, 5.54 & 0.040  & vortex \\

    6  & 0.05  &  1  &  1.7, 3.14 & 0.070  & spiral \\
    7  & 0.05  &  1  &  1.5, 2.77 & 0.080 & spiral\\
    
    \hline
  \end{tabular}
  \label{experiments}
\end{table}

\subsection{Vortex modes in weakly self-gravitating discs} 
We first compare Case 0 and Case 1. The setup for these runs are
identical except that the disc potential is neglected in Case 0, which 
corresponds to the standard approach to model disc-planet systems
\citep[e.g.][]{dangelo10}. Case 1 is the  
self-gravitating version of Case 0. We show below that, even with a
minimum Toomre parameter of $Q_0=8$, disc self-gravity affects the
evolution of the vortex instability.  

%
%

\begin{figure}
  \centering
  \includegraphics[width=\linewidth,clip=true,trim=0cm .2cm 0cm
    0cm]{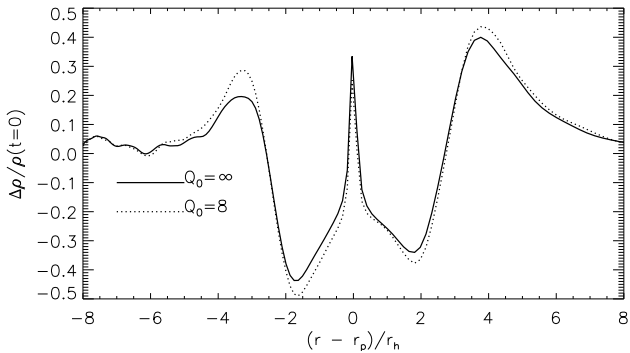} 
  \caption{Gap profile opened by a giant planet in a
    non-self-gravitating disc (solid, Case 0) and a self-gravitating
    disc (dotted, Case 1). The azimuthally-averaged relative density
    perturbation in the midplane is shown. Profiles for other cases
    are similar.
    \label{vortex8_gap}} 
\end{figure}

Fig. \ref{vortex8_gap} shows the midplane gap profile at $t=25P_0$ in
terms of the relative density perturbation. The snapshot is taken
before instabilities develop. Case 1 has a slightly deeper gap and
steeper gap edges than Case 0. This is because in a self-gravitating
calculation such as Case 1, fluid bound to the planet adds to its
mass. For both runs, the mass in the planet's Hill sphere is $\sim
0.02M_p$. 

%


\subsubsection{Development of instability}
\begin{figure*}
  \centering
  \includegraphics[scale=.5,clip=true,trim=.86cm 0.6cm 2.4cm
    .0cm]{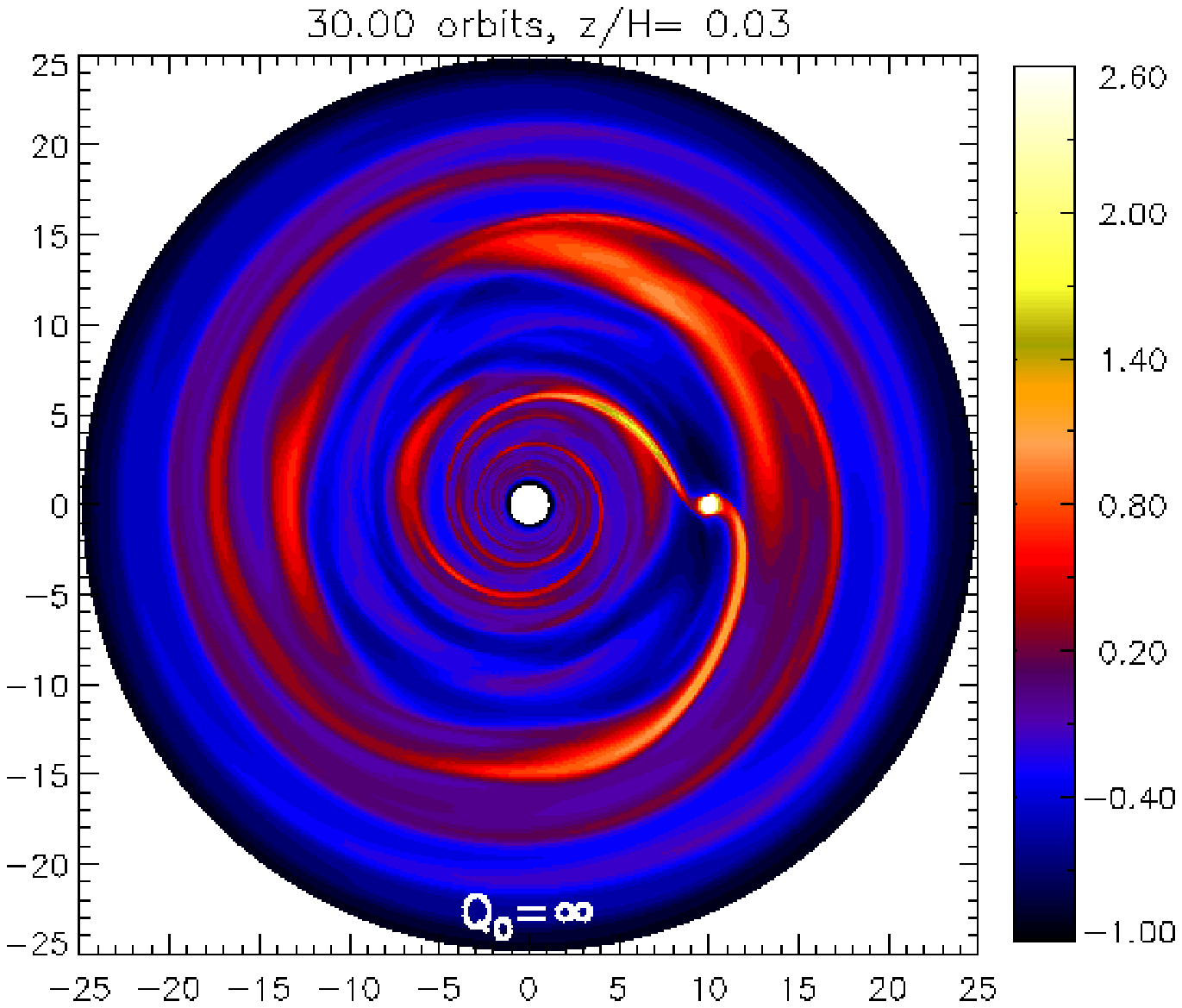}\includegraphics[scale=.5,clip=true,trim=0.75cm
    0.6cm 2.4cm
    0.0cm]{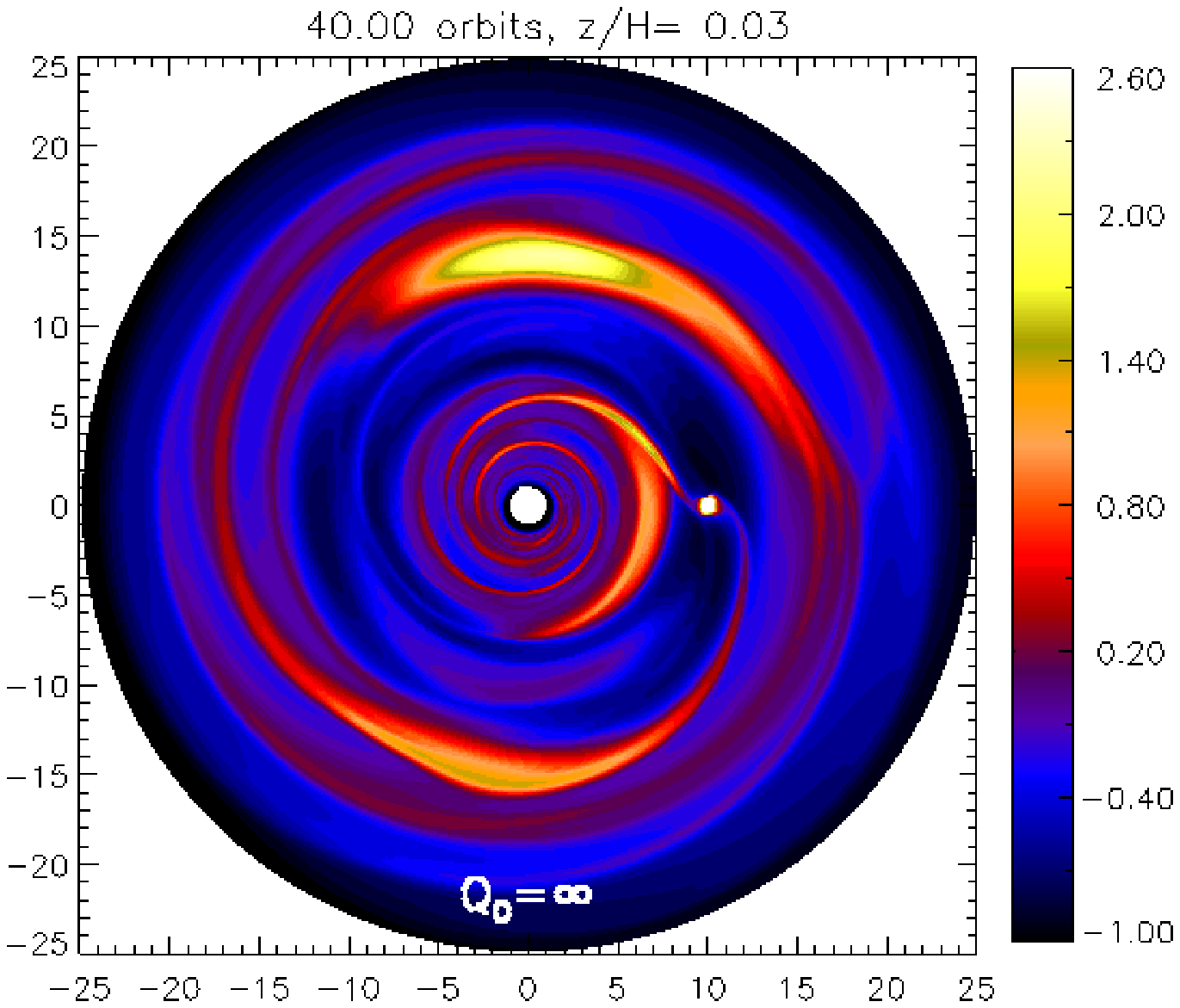}\includegraphics[scale=.5,clip=true,trim=0.75cm
    0.6cm 0cm .0cm]{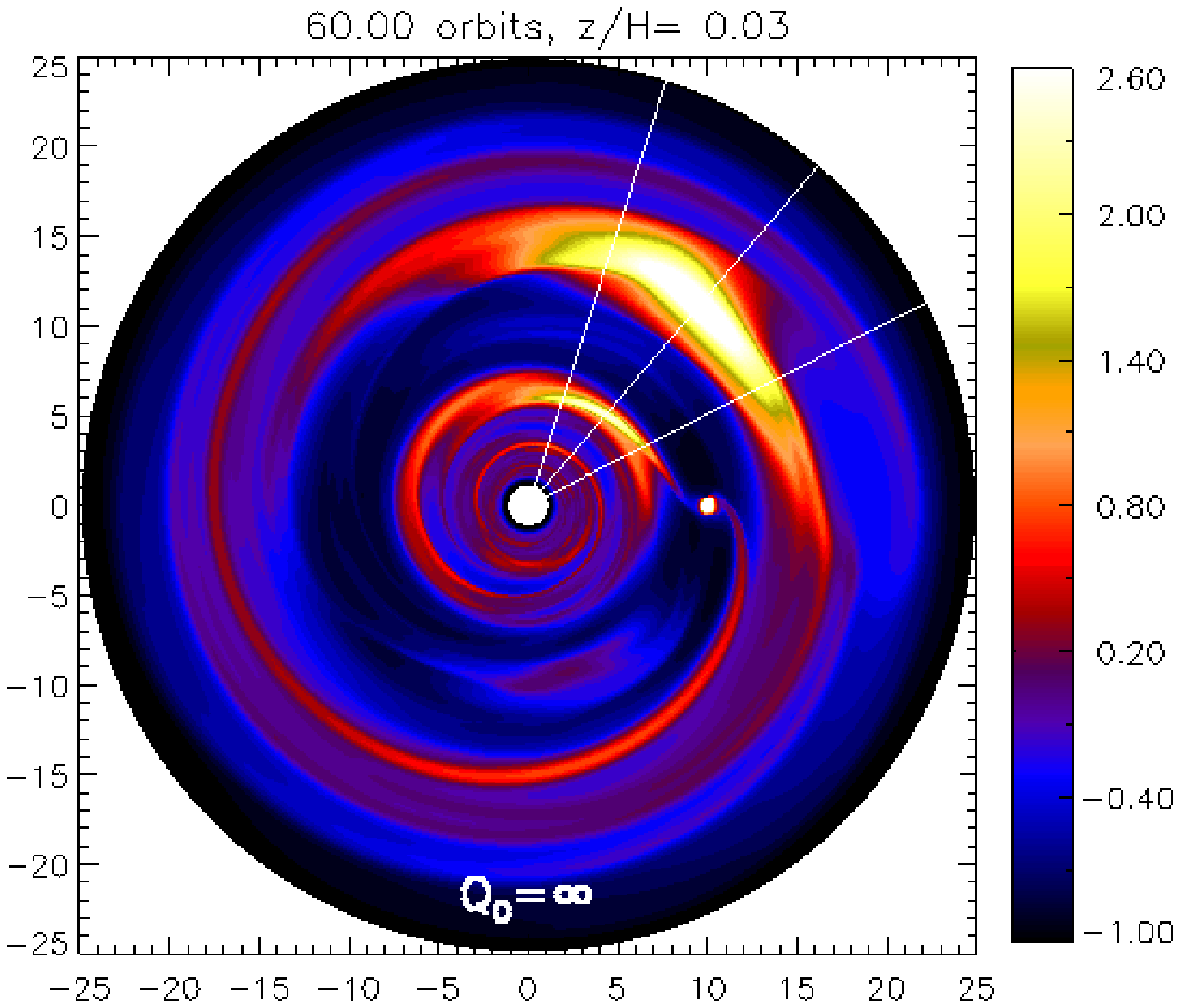}\\
  \includegraphics[scale=.5,clip=true,trim=.86cm 0.6cm 2.4cm
    .64cm]{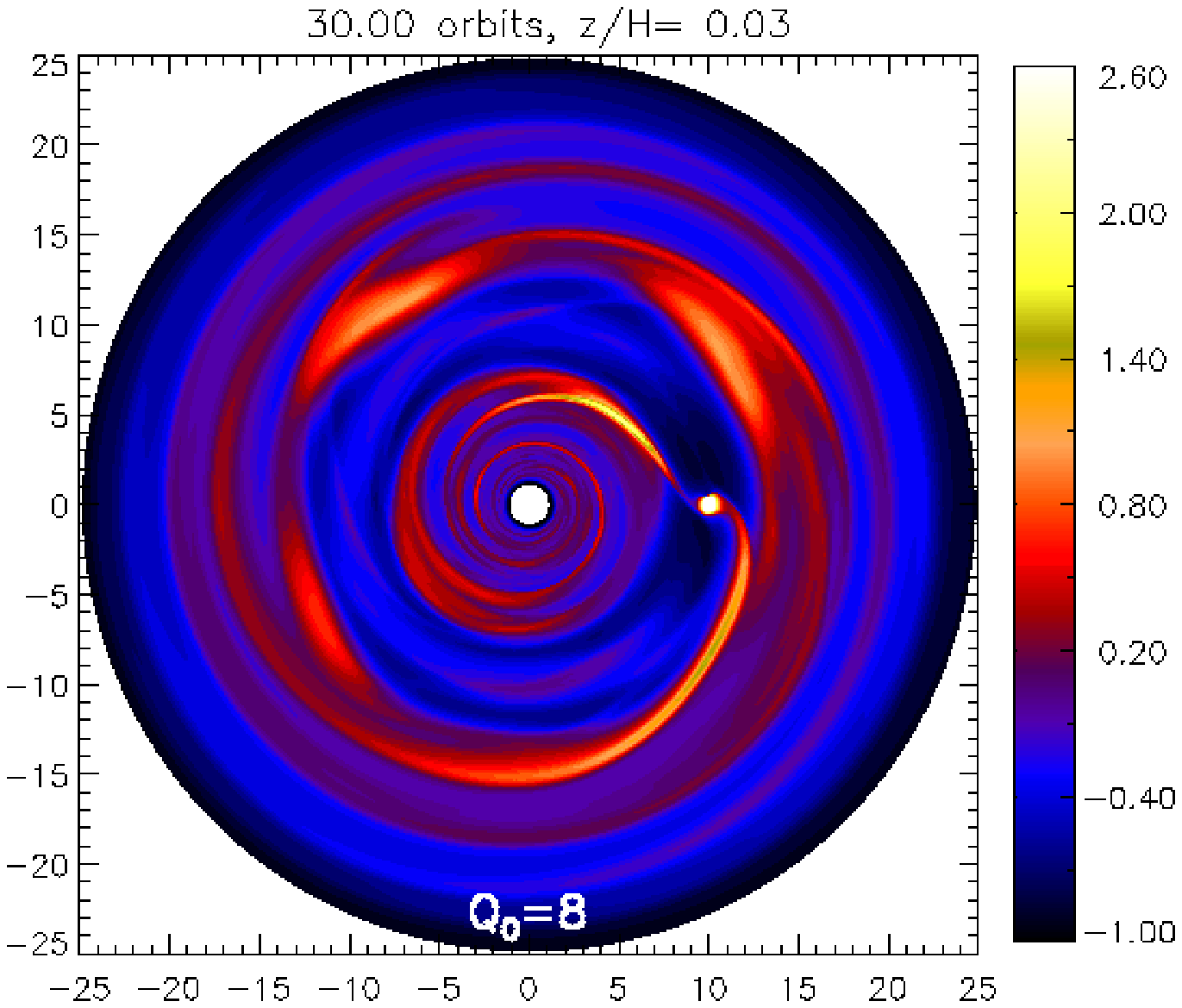}\includegraphics[scale=.5,clip=true,trim=0.75cm
    0.6cm 2.4cm
    .64cm]{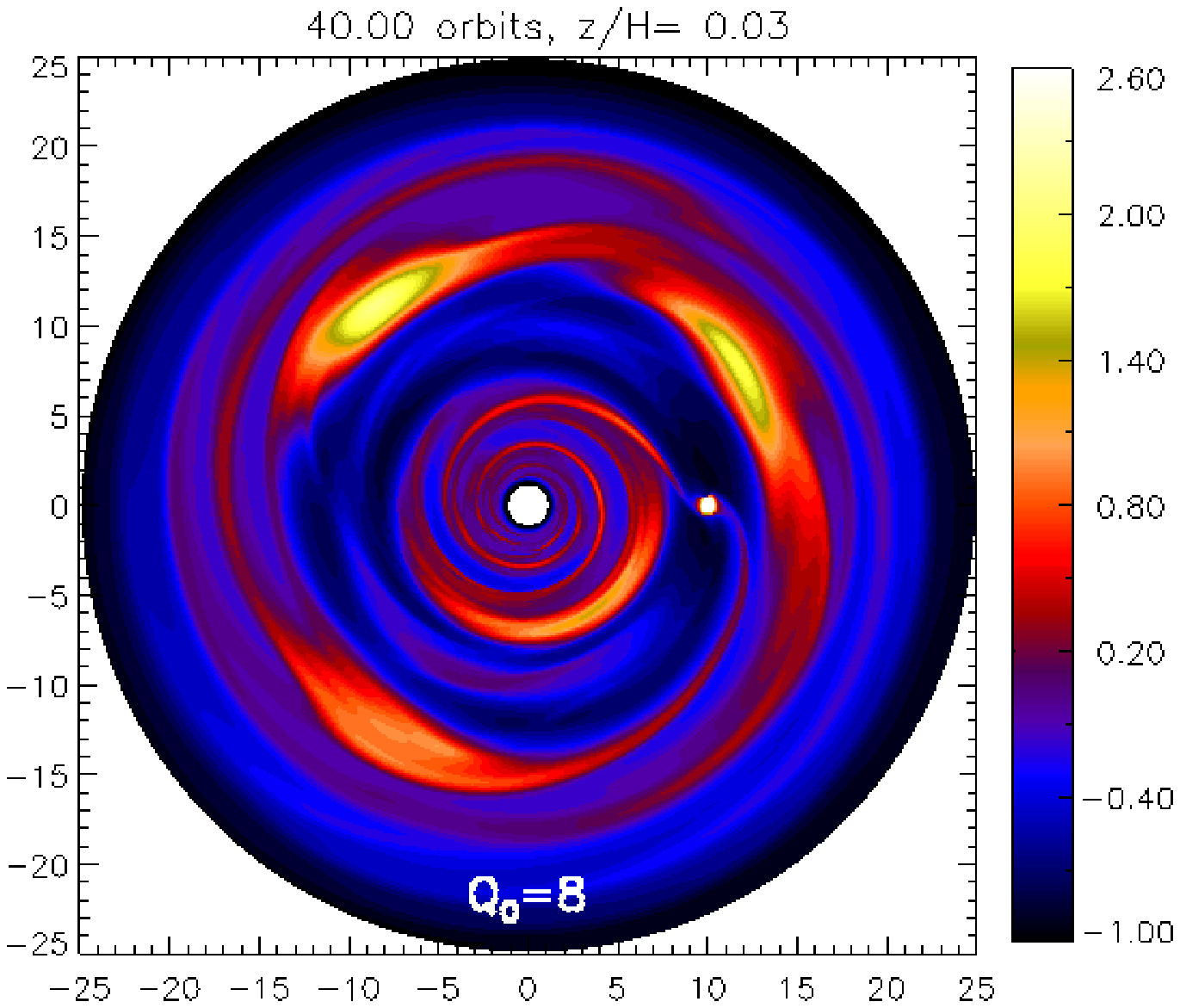}\includegraphics[scale=.5,clip=true,trim=0.75cm
    0.6cm 0cm .64cm]{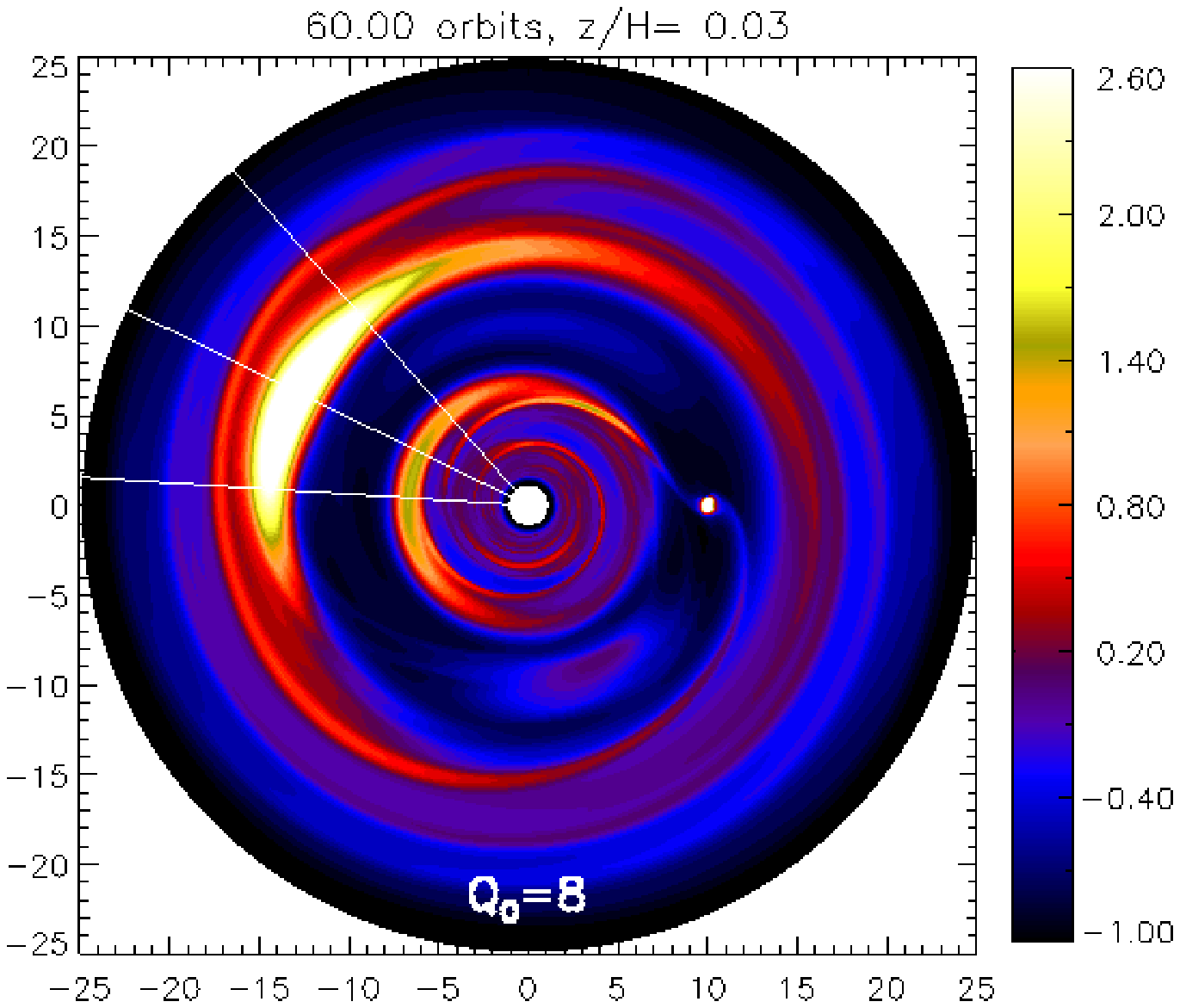}
  \caption{Development of the vortex instability in the
    non-self-gravitating Case 0 (top) and the weakly self-gravitating Case 1
    (bottom). The midplane relative density perturbation is shown. 
    White lines at intermediate azimuths corresponds to the vortex
    centroid in Fig. \ref{vortex8_stream}, while the azimuthal range
    marked by the lines was set for averaging in
    Fig. \ref{vortex8_vprofile}. 
    \label{vortex8_polar_dens}}
\end{figure*}

Fig. \ref{vortex8_polar_dens} shows the evolution of the gap. It is
clear that the vortex instability can develop in 3D with or without
self-gravity. At $t=30P_0$, two vortices are visible 
in Case 0 while 3 vortices are seen in Case 1 (in both cases there may
be another vortex coinciding with the outer planetary wake).

For the weakly self-gravitating discs considered here, vortex modes
with the same $m$ have been excited, but without self-gravity vortices
merge soon after formation. (In Case 0, the over-density at the outer
gap edge just ahead of the planet appears to be a merging vortex-pair,
rather than a single vortex from the 
instability.) At $t=40P_0$, only a vortex-pair remains in Case 
0 while merging is delayed in Case 1 with a 3-vortex configuration.

The differences between self-gravitating and
non-self-gravitating simulations shown in 
Fig. \ref{vortex8_polar_dens} are similar to those seen in 2D
simulations \citep{lyra09,lin11a}. \citeauthor{lin11a} demonstrated
that self-gravity sets a minimal inter-vortex distance so that merging
is resisted. Three-dimensionality does not affect the evolution
of the vortex instability in the $r\phi$ plane. However, at this level
of self-gravity merging is only delayed. The final state for both
cases is a single vortex circulating the gap edge.

\subsubsection{Effect of self-gravity on vortex vertical structure}  
\begin{figure}
  \centering
  \includegraphics[scale=.47,clip=true,trim=0cm 1.35cm .0cm
    0.65cm]{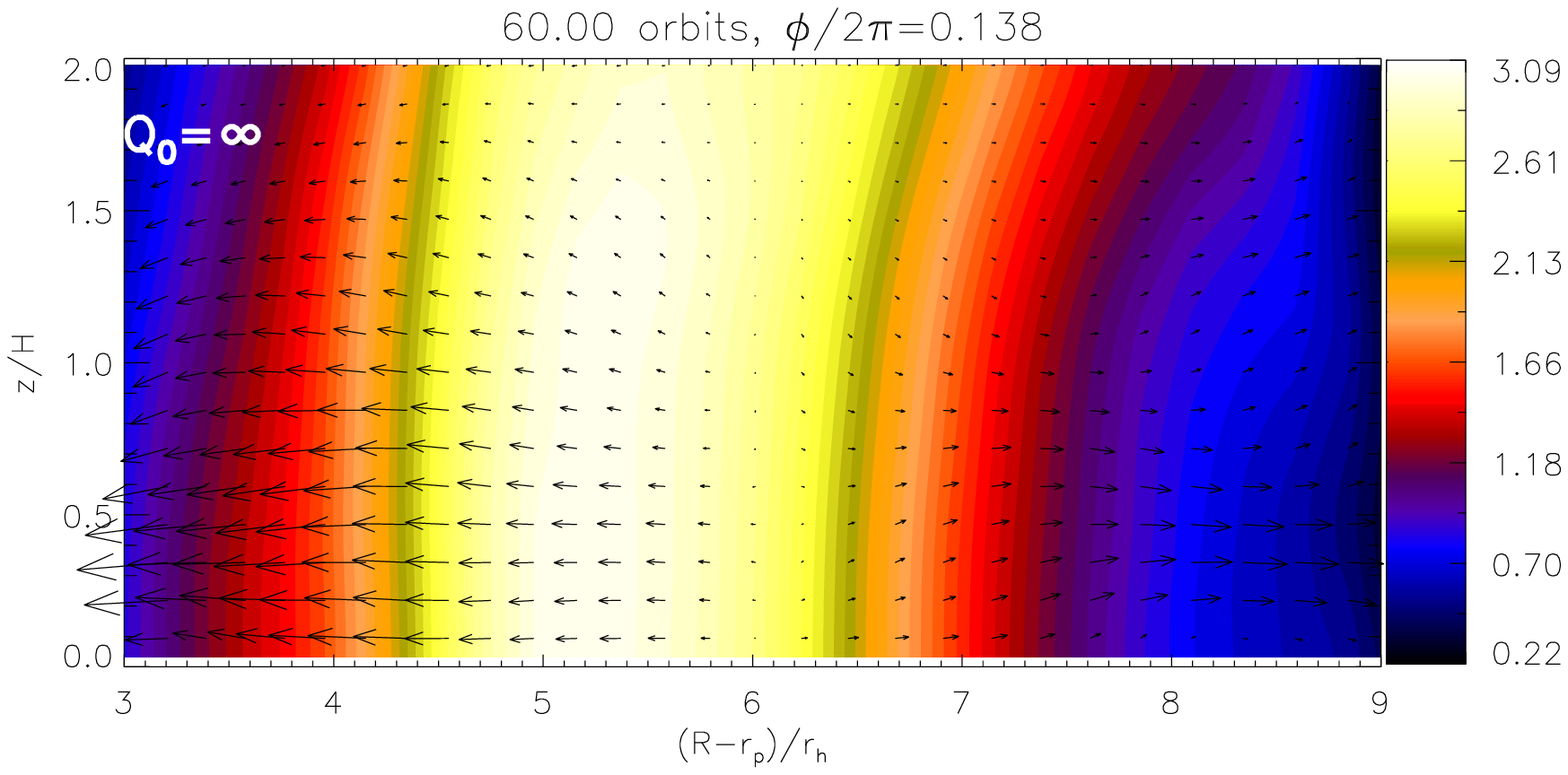}\\\includegraphics[scale=.47,clip=true,trim=0cm 
    0.0cm .0cm
    0.65cm]{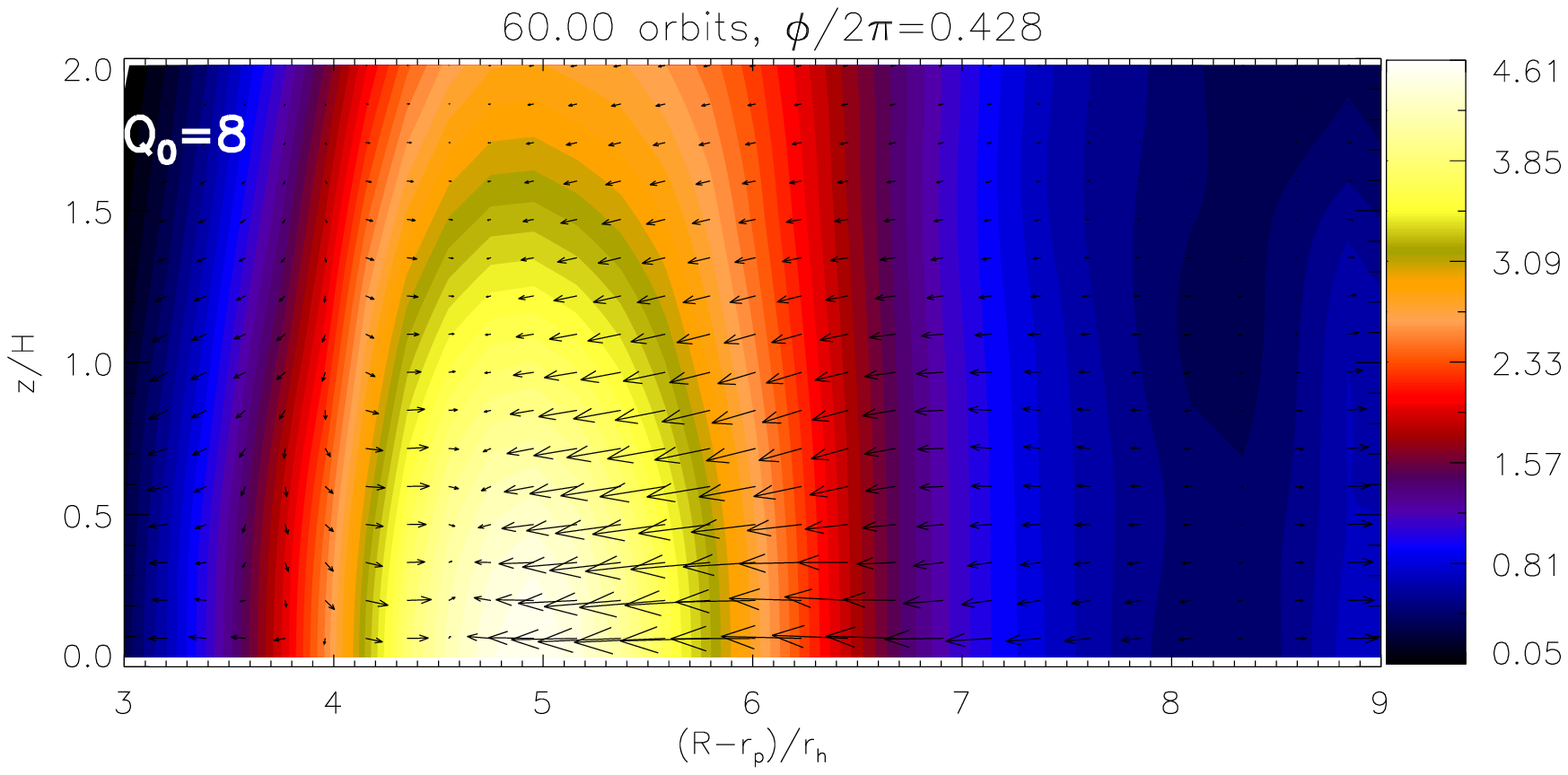}\\
  \caption{Vertical structure of a vortex in a non-self-gravitating
    disc (top, Case 0) and a weakly self-gravitating disc (bottom,
    Case 1). Contours show the relative density perturbation and the arrows
    are mass flux vectors $\rho\bm{u}$ projected onto 
    this plane. The snapshots correspond to the vortex centroids
    marked by white lines at intermediate azimuths in
    Fig. \ref{vortex8_polar_dens} (right panel).  
    \label{vortex8_stream}} 
\end{figure}

Here we compare the final vortex in Case 0 and Case
1. Fig. \ref{vortex8_stream} shows their vertical structure 
in the $Rz$ plane. The snapshots correspond
to the vortex centroid, marked by white lines at intermediate azimuths
in the right panel of Fig. \ref{vortex8_polar_dens}. Without
self-gravity the vortex instability produces predominantly columnar
disturbances in the relative density perturbation $W$.

Case 0 is consistent with recent linear calculations of the vortex
instability in non-self-gravitating 3D discs, which show that  
$W$ has essentially no vertical dependence at the radius where
vortex-formation is expected
\citep{meheut12a,lin12}. In Fig. \ref{vortex8_stream} the Case 0 vortex
does show very weak vertical dependence near the upper boundary. This
is likely due to the finite vertical domain adopted in our
model\footnote{Linear calculations of vertically isothermal discs
  usually assume an atmosphere of infinite extent.}.
  

By contrast, the self-gravitating vortex in Case 1 clearly display
stratification in the relative density perturbation. The vortex is
more concentrated toward the midplane.  With self-gravity, density
enhancement in the vortex can be $\sim 50\%$ higher than
without.   

\begin{figure}
  \centering
  \includegraphics[width=\linewidth]{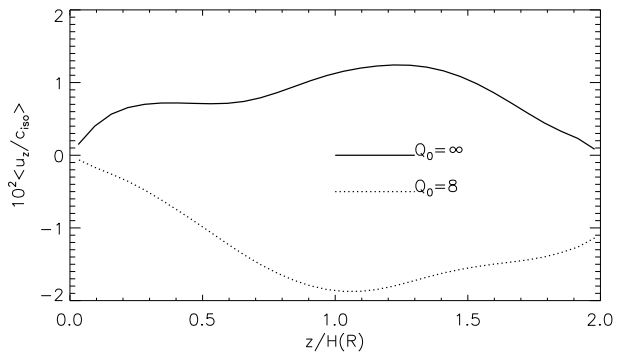} 
  \caption{Average vertical velocity $u_z$, normalised by the sound
    speed, in a non-self-gravitating vortex (Case 0, solid) and a
    self-gravitating vortex (Case 1, dotted).  
    The radial range for the average is 
    taken over $r-r_p\in[4,6.5]r_h$ for Case 0 and
    $r-r_p\in[4,6]r_h$ for Case 1 since the latter vortex is slightly
    smaller (see Fig. \ref{vortex8_stream}).  The azimuthal range 
    is that marked by white lines in Fig. \ref{vortex8_polar_dens}
    (right panel).   
    \label{vortex8_vprofile}} 
 \end{figure}

In Fig. \ref{vortex8_vprofile} we plot vertical velocities  
averaged over the vortices. The non-self-gravitating case typically
involves positive vertical velocity \citep[also seen
  in][]{meheut12a}. Perhaps not surprisingly, the self-gravitating
case has as strong over-density near the midplane to provide vertical
acceleration, so on average  the vertical velocity is negative. The
precise values in  Fig. \ref{vortex8_vprofile} depends on the
averaging procedure but the contrast in $\mathrm{sgn}(u_z)$ between
the two cases is robust (even when we consider the point in the
$R\phi$ plane where density perturbation is largest and do not perform
an average). The quantity $\avg{u_z^2/c_\mathrm{iso}^2}^{1/2}$
behave similarly.    

Although there is vertical motion, it is worth noting that the
vertical Mach number is only a few per
cent. Fig. \ref{vortex8_polar_dens} also indicate that the flow is
horizontal on average. This suggests approximate vertical hydrostatic
balance. If self-gravity is included, then just like in the set up of
the initial disc, the additional vertical force will enhance the
midplane density. Hence, we observe a more stratified vortex in Case
1. 


\subsection{Vortex modes with moderate self-gravity}

Cases 2---5 are all self-gravitating and develop the vortex
instability. Case 2 and Case 3 are continuations of Case 1 to 
more massive discs. Case 4 and Case 5 are
identical runs except a Jupiter-mass planet ($q=10^{-3}$) and a
thinner disc ($h=0.05$) are adopted.  

\begin{figure}
  \centering
  \includegraphics[scale=.27,clip=true,trim=0cm 1.84cm 0.0cm 
    0cm]{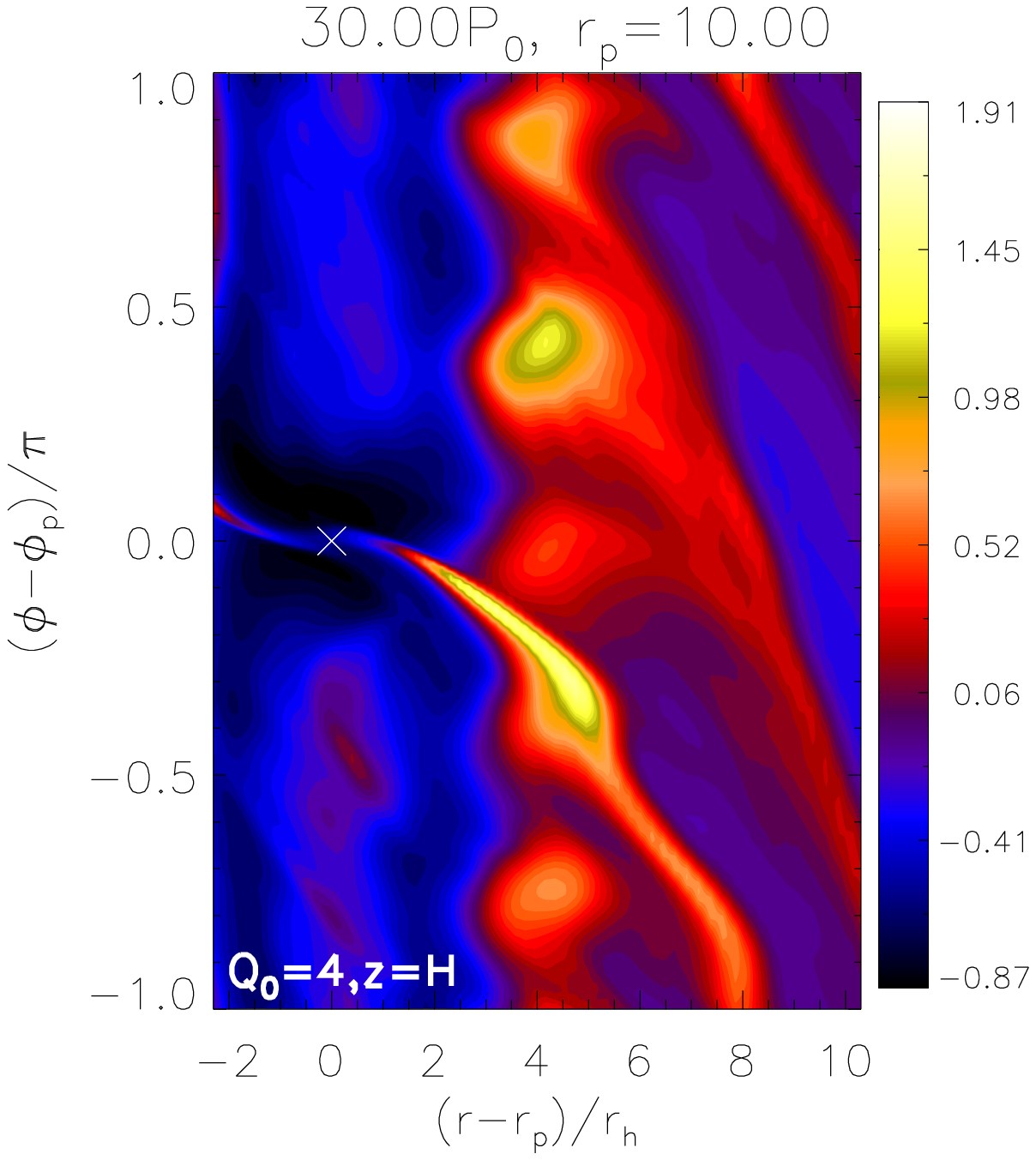}\includegraphics[scale=.27,clip=true,trim=2.3cm   
    1.84cm 0cm 
    0cm]{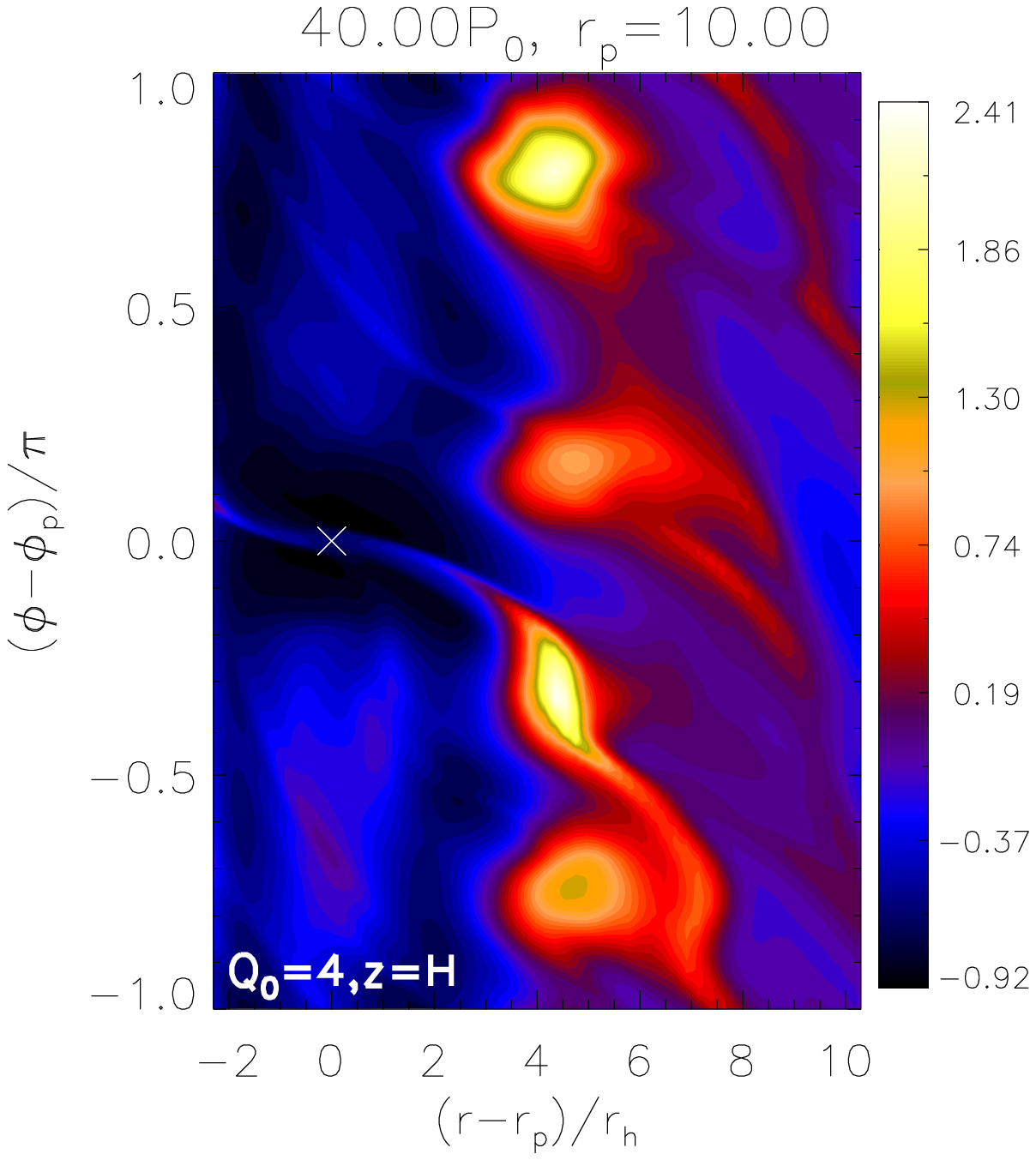}\includegraphics[scale=.27,clip=true,trim=2.30cm    
    1.84cm 0cm 0cm]{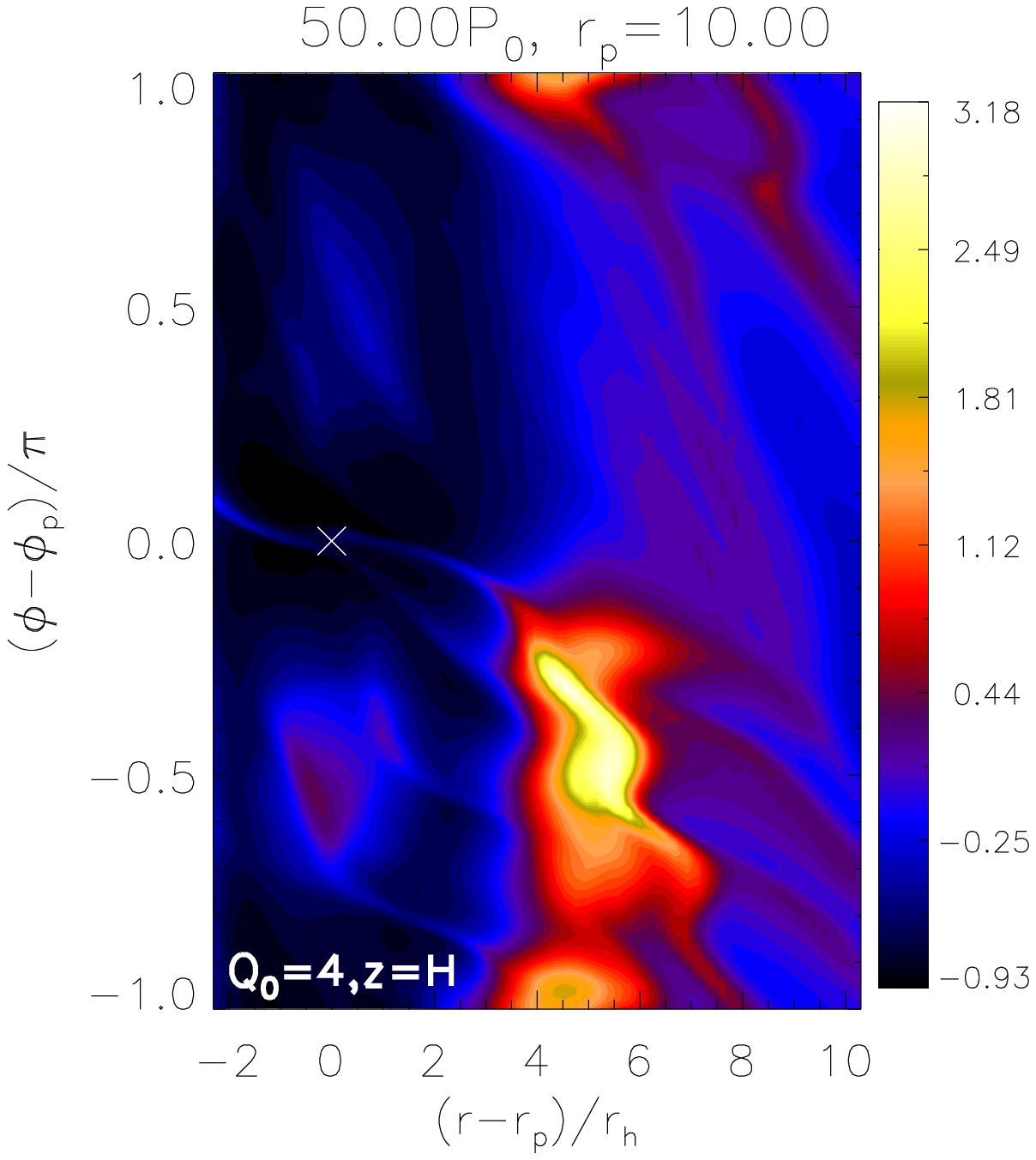}\\
  \includegraphics[scale=.27,clip=true,trim=0cm 0.0cm 0.0cm
    0.99cm]{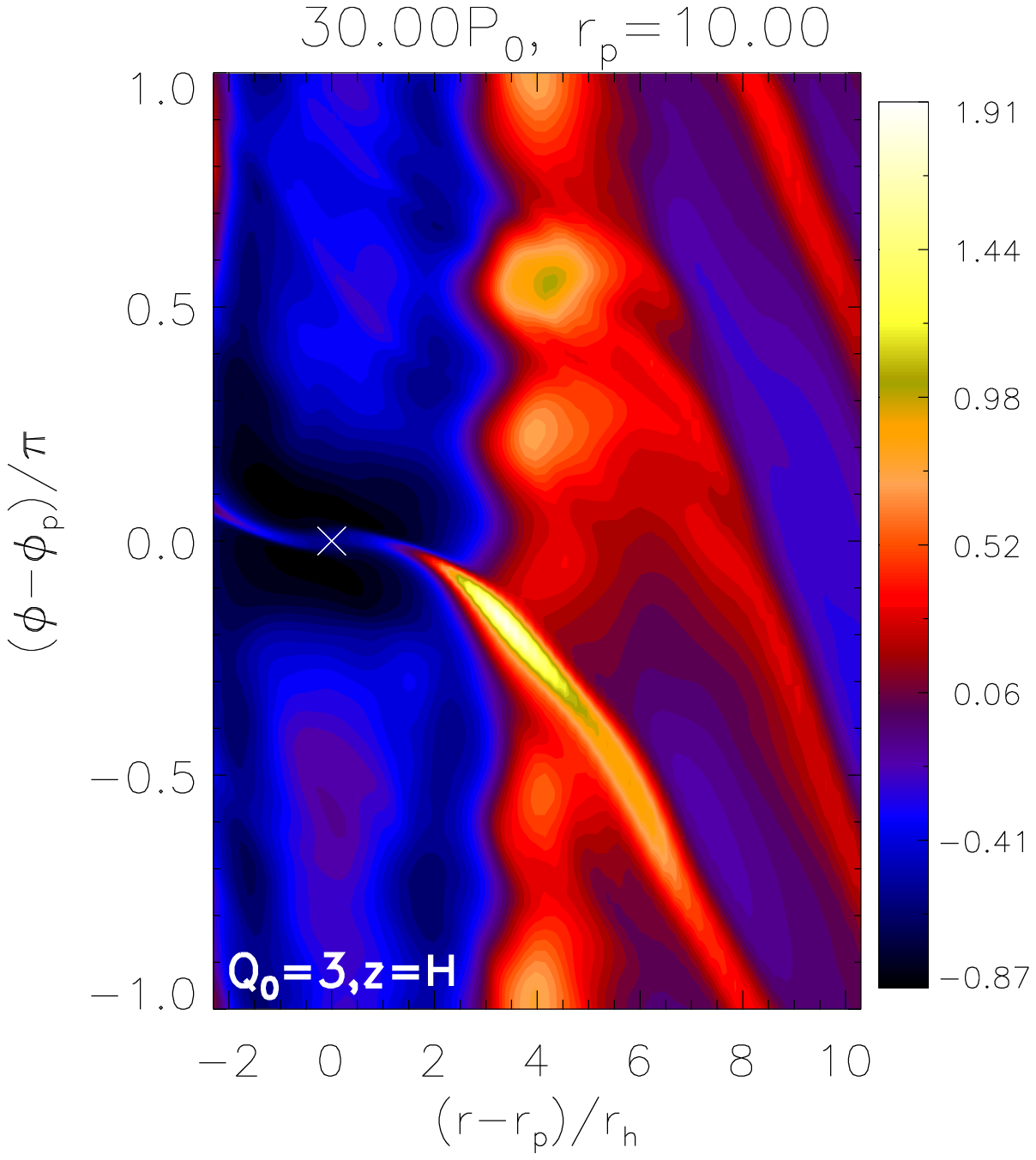}\includegraphics[scale=.27,clip=true,trim=2.30cm  
    0.0cm 0cm
    0.97cm]{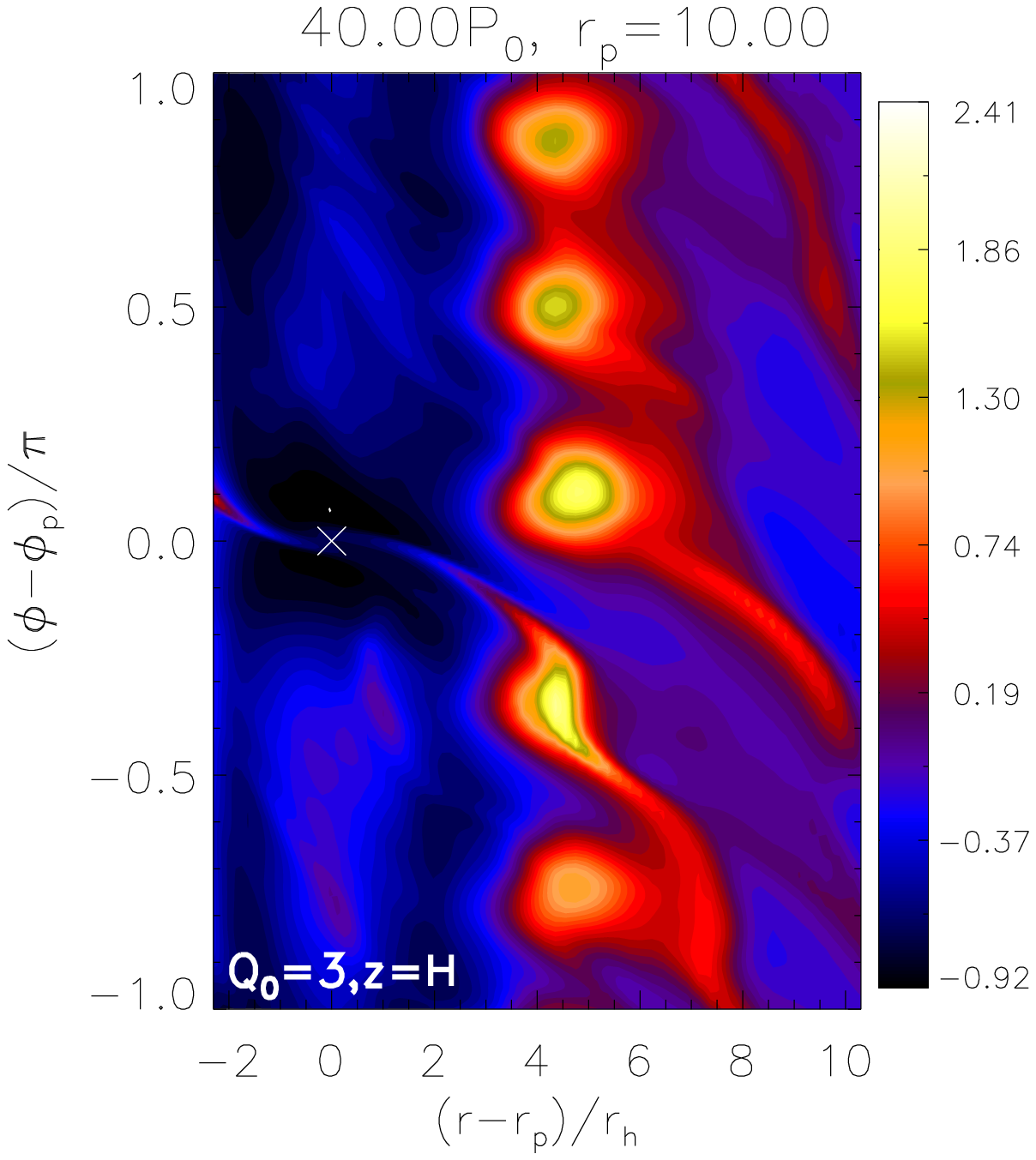}\includegraphics[scale=.27,clip=true,trim=2.30cm    
        0.0cm 0cm 0.97cm]{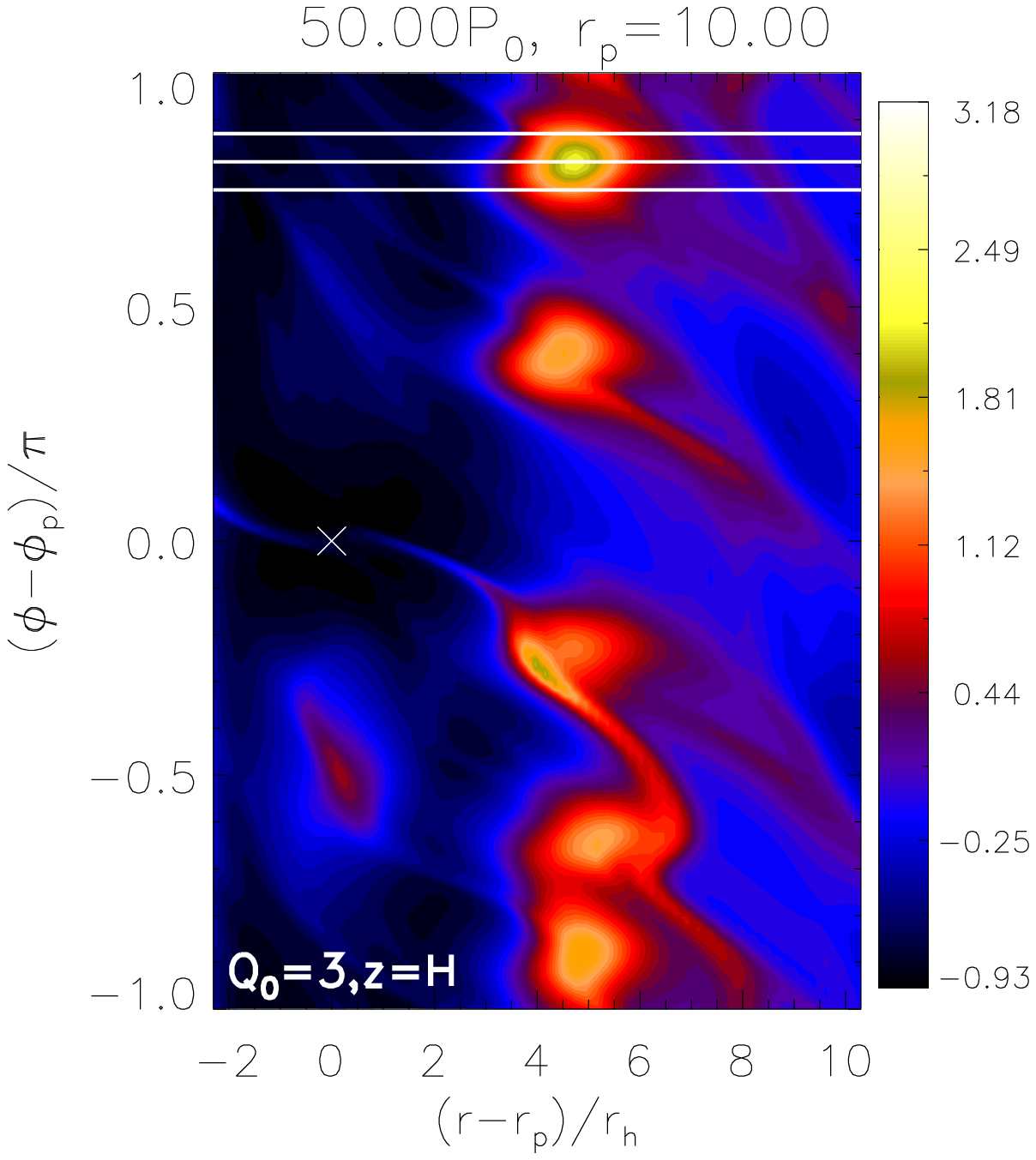}   
  \caption{Development of the vortex instability at gap edges in
    moderately self-gravitating discs. The relative density
    perturbation is shown for Case 2 (top) and Case 3 (bottom) at
    $z=H$. The planet position is marked by the cross. Horizontal
    lines in the plot for $Q_0=3,\,t=50P_0$ indicate azimuths
    taken in Fig. \ref{vortex10_stream}. 
    \label{vortex4_vortex10_overall}}
\end{figure}

We compare the relative density perturbation between Case 2 and
Case 3 in Fig. \ref{vortex4_vortex10_overall}. The snapshots are taken
at $z=H$ but are very similar to razor-thin disc simulations
\citep{li09,yu10,lin11a}. In both Cases the $m=5$ vortex mode is
excited (cf. $m=3$---4 in Case 1). At $t=30P_0$, vortices are just
beginning to  emerge in Case 3, whereas distinct blobs can already
be identified in Case 2, with larger over-densities than vortices in
Case 3. This indicates a stronger instability with decreasing strength
of self-gravity. Vortex merging ensues in Case 2 but does not occur in
Case 3 within the simulation time-scale (unlike Case 1).   
 
For razor-thin discs, \cite{lin11a} have shown that self-gravity has a
stabilising effect against vortex modes with low $m$. This contributes
to favouring higher $m$ modes and hence more vortices with increasing
self-gravity. Fig. \ref{vortex4_vortex10_overall}, together with
results for Case 1, show that this effect persist in
3D. Resisted-merging, seen in 2D models, also occur in 3D. Because
this is due to vortices executing mutual horseshoe turns, only
horizontal self-gravitational forces are important. We do not
expect the vertical dimension to significantly affect 
vortex-vortex gravitational interaction.

\subsubsection{A 3D vortex in Case 3}
Here we examine the vortex in Case 3 marked by horizontal white lines
in Fig. \ref{vortex4_vortex10_overall}. Note that no merging has
occurred. The vortex is radially located
about $r_v \sim r_p + 5r_h$. It has  
azimuthal and radial sizes 
$\Delta\phi_v\sim9h$ and $\Delta
r_v\sim 2.5H$, respectively. Its mass is $M_v\sim 
8.4\times10^{-4}M_*$\footnote{Since no merging has
occurred, $M_v$ can also be estimated by the mass contained in 
an annulus about $r_v$ in the unperturbed disc divided by $m=5$. This 
gives $8\times10^{-4}M_*$ if the annulus width is $2r_h$.}. The 
vortex Hill radius $r_{hv}\sim 0.9H$ is smaller than its
horizontal size but comparable to its vertical size at the vortex
centroid.   

\begin{figure}
  \centering
  \includegraphics[scale=.47,clip=true,trim=0cm 1.35cm .0cm
    0.65cm]{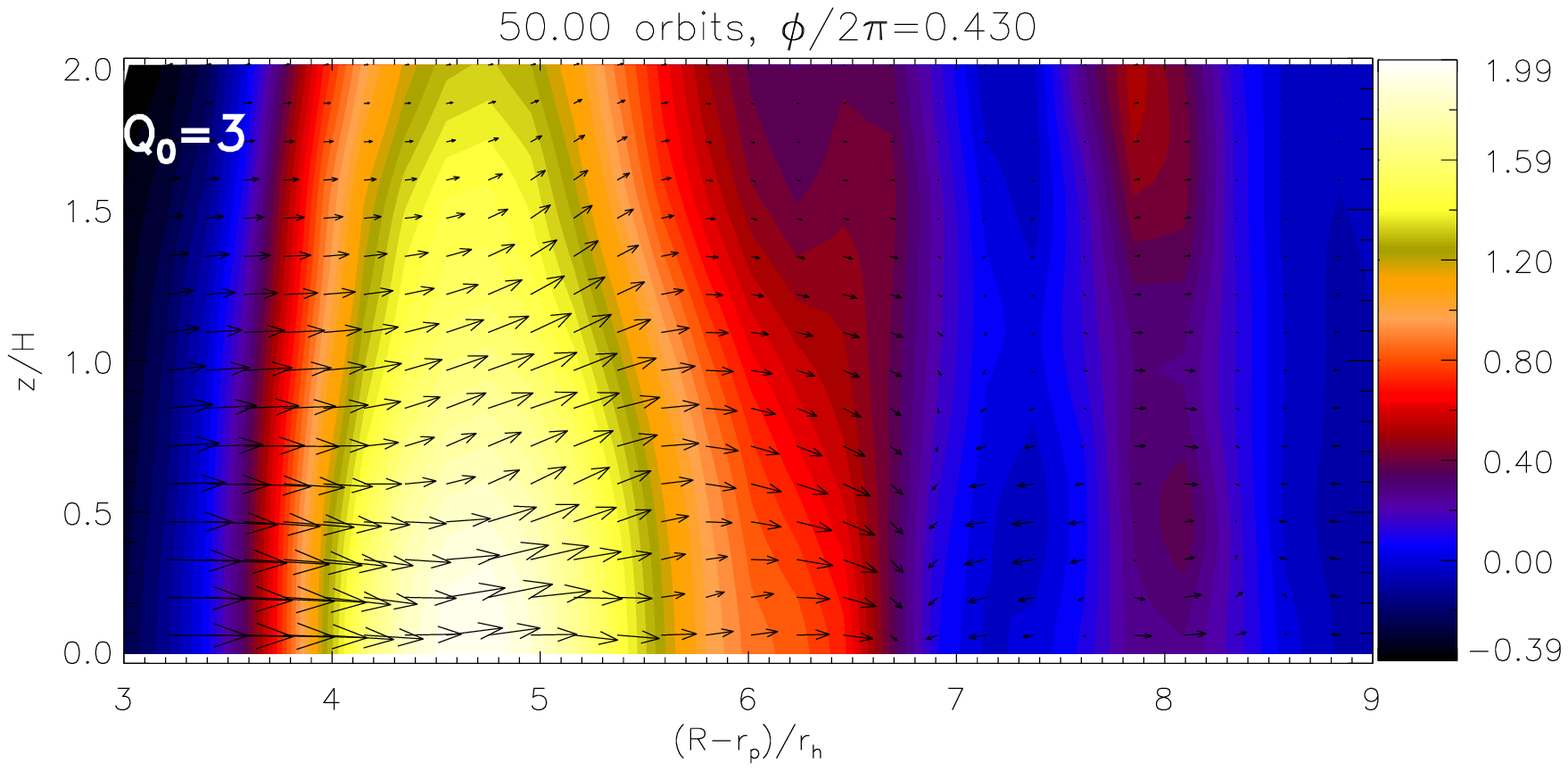}\\\includegraphics[scale=.47,clip=true,trim=0cm 
    1.35cm .0cm
    0.65cm]{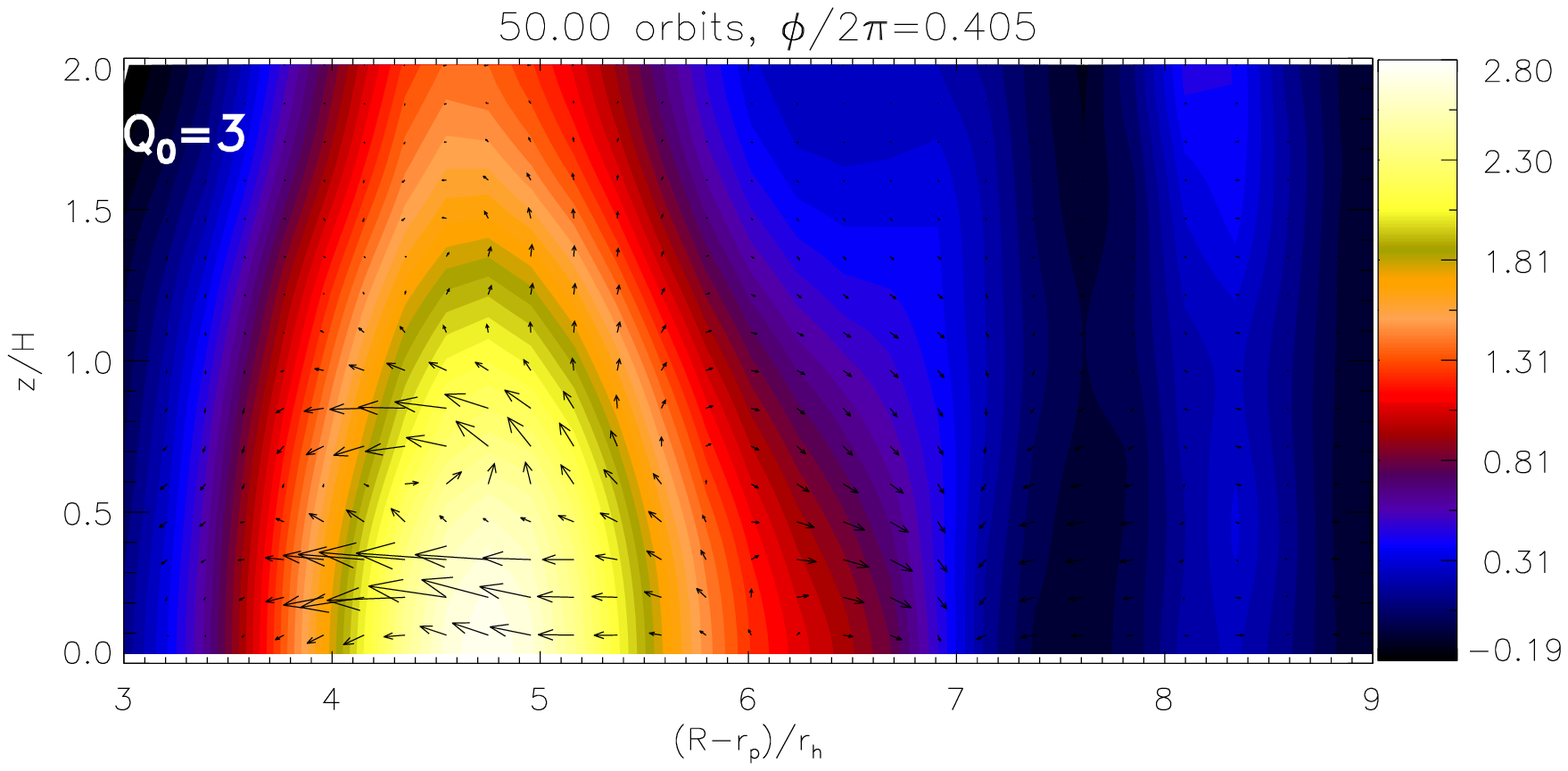}\\\includegraphics[scale=.47,clip=true,trim=0cm 
    0.0cm .0cm
    0.65cm]{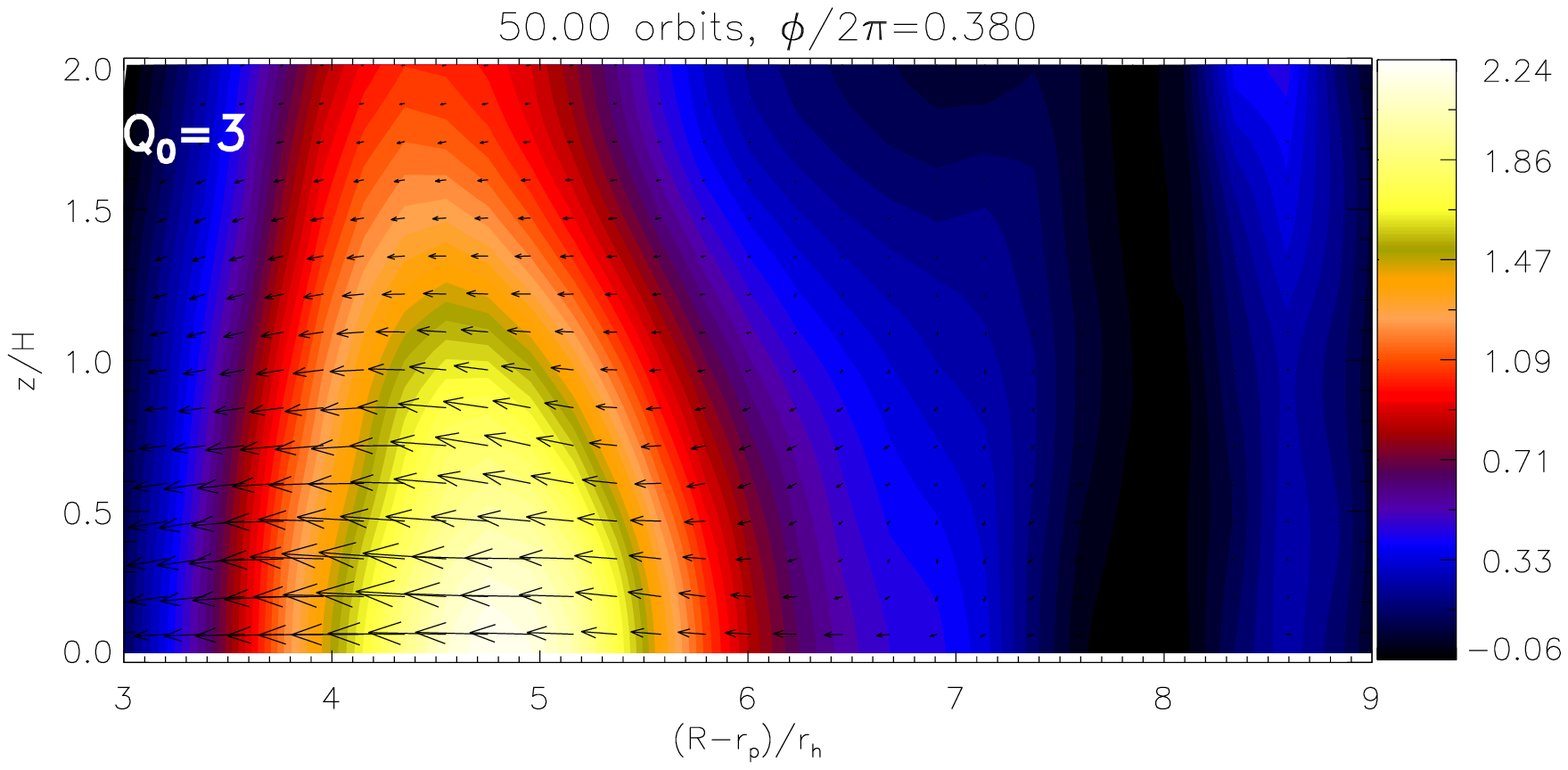}
  \caption{Vertical structure of a vortex in the moderately
    self-gravitating Case 3. The relative density perturbation are
    overlaid by the mass flux vectors $\rho\bm{u}$ projected onto
    this plane. The slices are taken at $t=50P_0$ and azimuths  
    $(\phi-\phi_p)/\pi=0.86$ (top), $0.81$ (middle) and $0.76$
    (bottom). These correspond to the horizontal lines marked
    in Fig. \ref{vortex4_vortex10_overall}.   
    \label{vortex10_stream}} 
\end{figure}

Fig. \ref{vortex10_stream} shows the vertical structure of the
vortex described above. As expected it is more stratified than in 
weakly self-gravitating discs (Fig. \ref{vortex8_stream}), especially
at the vortex centroid where the over-density is maximum. The initial
Keplerian Toomre parameter at $r=r_v$ is $Q=4.3$. A density
enhancement by a factor $\gtrsim 2$ easily gives $Q\lesssim 2$, so
that self-gravity in the perturbed state is dynamically important. 

\begin{figure}
  \centering
  \includegraphics[width=\linewidth]{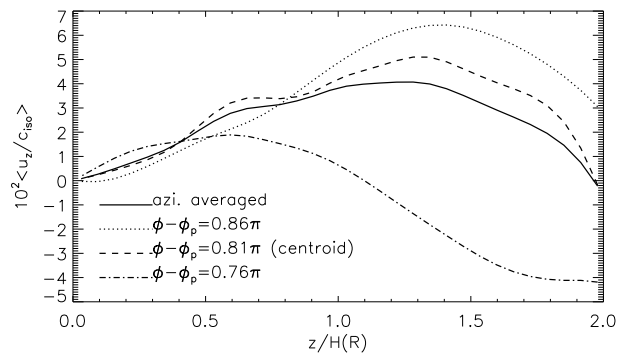} 
  \caption{Vertical velocities $u_z$, normalised by the sound
    speed, for the vortex in the self-gravitating Case 3 shown in
    Fig. \ref{vortex4_vortex10_overall}. The velocities are all
    radially averaged over $r-r_p\in[4.0,5.5]r_h$.
 The solid line is also
    averaged over the vortex azimuthal extent. Azimuthal slices ahead
    of the vortex (dotted), at the centroid (dashed) and behind the
    vortex (dashed-dot) are also shown. 
    \label{vortex10_vprofile}} 
 \end{figure}

Ahead and behind the vortex centroid, the flow field is mostly 
horizontal  (top and bottom panels in Fig. \ref{vortex10_stream}).  
It corresponds to the motion of an anti-cyclonic disc  vortex common
in 2D simulations \citep{li01}. In this plane, vertical motions only become  
significant compared to $u_R$ close to the vortex centroid (middle
panel). We plot the average $u_z$ for the vortex 
in Fig. \ref{vortex10_vprofile}. Azimuthal slices at the vortex centroid,
behind it and ahead of it are also shown. The average vertical
velocity is positive (solid line). This is qualitatively different to  
the merged vortex in the weakly self-gravitating Case 1
(Fig. \ref{vortex8_stream}). We
have examined other vortices but could not identify a `typical'
vertical flow structure (c.f. anti-cyclonic horizontal
flow is generic). Although the vertical velocities appear 
to display a range of behaviour, the vertical Mach number is still
very small.


\begin{figure}
  \centering
  \includegraphics[scale=.47,clip=true,trim=0cm 1.4cm .0cm
    0.6cm]{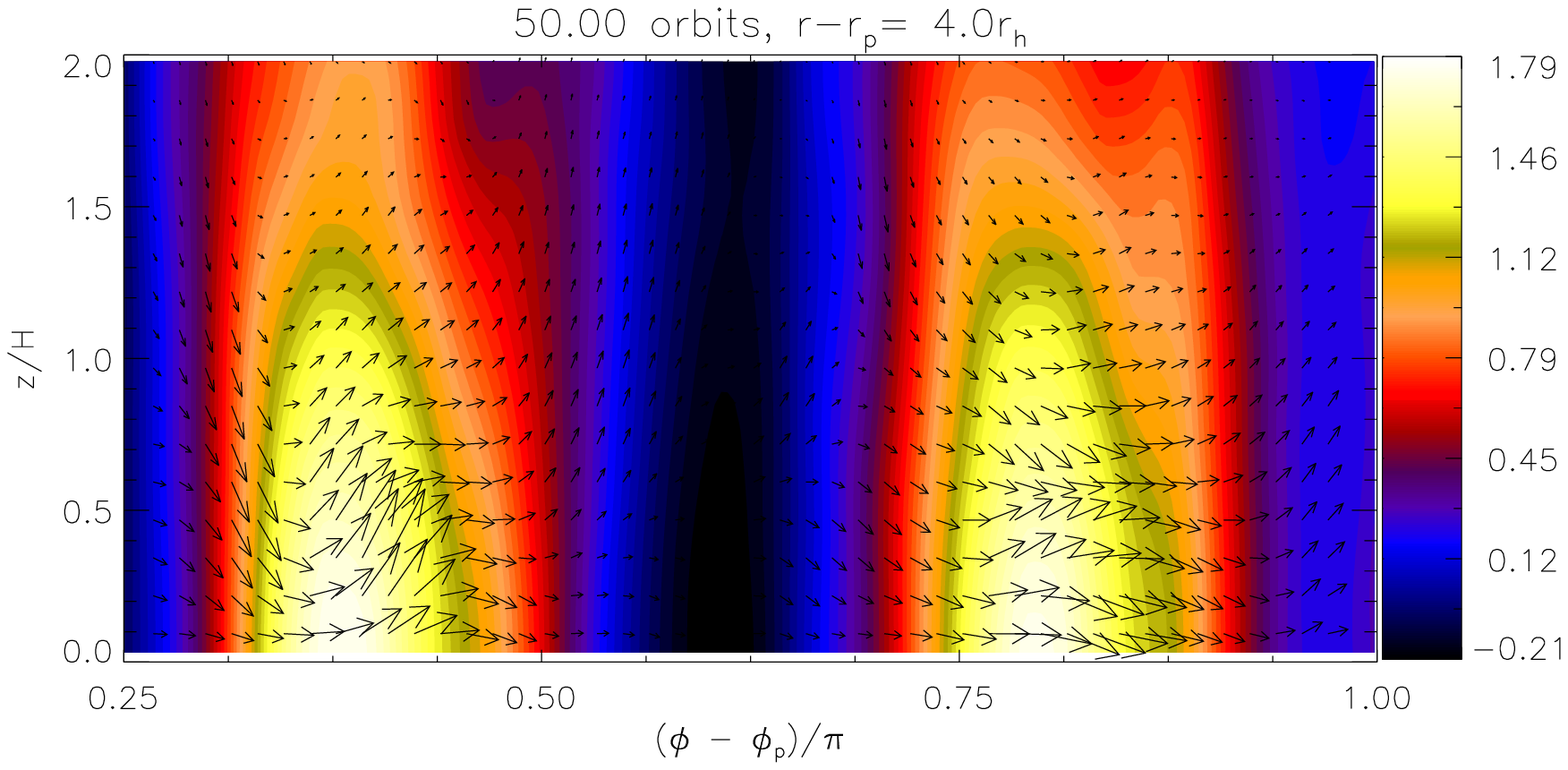}\\\includegraphics[scale=.47,clip=true,trim=0cm 
    1.4cm .0cm
    0.6cm]{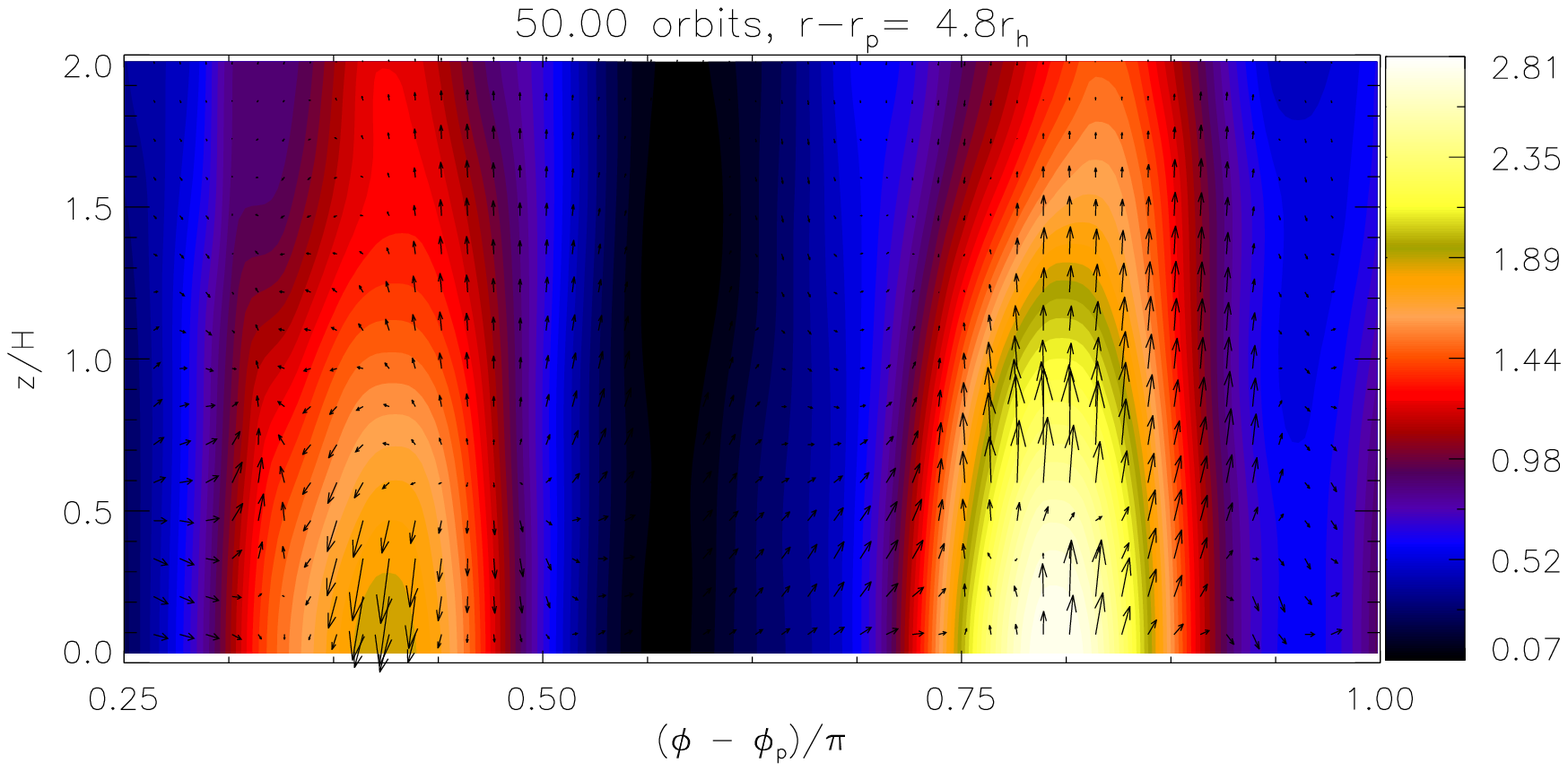}\\\includegraphics[scale=.47,clip=true,trim=0cm 
    0.0cm .0cm
    0.6cm]{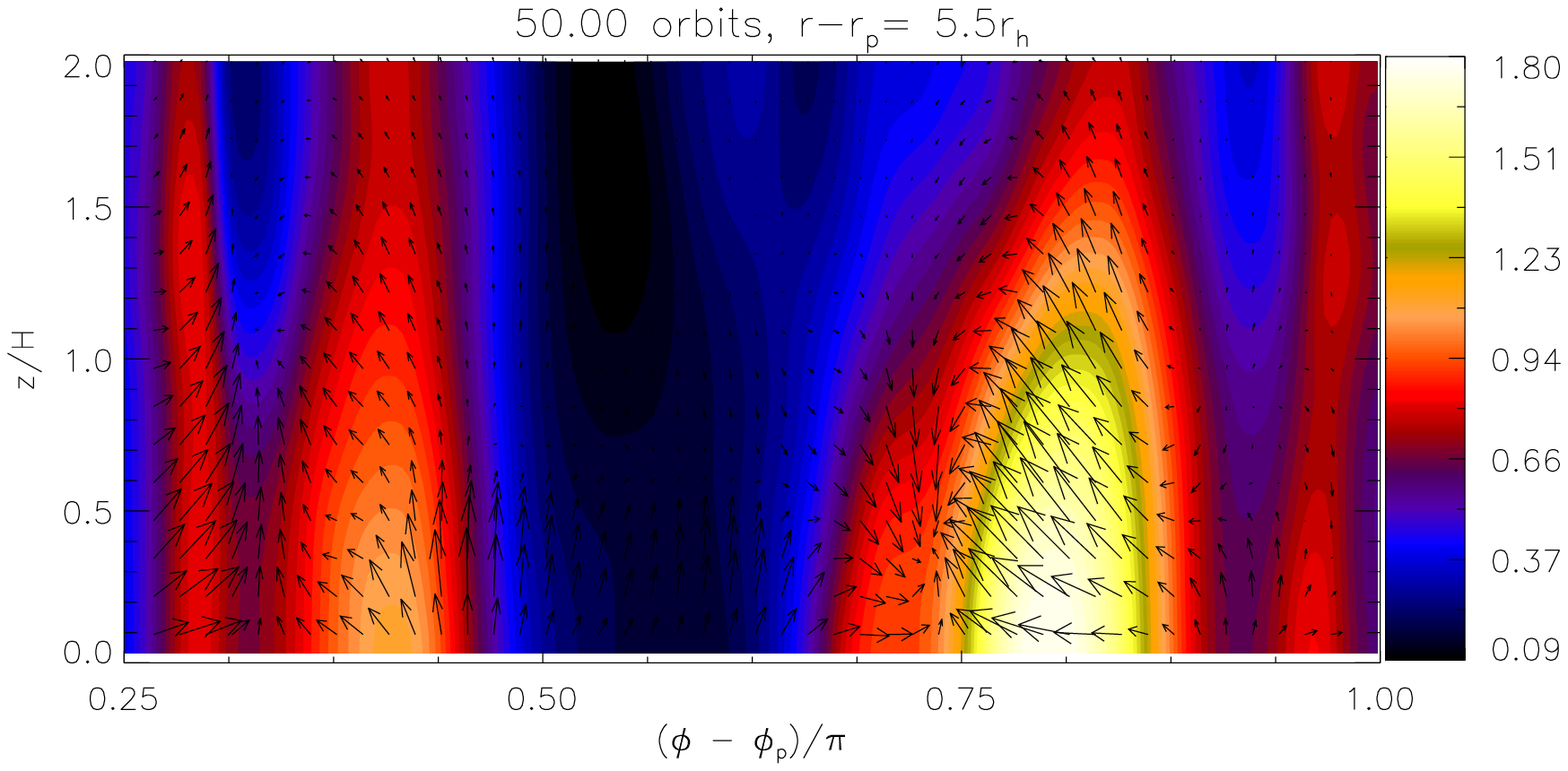}
  \caption{Azimuthal vortex structures in the 
    self-gravitating Case 3 at $t=50P_0$. The background flow is from
    left to right. 
    The relative density perturbation are
    overlaid by the perturbed mass flux vectors $\rho(\bm{u} -
    R\Omega_k\hat{\bm{\phi}} )$ projected onto this plane.  
    The slices are taken at radii 
    $(r-r_p)/r_h=4.0$ (top), $4.75$ (middle) and $5.5$
    (bottom). These correspond to the flow just interior, at, and just
    exterior to the vortex centroid on the right. 
\label{vortex10_stream_pz}} 
\end{figure}

We also find the vortices have height-dependent azimuthal 
structure. Fig. \ref{vortex10_stream_pz} shows several slices in
the $\phi z$ plane. The choice of radii for these plots were based on 
the vortex described above (corresponding to the right vortex in the
figure). The vortices are more columnar closer to the gap edge 
($r-r_p=4r_h$, top slice). Moving away from the gap edge, the vortices
develop front-back asymmetry and are thinner with increasing
height ($r-r_p=5.5r_h$). 

The increased three-dimensional structure away from the gap edge could
be related to density waves emitted by the vortex
\citep{paardekooper10}. Back-reaction of these waves on the vortex
vertical structure, if any, is expected to be weaker on the side of
the vortex adjacent to the low-density gap edge (top slice in
Fig. \ref{vortex10_stream_pz}). 

The perturbed azimuthal velocities away from the vortex centroids again 
follow the expected anti-cyclonic motion in the horizontal plane (being
positive interior and negative exterior to the centroid). However,
unlike the previous $Rz$ plots, here the vertical velocities can often
be comparable or larger than the perturbed azimuthal velocities
(e.g. $r-r_p=5.5r_h$, bottom panel). Of course, the total azimuthal
velocity is supersonic and therefore much larger than vertical
velocities. As remarked above, the vortices do not share the same 
structure. For example, the left vortex centroid involves negative
vertical velocity while $u_z>0$ for the right vortex ($r-r_p=4.75r_h$,
middle panel). 


\subsubsection{Case 4 and Case 5}  
Cases 4---5 are additional examples of the
vortex instability in 3D discs with smaller $h$ ($=0.05$) than the
above runs (with $h=0.07$).  However, initial Toomre
profile is nearly independent of $h$, so the strength of self-gravity
is unchanged. We also used a smaller planetary mass, $q=10^{-3}$ so
the ratio $q/h^3=8$ is not significantly larger than Cases 2---3
($q/h^3=5.8$). We checked that prior to instability, the outer gap
edge profiles of these cases are similar.

We found the instability grows slower with $h=0.05$ as vortices become
identifiable at $t=35P_0$, compared to $t=30P_0$ for Cases 2 and 3. A
decrease in growth rate with sound-speed (which is proportional to
$h$) was already noted in 2D \citep{li00}. Thus, the effect of
sound-speed on the vortex instability remain unchanged by the 3D
geometry.      

\begin{figure}
  \centering
  \includegraphics[scale=.43,clip=true,trim=0cm 1.84cm 1.9cm 
    0cm]{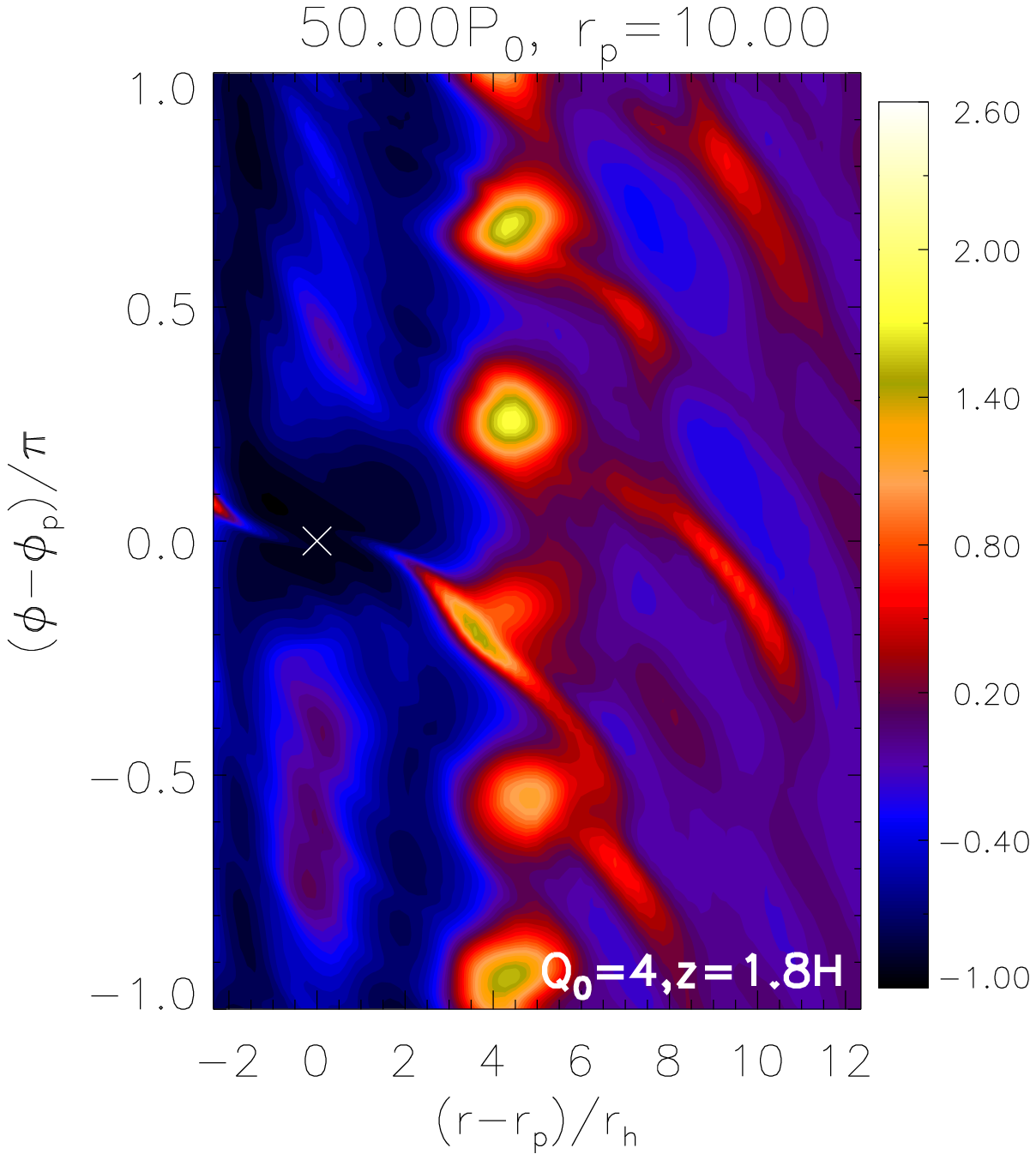}\includegraphics[scale=.43,clip=true,trim=2.3cm   
    1.84cm 0cm 
    0cm]{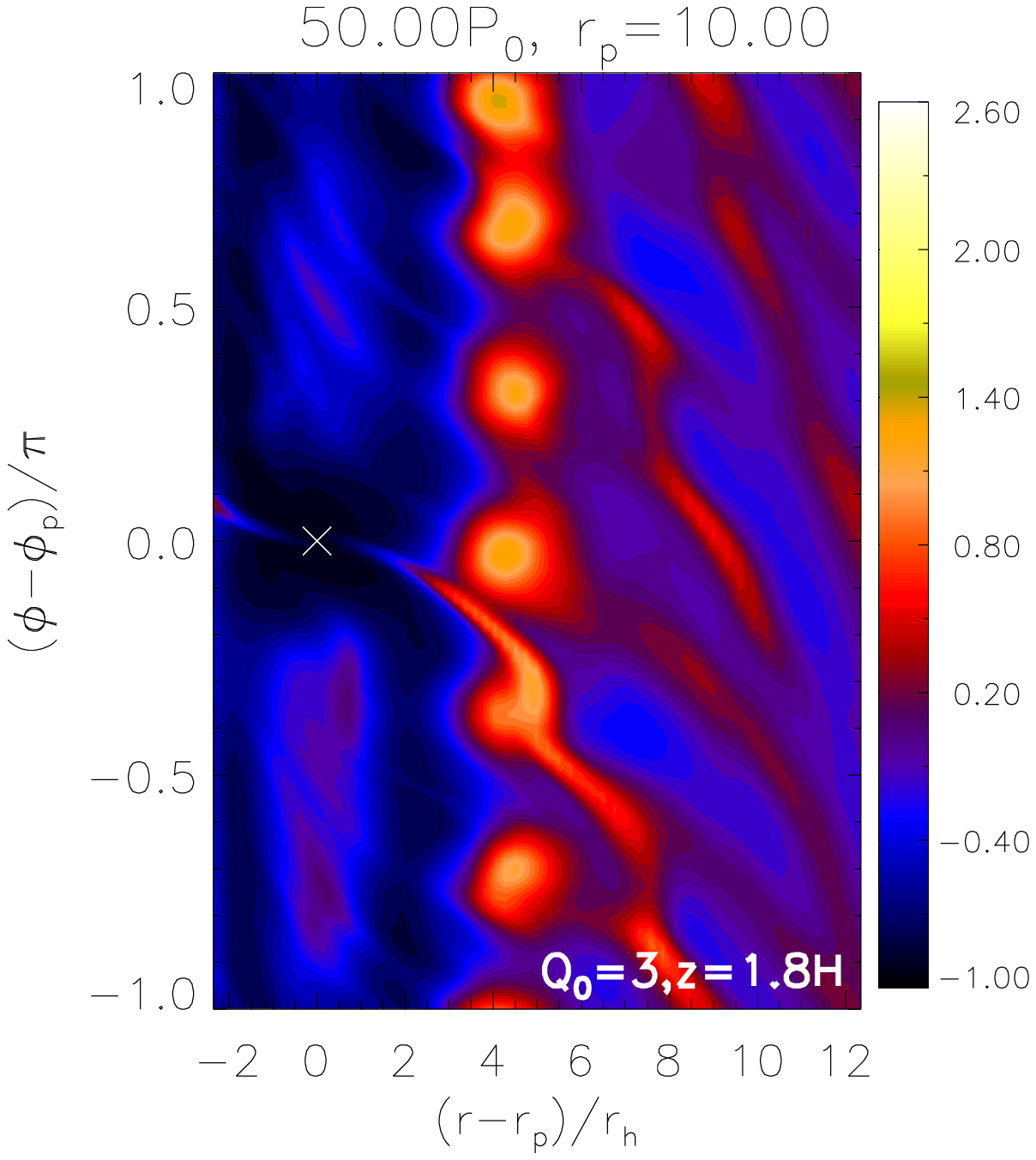}
  \caption{Vortex instability in Case 4 (left) and Case 5 (right). The
    relative density perturbation near the upper vertical boundary
    . These runs are
    similar to Cases 2---3, except a colder disc is employed
    ($h=0.05)$ and a smaller planetary mass is used to open the gap
    $q=10^{-3}$. 
    \label{vortex2_vortex2b}}
\end{figure}

Fig. \ref{vortex2_vortex2b} show several
snapshots of Case 4 and Case 5 at $t=50P_0$ near the upper disc
boundary. As before, high in the atmosphere the density enhancement
is weaker with increasing self-gravity. Consistent with 2D simulations, 
the more self-gravitating Case 5 develops the $m=6$ vortex mode
whereas $m=5$ vortices develop in Case 4 with weaker
self-gravity. Merging is strongly resisted in these runs. In fact,
Case 5 was extended to $t=135P_0$ and only one vortex-pair merged.   

 
\subsection{Edge-spiral modes in massive discs}

\cite{lin11a,lin11b} found that as the strength of self-gravity is
increased, instability eventually shifts from localised vortices to 
low $m$ global spirals extending from the outer gap edge to the outer
disc. The vortex instability is suppressed because they are associated
with  vortensity minima (which we will check in \S\ref{vortensity}).  
\cite{lin11a} gives a simple energy argument which show that such 
association is not possible for sufficiently strong 
self-gravity. Association with vortensity maximum is favoured
instead.  These are the spiral instabilities. \citeauthor{lin11b}
called them \emph{edge} modes since they can still be considered 
associated with the gap edge.   

\begin{figure*}
  \centering
 \includegraphics[scale=.39,clip=true,trim=.86cm 0.6cm 2.4cm 
    .7cm]{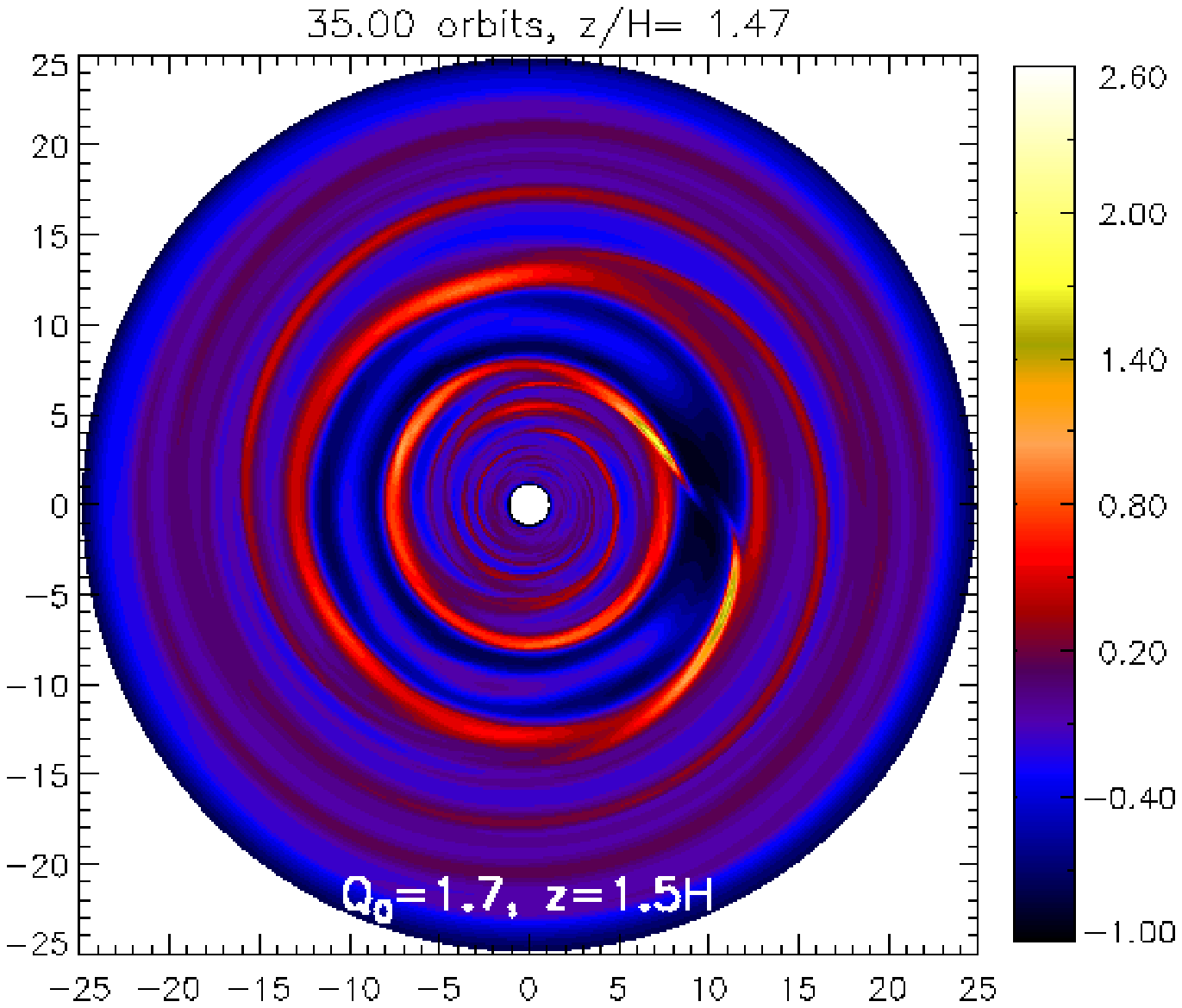}\includegraphics[scale=.39,clip=true,trim=0.75cm 
    0.6cm 2.4cm
    0.7cm]{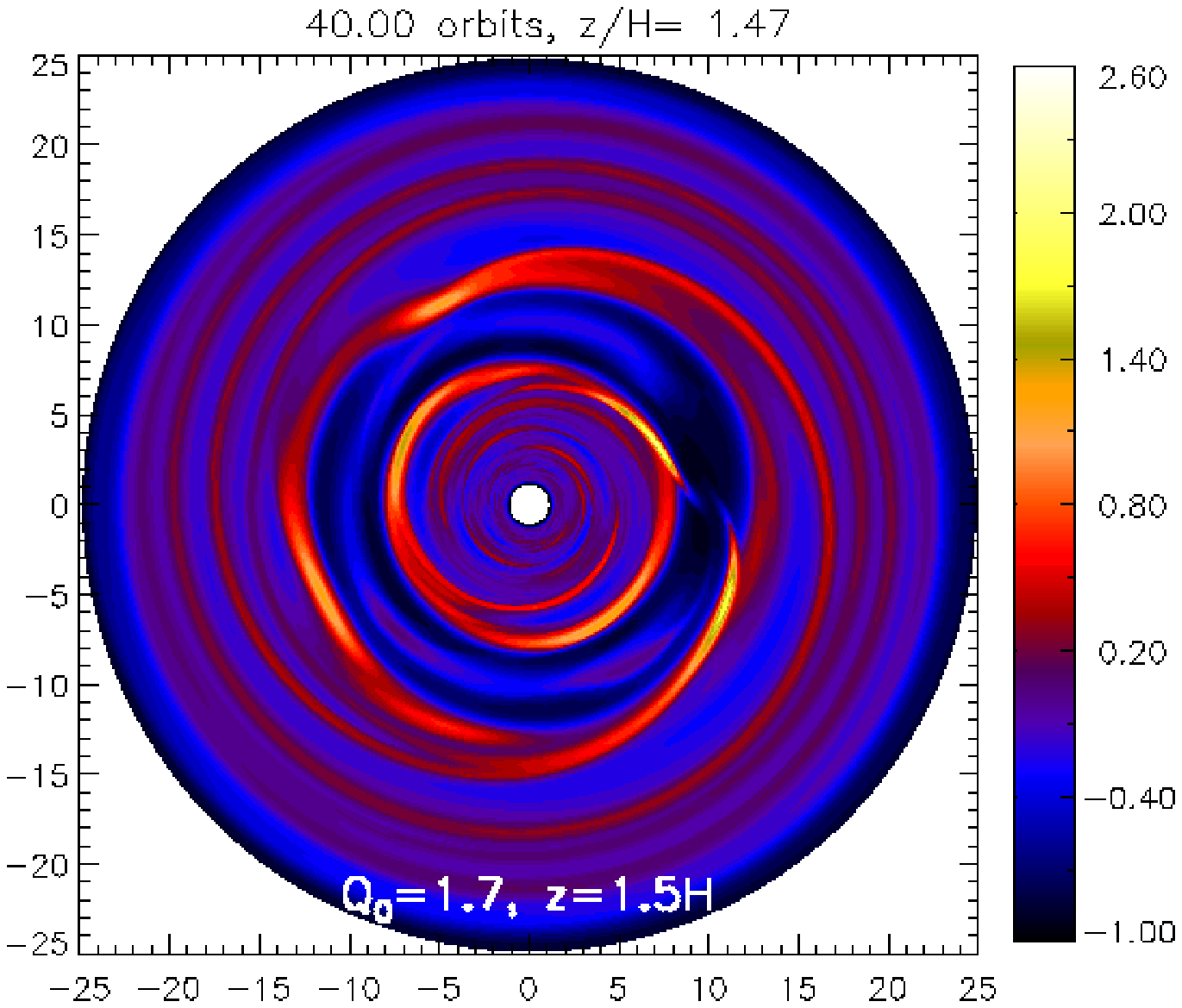}\includegraphics[scale=.39,clip=true,trim=0.75cm  
    0.6cm 2.4cm
    0.7cm]{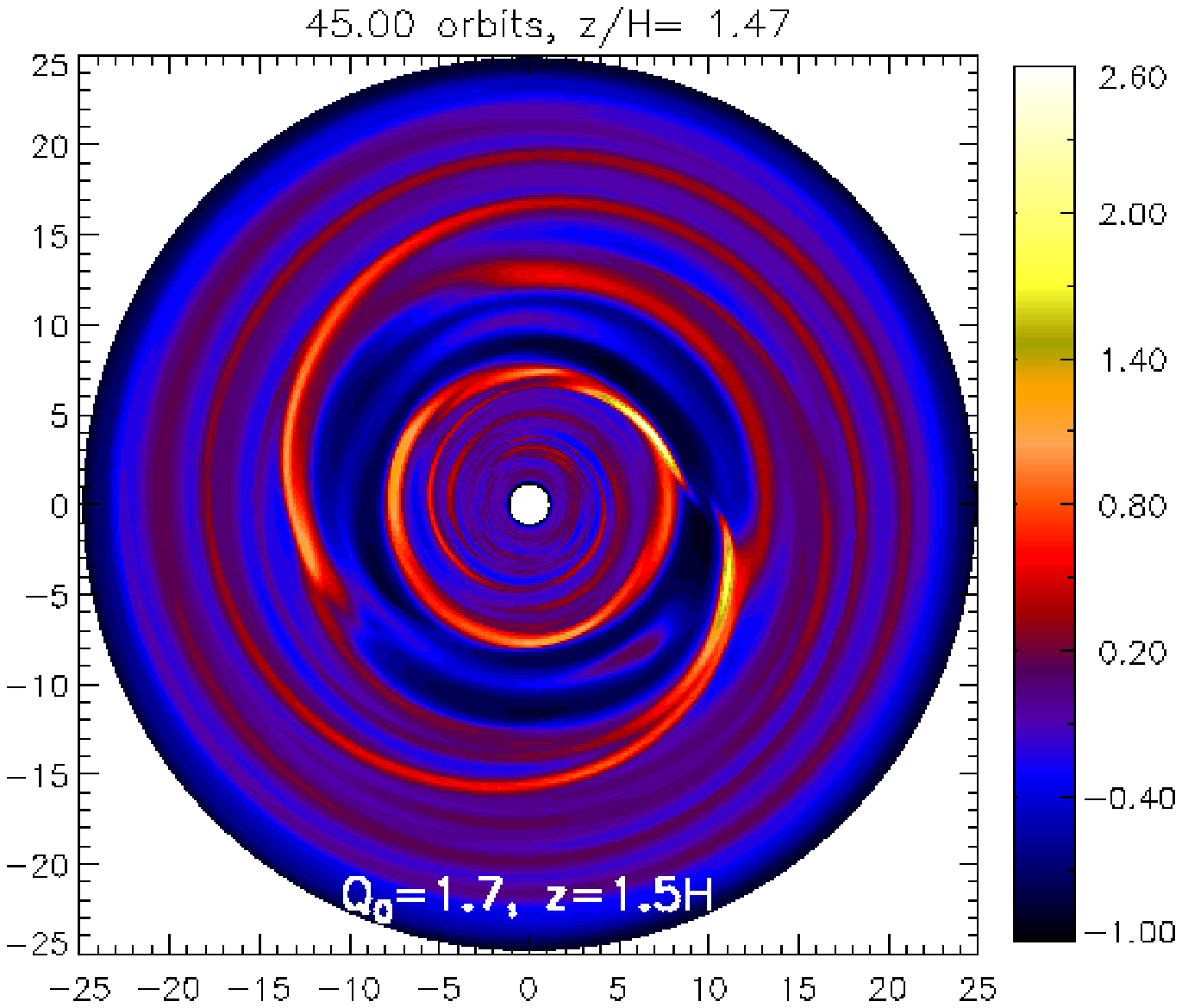}\includegraphics[scale=.39,clip=true,trim=0.75cm
    0.6cm 0cm .7cm]{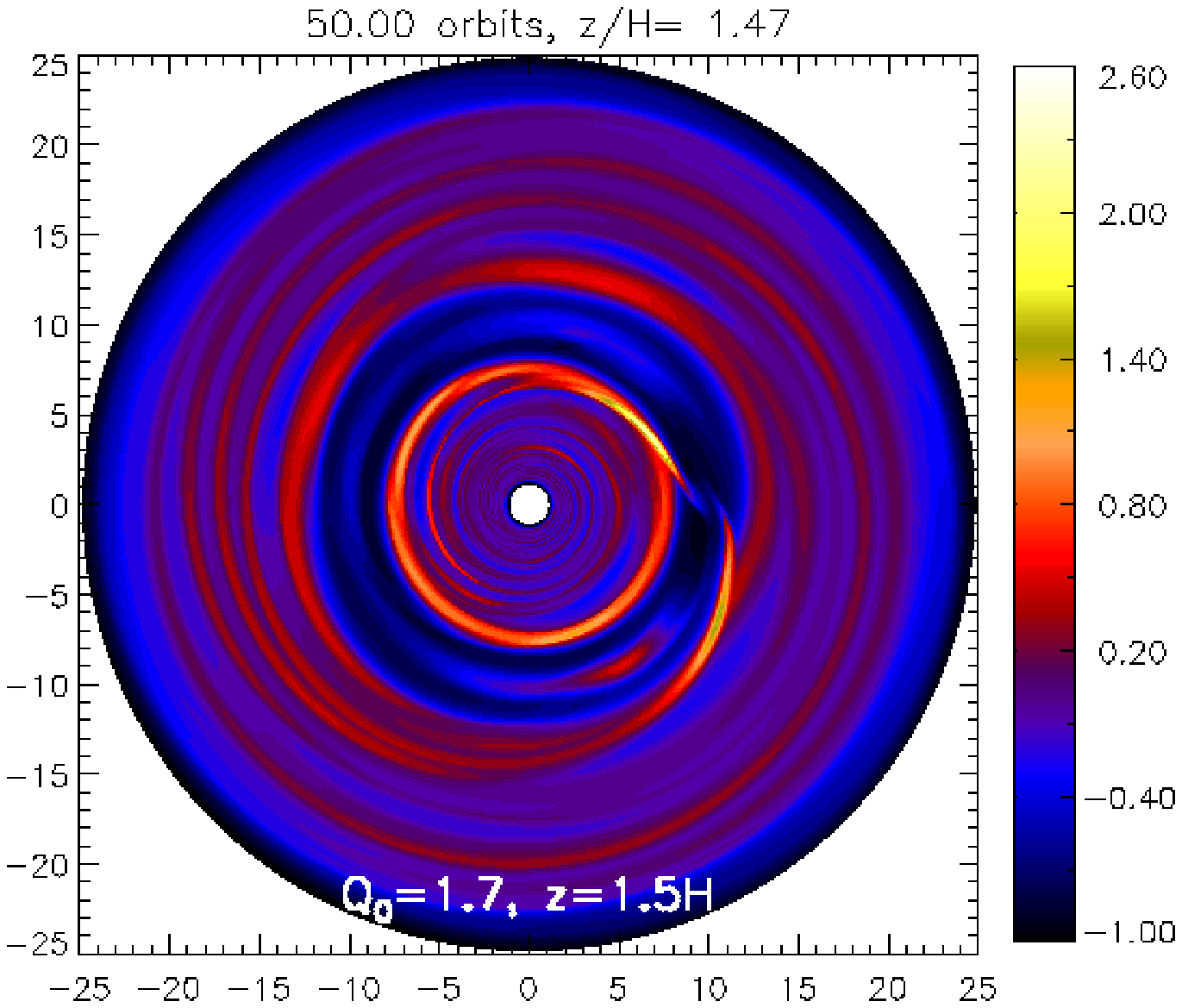}\\ 
  \includegraphics[scale=.39,clip=true,trim=.86cm 0.6cm 2.4cm 
    .7cm]{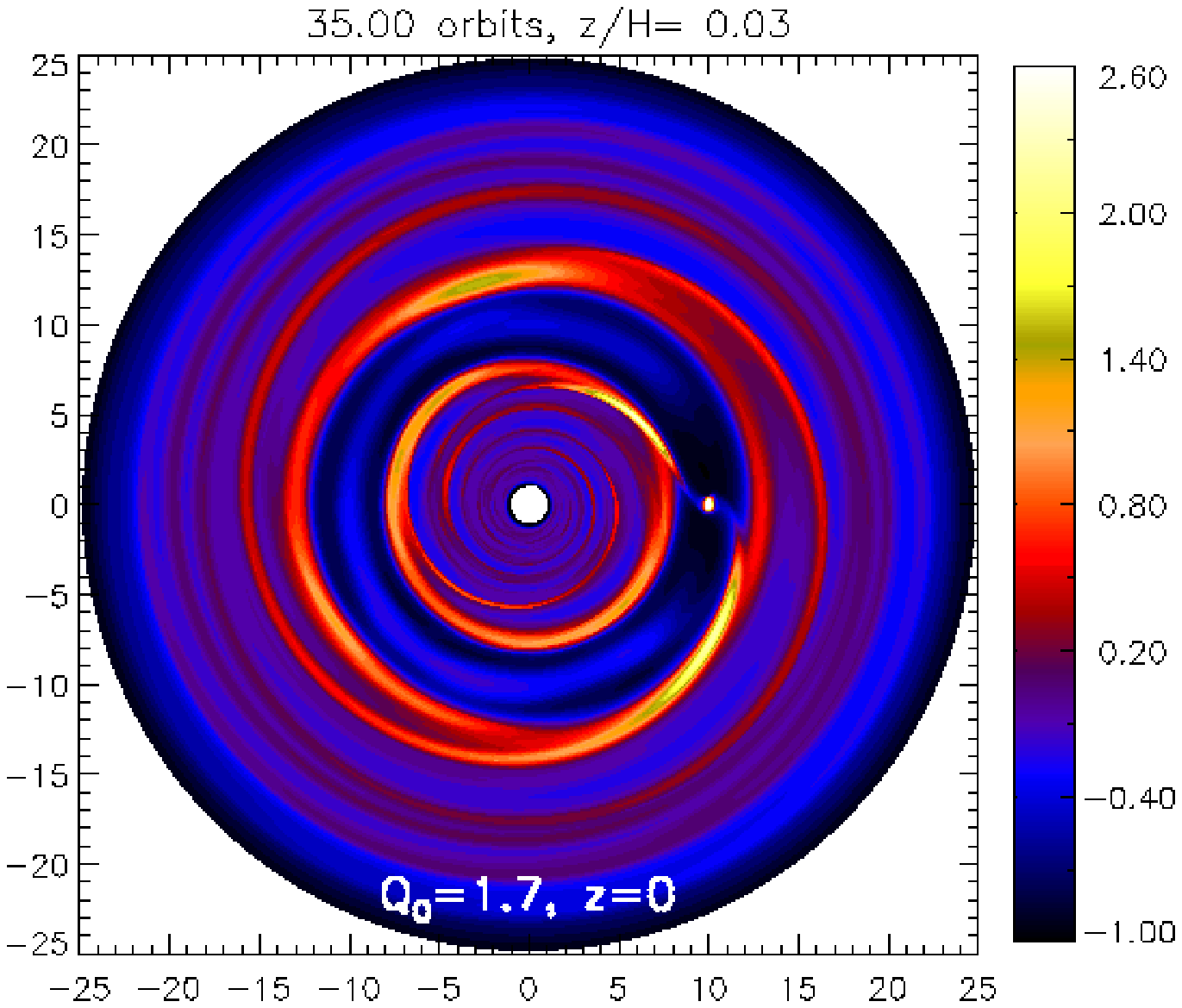}\includegraphics[scale=.39,clip=true,trim=0.75cm 
    0.6cm 2.4cm
    0.7cm]{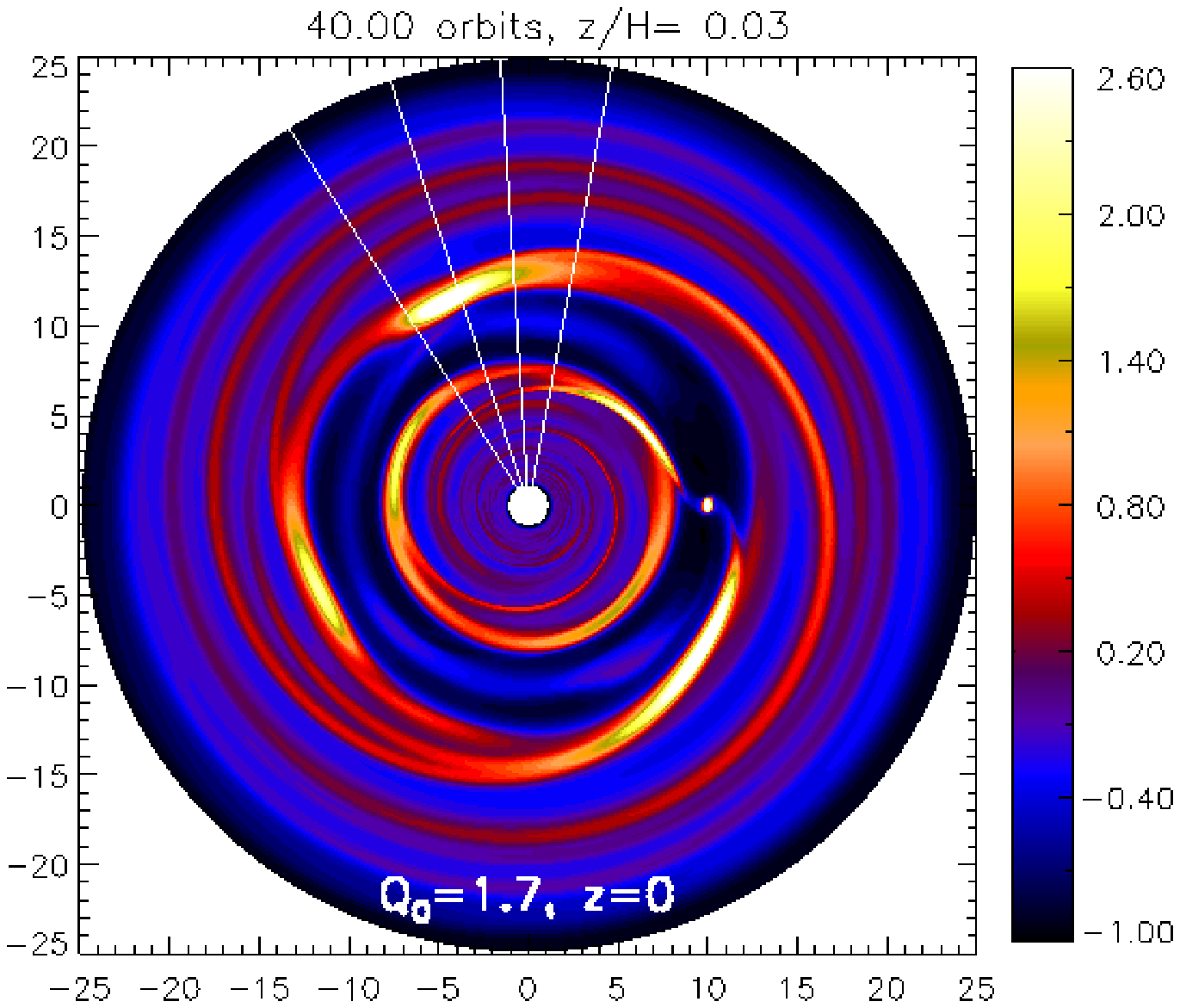}\includegraphics[scale=.39,clip=true,trim=0.75cm  
    0.6cm 2.4cm
    0.7cm]{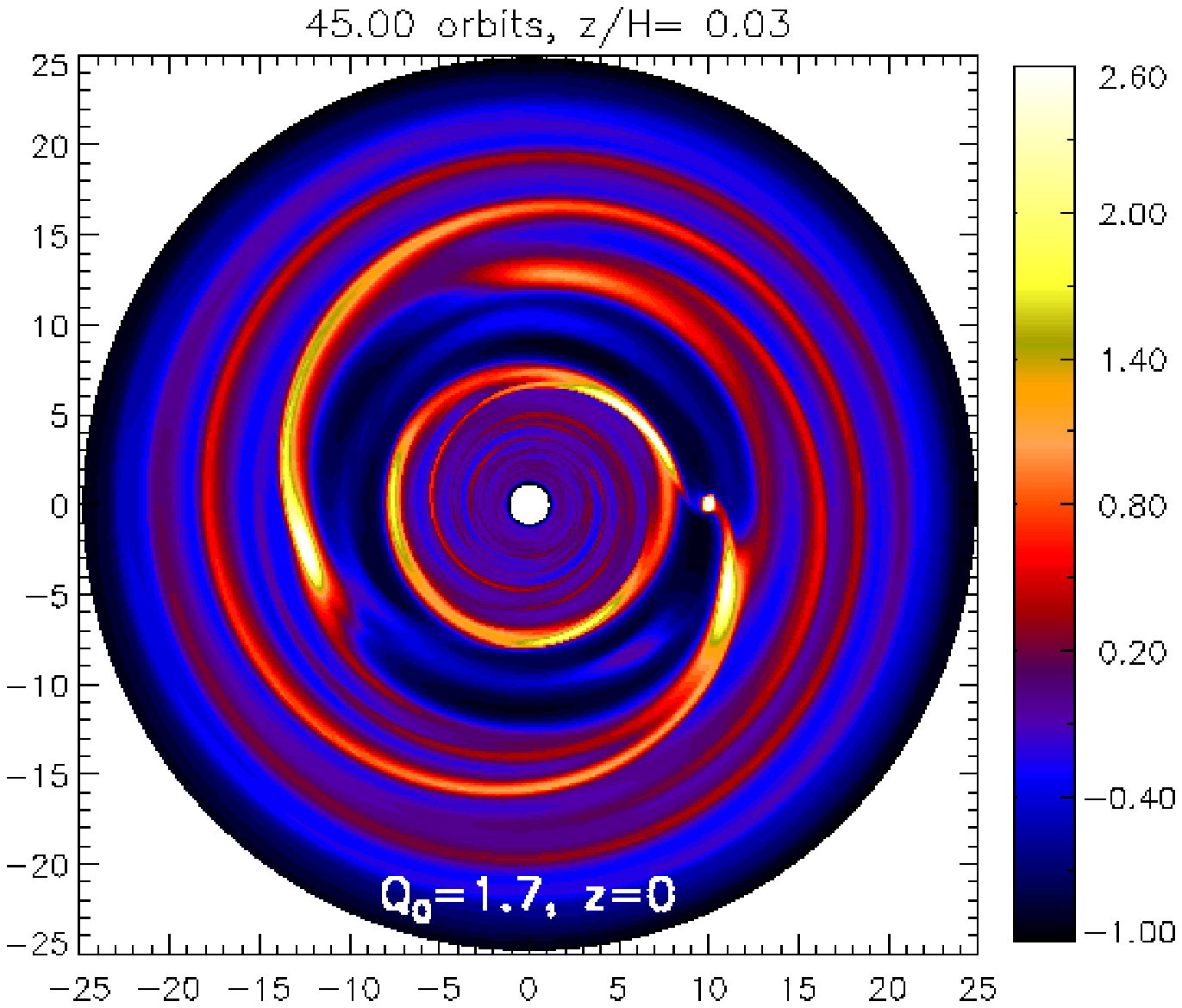}\includegraphics[scale=.39,clip=true,trim=0.75cm
    0.6cm 0cm .7cm]{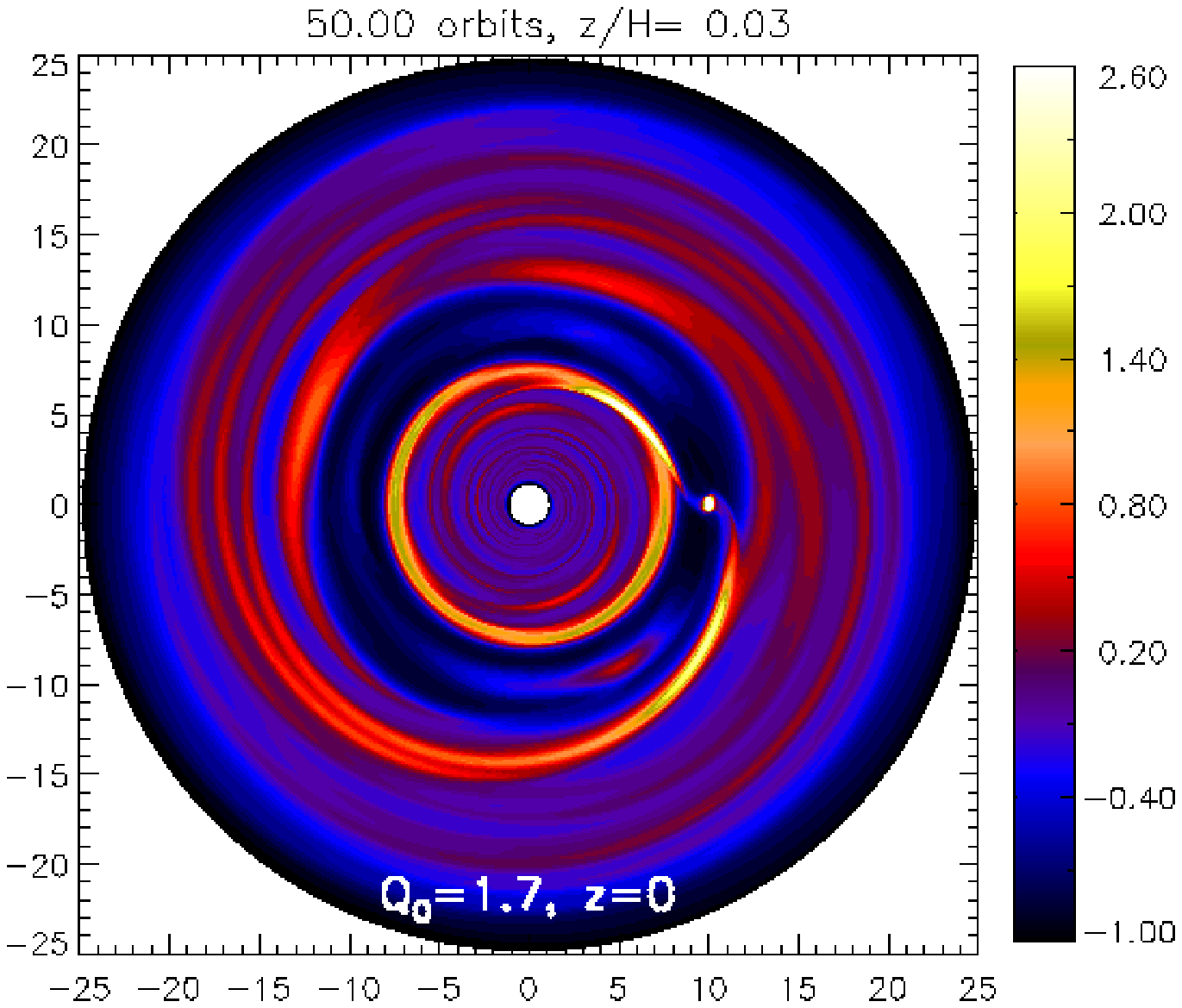} 
  \caption{Case 6: development of the edge mode instability at the gap
    edge of a Jovian mass planet. The relative density perturbation 
    in the midplane (bottom) and in the atmosphere (top) is shown at
    times (left to right) $t=35P_0,\,40P_0,\,45P_0,\,50P_0$. White
    lines indicate azimuthal cuts taken in Fig. \ref{vortex7_stream}. 
    \label{vortex7_polar_dens}}
\end{figure*}

We begin to observe edge modes in our 3D models when $Q_0=1.7$ (Case
6)\footnote{Note that the transition from vortex to spiral modes with
  decreasing $Q_0$ is not abrupt. It is possible 
  to have a mixture.}. Its evolution is depicted in
Fig. \ref{vortex7_polar_dens}.  
A $m=2$---3 disturbance develops at the outer gap edge at
$t=35P_0$ and induces spiral density waves in the exterior disc through
self-gravity. Interaction between the edge disturbance
and the wider disc leads to the global spiral pattern seen at
$t=40P_0$ and $t=45P_0$. This coupling is necessary for instability. 
The edge mode is associated with a local $\mathrm{max}(Q)$
just inside the unperturbed outer gap edge. In Case 6,
$\mathrm{max}(Q)\sim 8$ so a
\emph{local} gravitational instability is not possible. However, we 
can still consider the global edge mode as being composed of two
parts: an edge disturbance where density perturbation is largest, and
the spiral arm it induces.

Comparison between the two heights in Fig. \ref{vortex7_polar_dens}
show that edge modes are significantly vertically stratified. Most of
the density perturbation is confined near the midplane. Unlike  
vortex modes the spirals appear transient. Their 
amplitudes are much reduced by  $t=50P_0$, but are still visible.
This is likely a radial boundary effect. The transition to a low
density annulus  towards the outer disc edge is a result of the
standard outflow boundary condition applied there. While the linear 
edge mode instability is insensitive to boundary conditions, its long
term evolution is affected. The outer disc edge can reflect waves back to
to gap edge to stabilise it, causing saturation \citep{lin11b}.    

\subsubsection{Vertical structure of an edge mode}  
\begin{figure}
  \centering
  \includegraphics[scale=.47,clip=true,trim=0cm 1.35cm .0cm
    0.65cm]{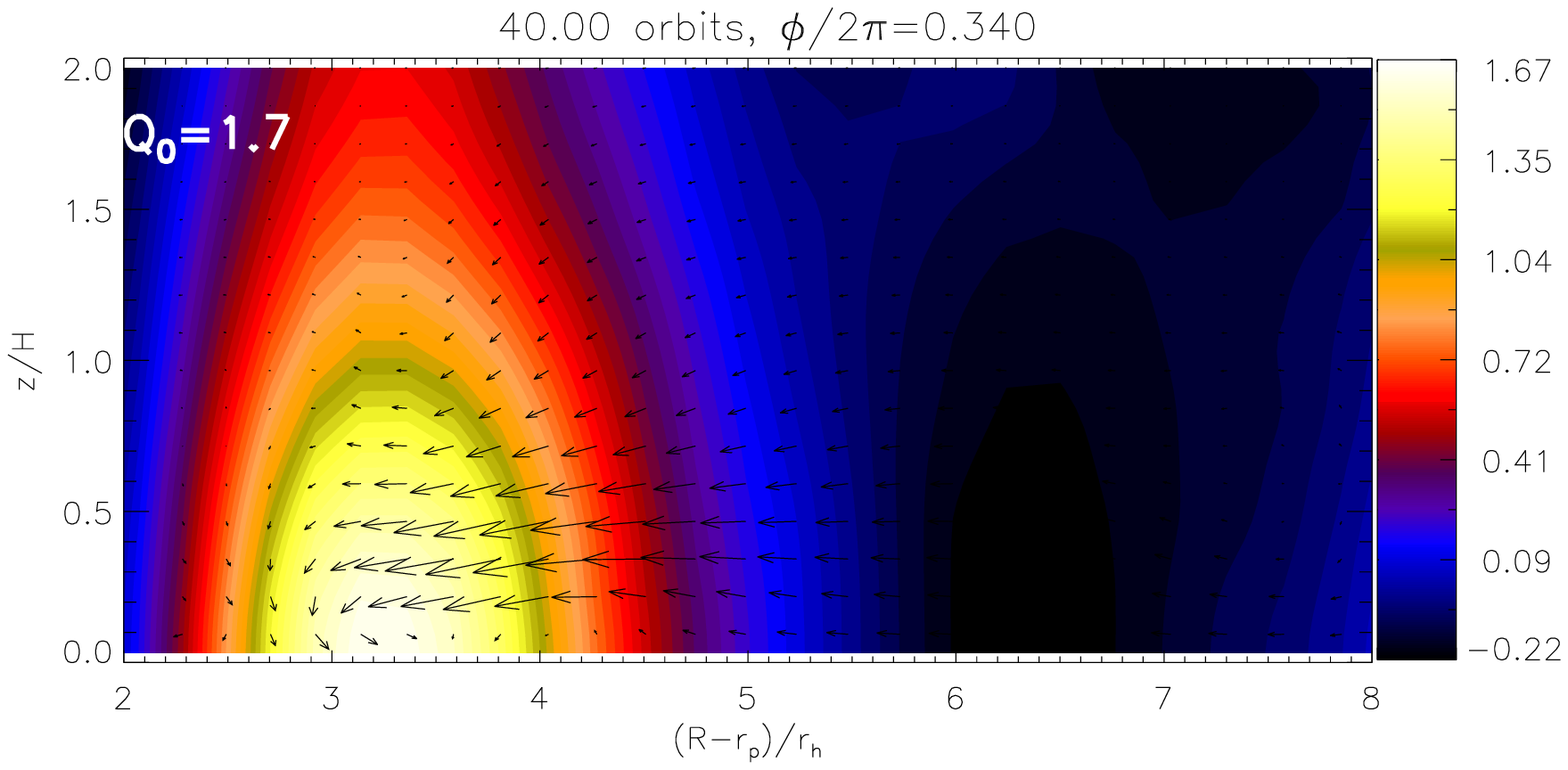}\\\includegraphics[scale=.47,clip=true,trim=0cm 
    1.35cm .0cm
    0.65cm]{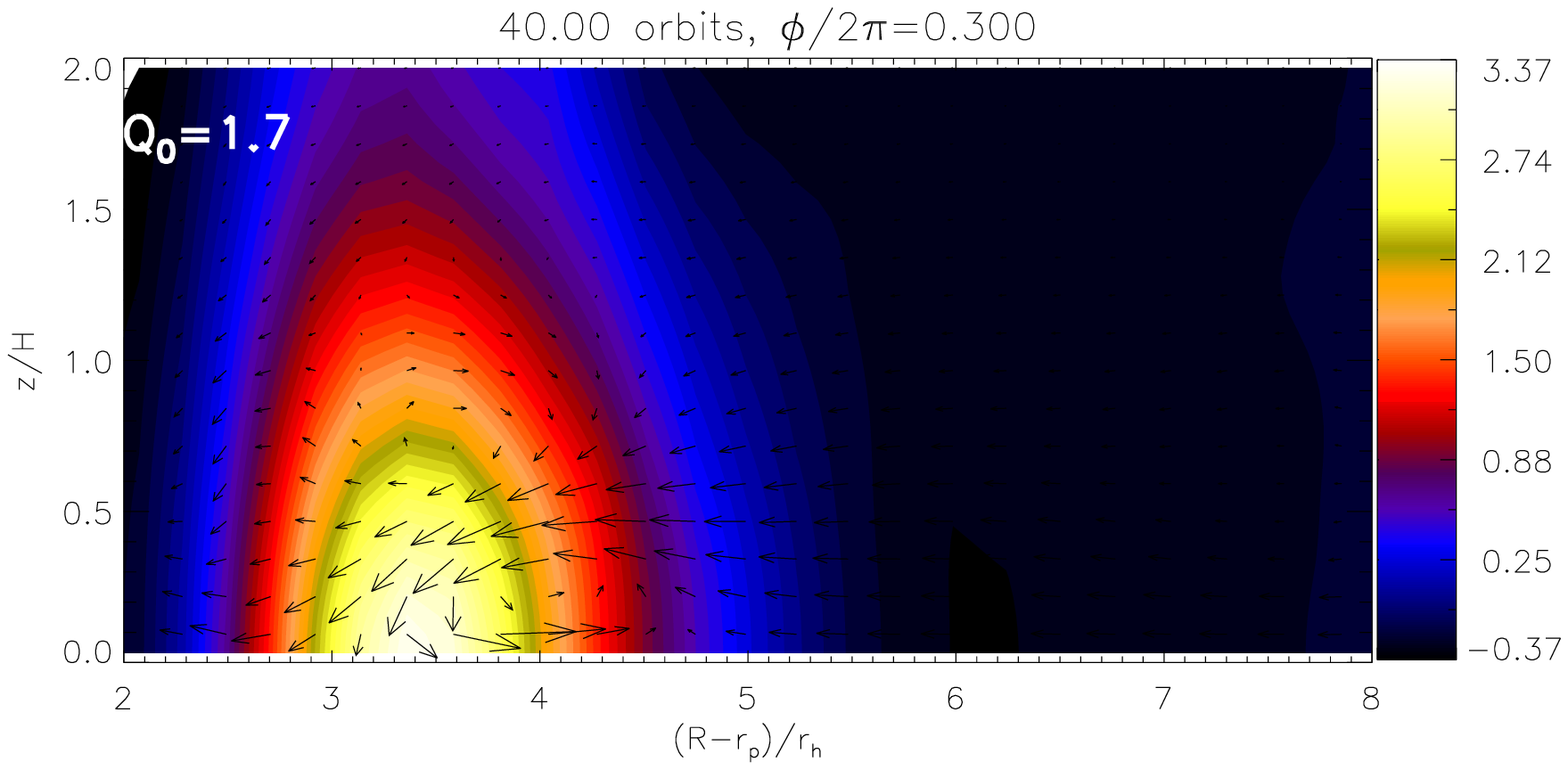}\\\includegraphics[scale=.47,clip=true,trim=0cm 
    1.35cm .0cm
    0.65cm]{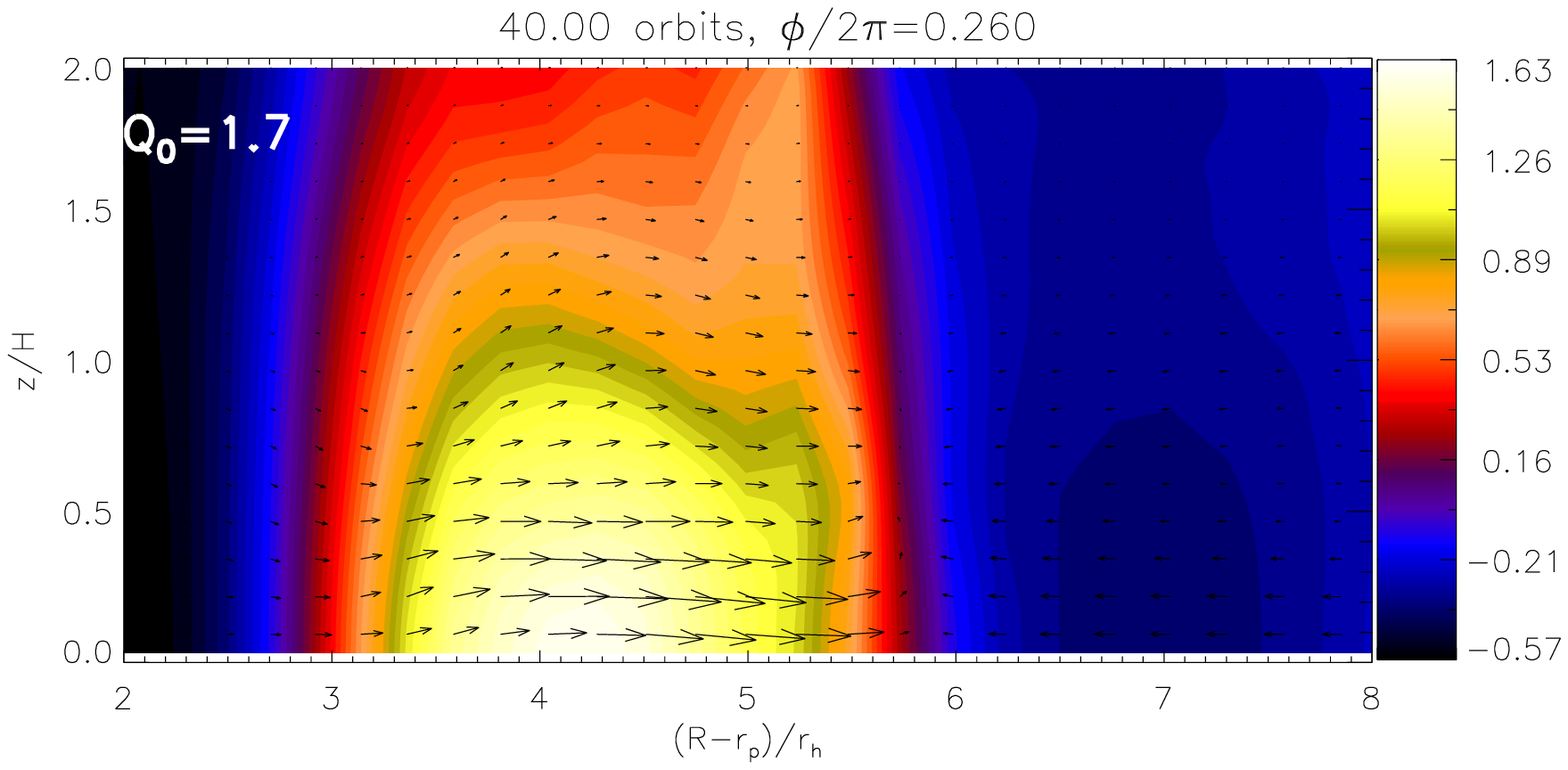}\\\includegraphics[scale=.47,clip=true,trim=0cm  
    0.0cm .0cm
    0.65cm]{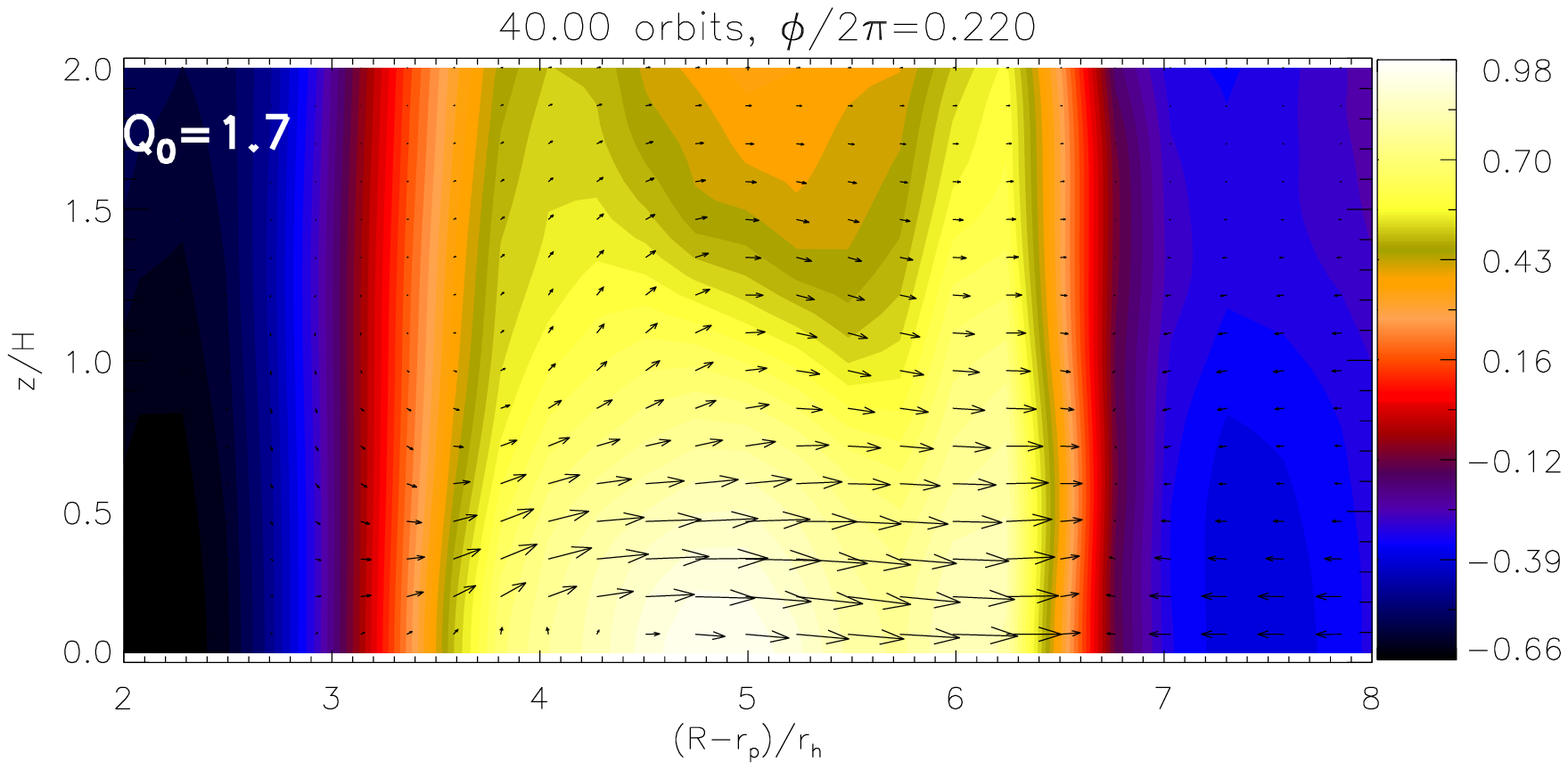}
  \caption{Vertical structure of the spiral instability at the gap
    edge in Case 6. The relative density perturbation are
    overlaid by the mass flux vectors $\rho\bm{u}$ projected onto
    this plane. The slices are taken at $t=40P_0$ and azimuths
    marked by white lines in Fig. \ref{vortex7_polar_dens} (with
    decreasing $\phi$ from top to bottom). 
    \label{vortex7_stream}} 
\end{figure}

Several vertical cuts of the edge mode in Case 6 are shown in
Fig. \ref{vortex7_stream}. The slices are taken at azimuths marked by
white lines in Fig. \ref{vortex7_polar_dens}. The top three plots are
associated with the edge disturbance, while the bottom plot is taken
at the transition between the edge disturbance and its spiral arm
extending to the outer disc. 

Fig. \ref{vortex7_stream} shows that the horizontal flow in the 
edge disturbance differ significantly from the vortex mode. If 
the second plot is considered the `centroid', then the inward
(outward) flow ahead (behind) it is in the opposite sense to
anti-cyclonic motion associated with a vortex mode. 
The centroid is the most stratified region. Its has midplane 
over-density $\sim 3.37$ corresponds to a Toomre $Q\sim 1$ in the
centroid, but we do not observe fragmentation. 

As we move away from the centroid in the decreasing $\phi$ direction, 
the edge mode decreases in amplitude but occupies more of the vertical 
domain. The bottom plot in Fig 
\ref{vortex7_stream} shows a radial split: the columnar disturbance
in $R-r_p\in[5.8,6.5]r_h$ is the beginning of the spiral wave excited
by the edge disturbance. The spiral arm is fully three-dimensional. 


\begin{figure}
  \centering
  \includegraphics[width=\linewidth]{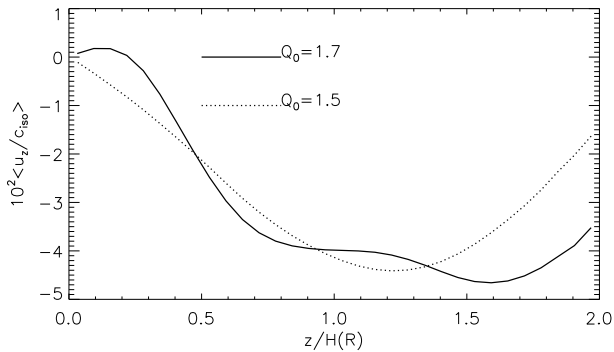} 
  \caption{Average vertical velocity of the edge disturbance in Case 6
    (solid) and Case 7 (dotted). The average is taken over
    $r-r_p\in[2.5,5.5]r_h$. 
    The azimuthal range is indicated by white lines
    in Fig. \ref{vortex7_stream} for Case 6 (first and third azimuth
    in the clockwise direction) and in 
    Fig. \ref{vortex1_polar_dens} for Case 7. 
    \label{vortex1_vortex7_vprofile}}
 \end{figure}

\begin{figure}
  \centering
  \includegraphics[width=\linewidth,clip=true,trim=0.75cm
    0.6cm 2.4cm
    .7cm]{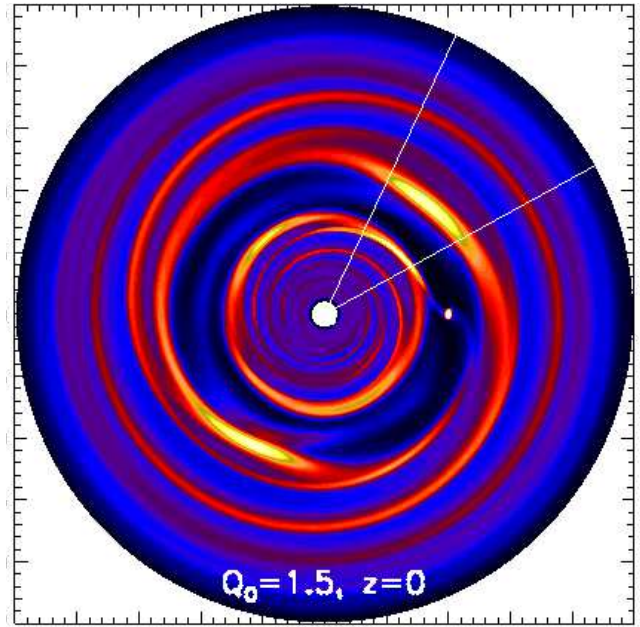}
  \caption{The $m=2$ edge mode in Case 7. The colourbar
    range is the same as in Fig. \ref{vortex7_polar_dens}. White lines
    indicate the azimuthal range taken for averaging in
    Fig. \ref{vortex1_vortex7_vprofile} (dotted line). The relative
    density perturbation of the edge disturbance in the $Rz$ plane is
    very similar to Case 6 (Fig. \ref{vortex7_stream}). 
    \label{vortex1_polar_dens}}
\end{figure}

In Fig. \ref{vortex1_vortex7_vprofile} we plot the average vertical velocity
inside an edge disturbance of the spiral mode. A simulation with $Q_0=1.5$ (Case 7) 
is also plotted for comparison (its midplane density perturbation is shown in
Fig. \ref{vortex1_polar_dens}). The flow is typically downwards toward
the midplane. This is expected because of strong vertical
self-gravity, as these are massive discs and the midplane density
enhancement is large. 
Note that $u_z$ approaches zero again beyond $z/H\sim 1.5$  because of
the imposed reflective upper boundary.

\section{Additional results analysis}\label{additional}
In this section we examine some secondary quantities derived
from the hydrodynamic simulations above. To keep this discussion
concise, we will use selected simulations from above for illustration.   

\subsection{Three-dimensionality}
A simple measure of three-dimensionality of the flow is to compare vertical 
to horizontal motion. Since we are interested in non-axisymmetric
perturbations to the gap edge, we first Fourier transform the
meridional momentum densities 

\begin{align}
  (v_{Rm}, v_{zm} ) \equiv\int_0^{2\pi}\rho\times(u_R,
  u_z)\exp{(-\ii m\phi)}d\phi. 
\end{align}
We define the three dimensionality as $\Theta_m(z/H)$, where 
\begin{align}
  \Theta_m^2 \equiv \frac{\avg{|v_{zm}|^2}}{\avg{|v_{zm}|^2}
    +\avg{|v_{Rm}|^2}},  
\end{align}
and $\avg{\cdot}$ denotes a radial average. 
Admittedly, this is a crude measure, and exact values of $\Theta_m$
varies somewhat with details of the average. However, we have
experimented with different averaging domains and 
found the features described below are robust.

\begin{figure}
  \centering
  \includegraphics[scale=.425,clip=true,trim=0.2cm 1.7cm 0cm
    0cm]{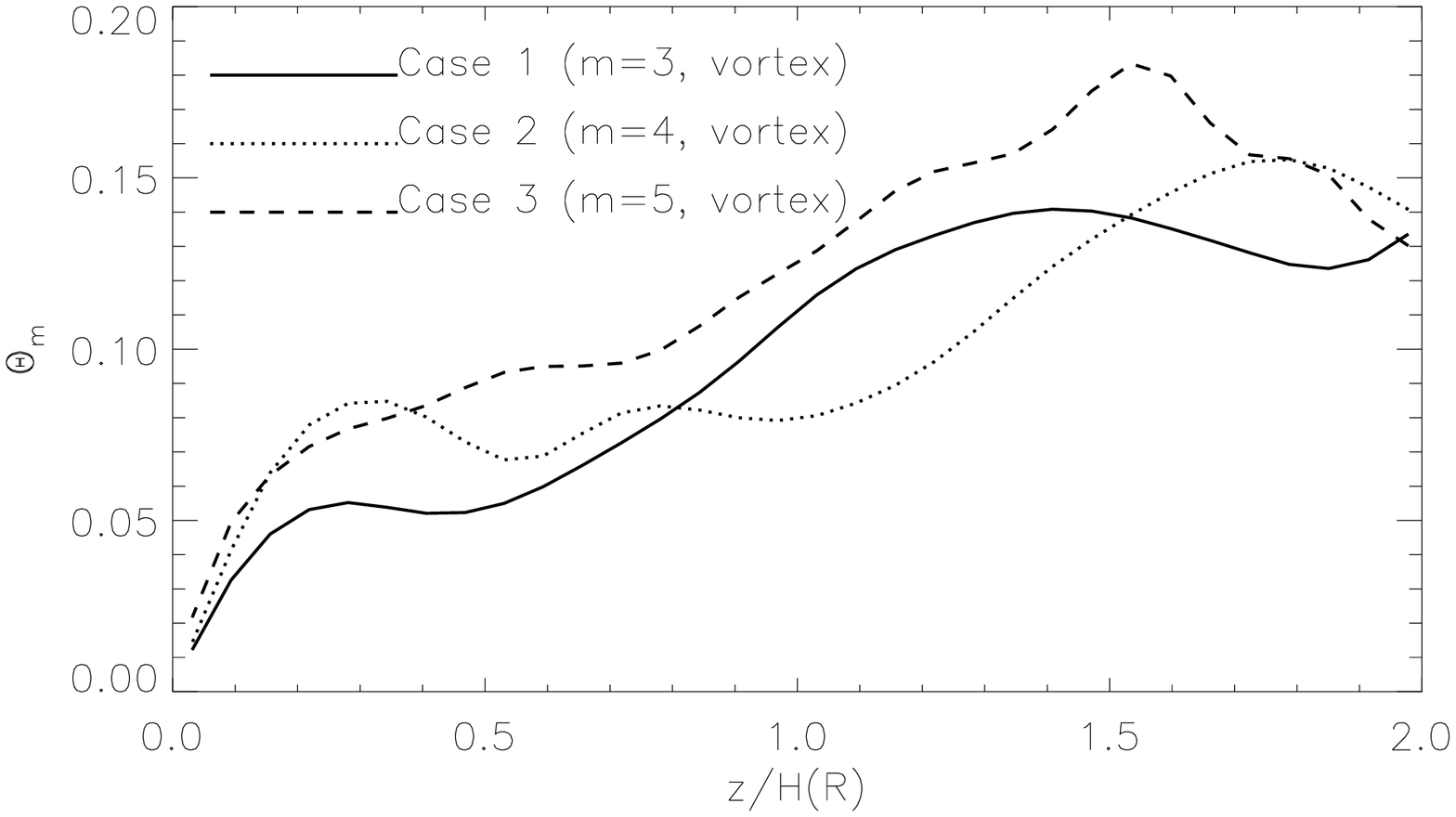}  
  \includegraphics[scale=.425,clip=true,trim=0.2cm 0cm 0cm
    0.2cm]{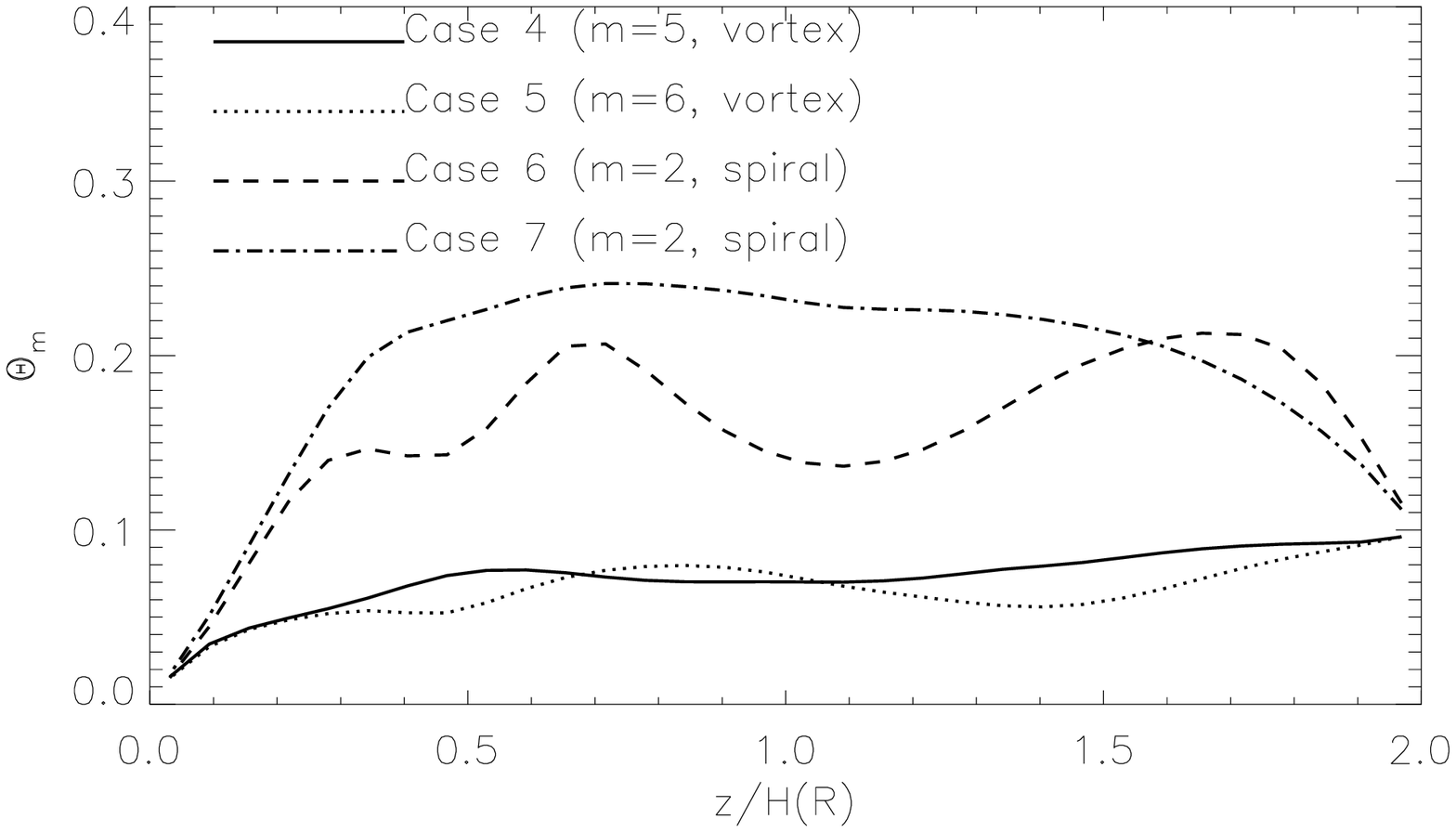}  
  \caption{Three-dimensionality of the non-axisymmetric
    flow near the outer gap edge. Top: Cases 1---3 (vortex modes). 
    Bottom: Cases 4---5 (vortex modes) and Cases 6---7 (spiral
    modes). The azimuthal wavenumber $m$ is chosen to match the number
    of vortices or large-scale spirals observed. 
    \label{compare_vprofiles_3d008}}
\end{figure}

The top panel in Fig. \ref{compare_vprofiles_3d008} shows $\Theta_m$
for Cases 1--3 at $t=40P_0$. The radial average is taken over
$r-r_p\in[3,7]r_h$. These are all vortex modes (see
Fig. \ref{vortex8_polar_dens} and
Fig. \ref{vortex4_vortex10_overall}). The flow becomes increasingly
three-dimensional away from the midplane but 
$\Theta_m = O(10^{-1})$ is small. In an averaged sense the flow is
mostly horizontal. At the end of the  
simulation for Case 1, an azimuthally extended vortex dominates the
flow, for which we measured $\Theta_1\sim
0.2$---0.3. Thus, although vertical motion can become an appreciable
fraction of horizontal motion, the former never dominates.  

$\Theta_m$ for Cases 4---7 are shown in in bottom panel of
Fig. \ref{compare_vprofiles_3d008}. The radial average is performed
over $r-r_p\in[2,6]r_h$ 
because the global spirals in Cases 6---7 significantly protrude the
gap edge. The snapshot is taken at $t=50P_0$ for Cases 4---5, at
$t=40P_0$ for Case 6 and at $t=30P_0$, so that the
vortices and spirals have comparable over-densities at the
gap edge. It also reflects the fact that spiral modes are more
unstable than vortex modes and develop earlier  
\citep{lin11b}. $\Theta_m \sim 0.2$ is again not particularly
large, but the spiral modes are distinctly more three-dimensional than
vortex modes. This is likely due to additional vertical
acceleration provided by the strong self-gravity in those cases.

%

%

\subsection{Vortensity field}\label{vortensity}
A fundamental distinction between the linear vortex and edge mode
instability is their association with local vortensity minimum and
maximum, respectively. In this section we compare vortensity fields of 
discs with vortex modes (Case 2) and edge modes (Case 7). 
More specifically, we examine the relative
perturbation to the vertical component of vortensity,  
\begin{align}
  \Delta\eta_z\equiv \frac{\eta_z - \eta_z(t=0)}{\eta_z(t=0)},
\end{align}
where
\begin{align}
  \eta_z \equiv \frac{\bm{\hat{z}}\cdot\nabla\times\bm{u}}{\rho}. 
\end{align}

\begin{figure}
  \centering
  \includegraphics[scale=.425,clip=true,trim=0cm 0cm 1.9cm 
    0.9cm]{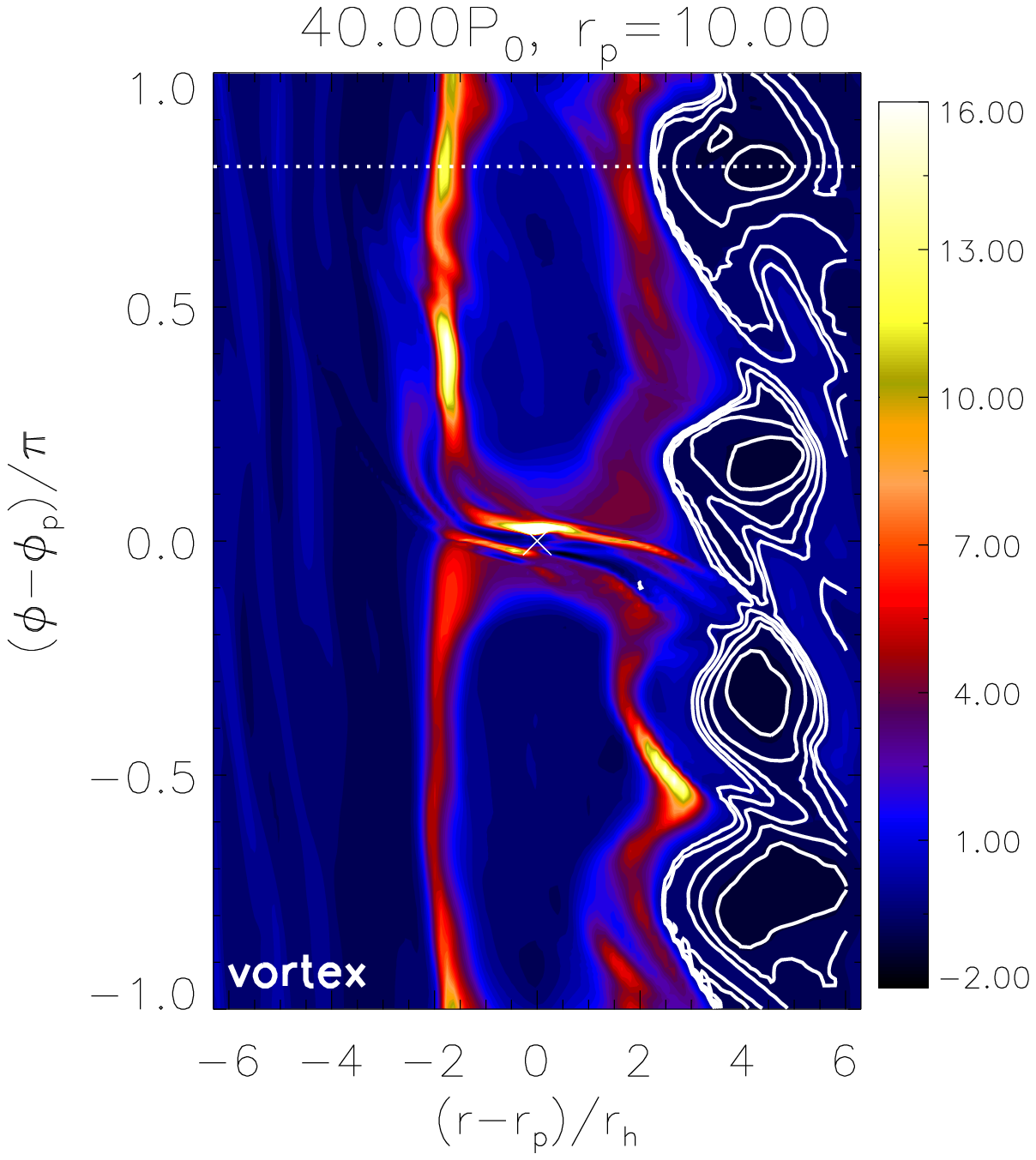}\includegraphics[scale=.425,clip=true,trim=2.3cm    
    0.0cm 0cm 
    0.9cm]{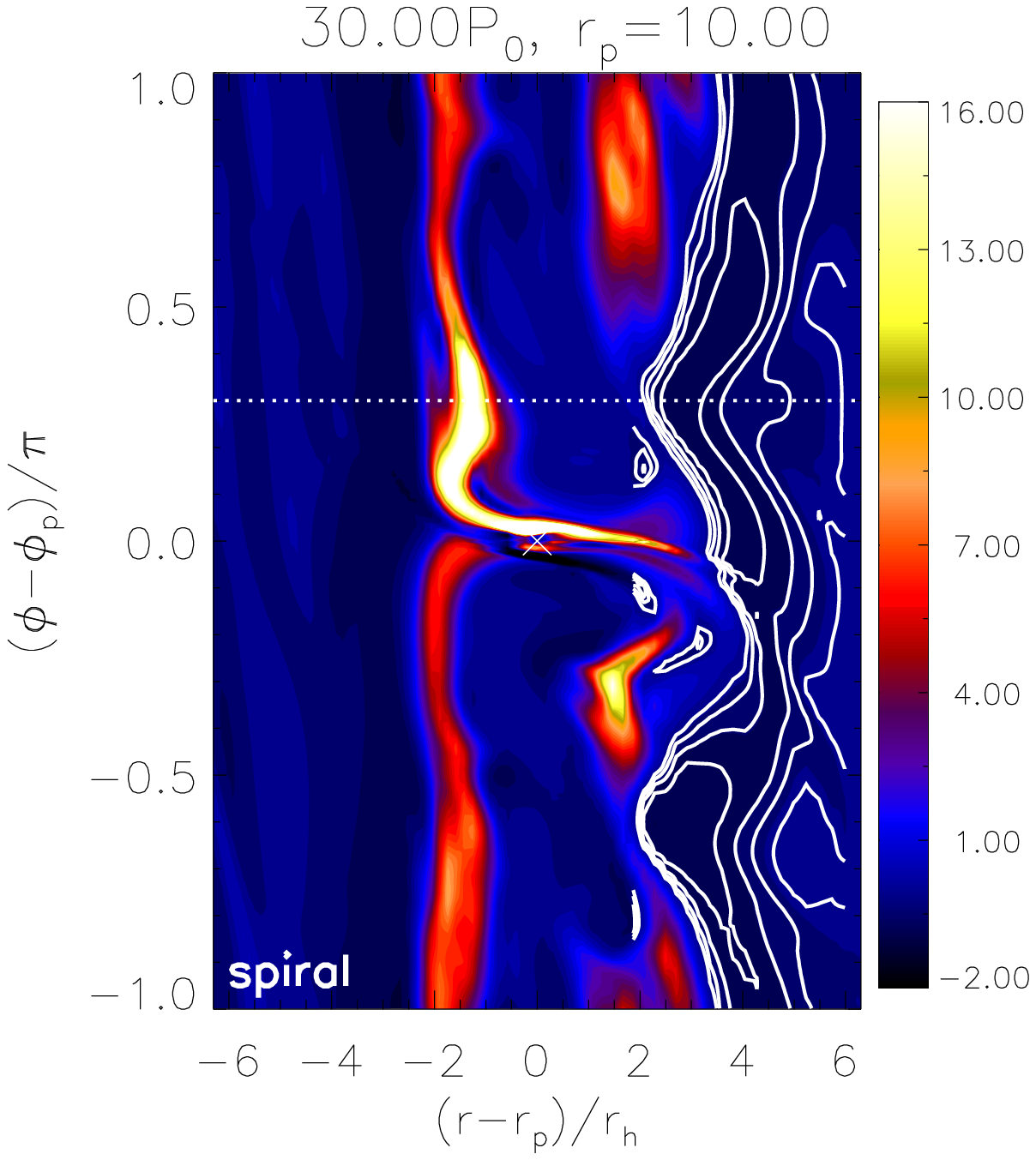}
  \caption{Relative perturbation to the vertical component of
    vortensity at the midplane in a disc with the vortex instability
    (left, Case 2 at $t=40P_0$) and the 
    spiral instability (right, Case 7 at $t=30P_0$). Negative
    perturbations in the region $r-r_p\in[2,6]r_h$ are outlined by white lines. 
    Dotted horizontal lines indicate azimuthal cuts taken in
    Fig. \ref{vortex4_vortex1_vortRZ}.   
    \label{vortex4_vortex1_vortxy}}
\end{figure}

Fig. \ref{vortex4_vortex1_vortxy} compares $\Delta\eta_z$ in the
midplane when vortices and spirals develop. For planetary gaps, 
vortensity maxima and minima are both located near the gap edges with 
characteristic separation of the local scale-height. The vortensity
ring at the inner gap edge ($r-r_p\simeq - 2r_h$) remain
well-defined. The vortex instability is associated with the local
vortensity minimum near the outer gap edge --- seen as localised closed
contour lines centred about $r - r_p \sim 4r_h$. The vortensity ring at
$r-r_p\sim +2r_h$ becomes distorted as a \emph{consequence} of
large-scale vortex formation just exterior to it. By contrast, the
edge-spiral mode is associated with the local vortensity
maximum. Their development inherently disrupts the vortensity
rings. This is seen in the right panel as the outer ring is broken up.    
    
\begin{figure}
  \centering
  \includegraphics[scale=.47,clip=true,trim=0cm 1.32cm .0cm
    .6cm]{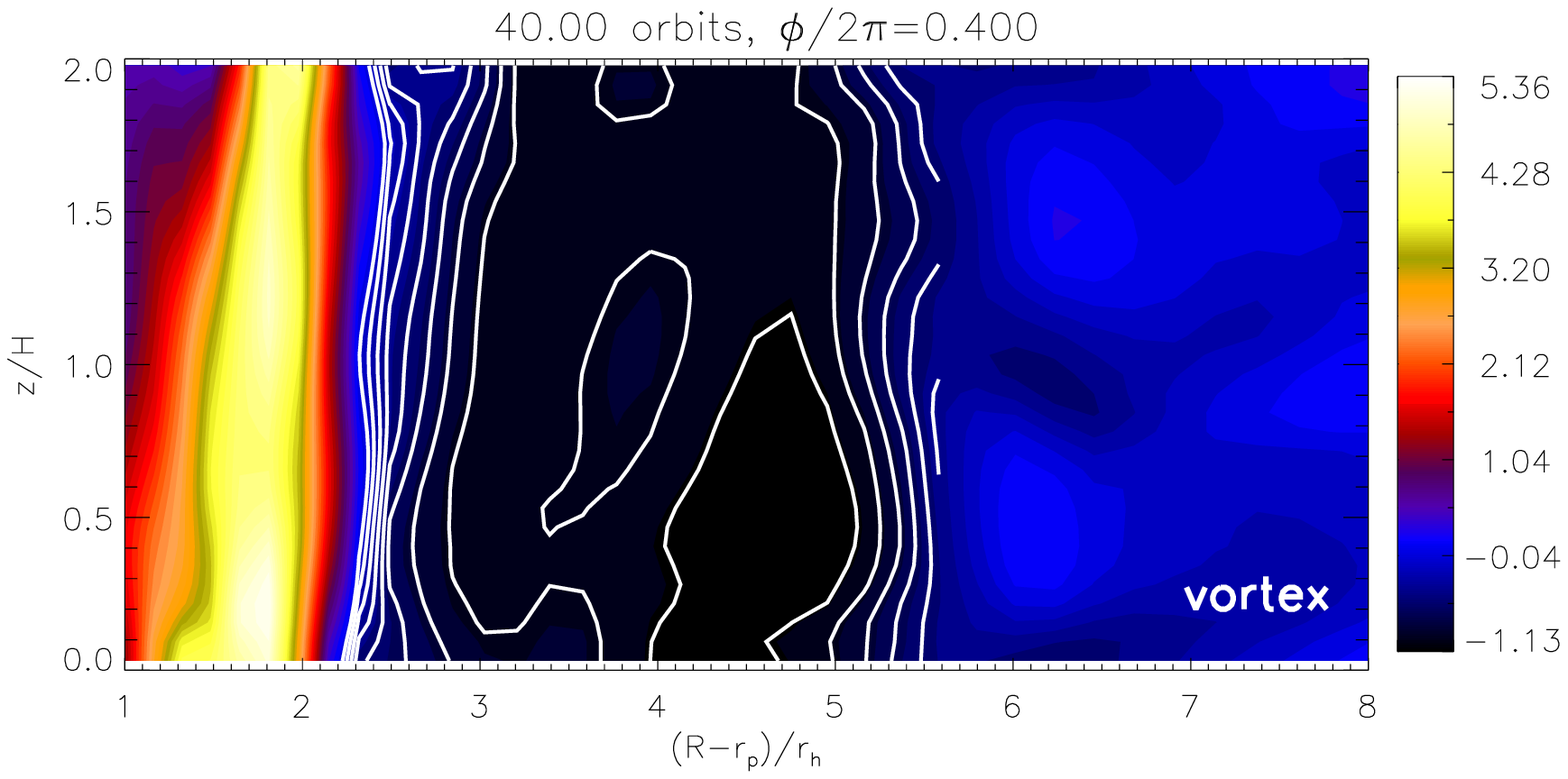}\\\includegraphics[scale=.47,clip=true,trim=0cm 
    0.cm .0cm
    .6cm]{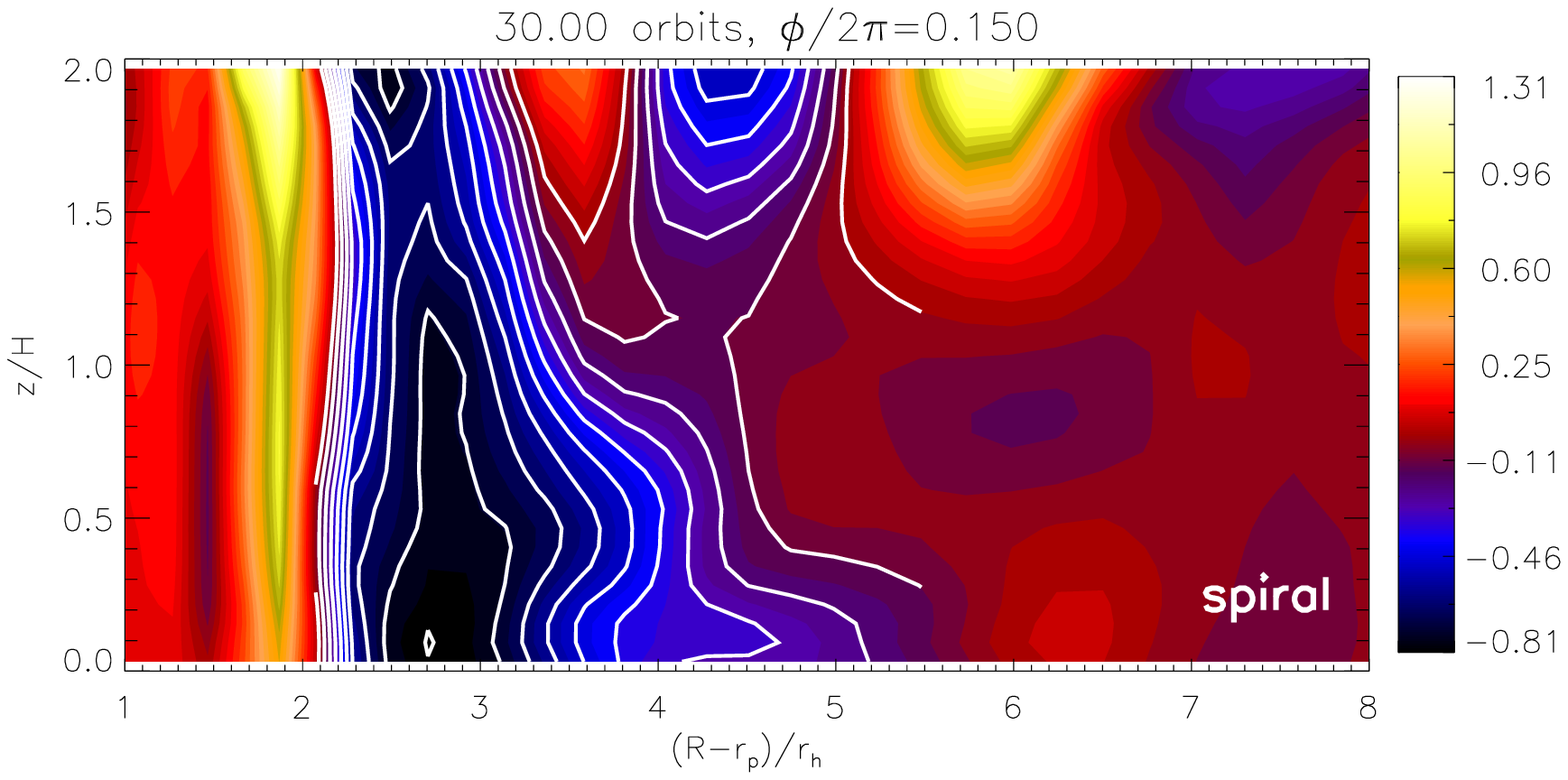}
  \caption{Relative vertical vortensity perturbation
    associated with the vortex instability (top) and spiral
    instability (bottom). The slices are taken at azimuths shown by
    white dotted lines in Fig. \ref{vortex4_vortex1_vortxy}. Negative
    perturbations in the region $r-r_p\in[2,5.5]r_h$ are outlined by
    white lines. \label{vortex4_vortex1_vortRZ}} 
\end{figure}

The vertical structures also differ. Fig. \ref{vortex4_vortex1_vortRZ}
compares $\Delta\eta_z$ at azimuths coinciding with 
a vortex or the edge disturbance of the spiral mode. Both
instabilities involve $\Delta\eta_z<0$. It is clear that 
the spiral mode has stronger vertical dependence. Its 
region of $\Delta\eta_z<0$ becomes thinner away from
the midplane. In the vortex case this region remains 
about the same width and  $\Delta\eta_z$ is approximately uniform
within it. 
While $\mathrm{min}(\Delta\eta_z)$ is of comparable
magnitude, the vortensity ring at $r-r_p=2r_h$is much weaker and
thinner in the spiral case ($\Delta\eta_z$ being a factor $\sim 4$
smaller than the vortex case).

\subsection{Disc-planet torques}
The presence of non-axisymmetric disturbances at gap edges is
expected to significantly affect disc-planet torques. It has been 
confirmed in 2D simulations that both vortex and spiral modes lead to
oscillatory torques of either sign \citep{li05,lin11b}. 
It this section we measure the disc-on-planet torques in several of the above simulations
to confirm the main features found in 2D. 

We calculate the specific torque acting on the planet due to a mass
element as 
\begin{align}
d\bm{T}(\bm{r}) &=
\frac{\bm{r}_p\times\bm{r}G\rho(\bm{r})d^3\bm{r}}{d_p^3}f(\bm{r},\bm{r}_p),\\
f(\bm{r},\bm{r}_p)&\equiv 1
-\exp{\left(-\frac{1}{2}\left|\frac{\bm{r}-\bm{r}_p}{\epsilon_c
      r_h}\right|^2\right)}.  
\end{align} 
The tapering function $f$ reduce contributions from close to the
planet, thereby reducing numerical artifacts arising from this region
because of the diverging potential and limited resolution. We set  
the parameter $\epsilon_c=1$ so that tapering does not significantly
reduce contributions from the instabilities, since they develop
at $\gtrsim 2r_h$ away from the planet's orbital radius.  

\begin{figure}
  \centering
  \includegraphics[scale=.425,clip=true,trim=0.2cm 1.cm 0cm
    0cm]{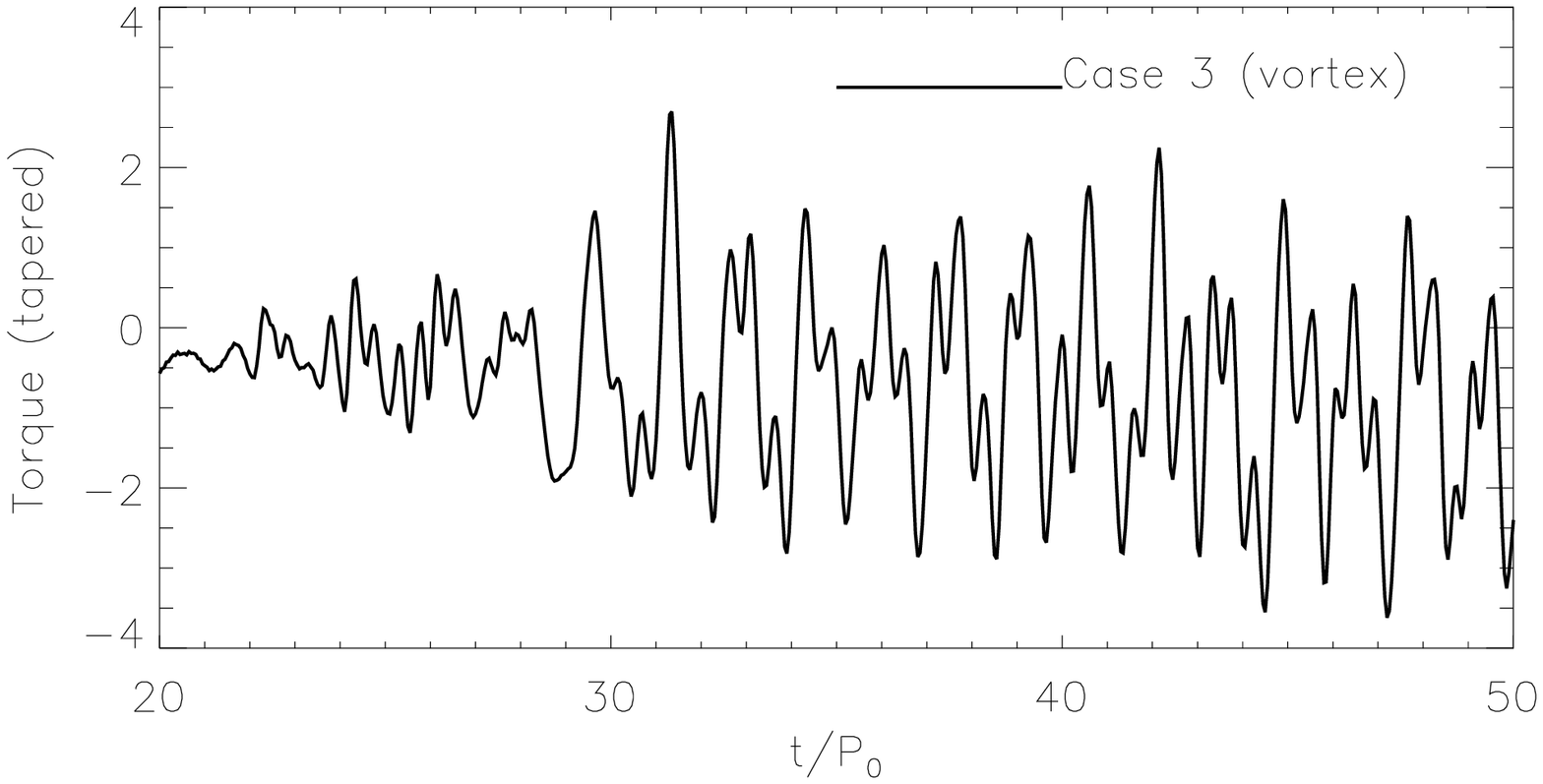}  
  \includegraphics[scale=.425,clip=true,trim=0.2cm 0cm 0cm 0.2cm]{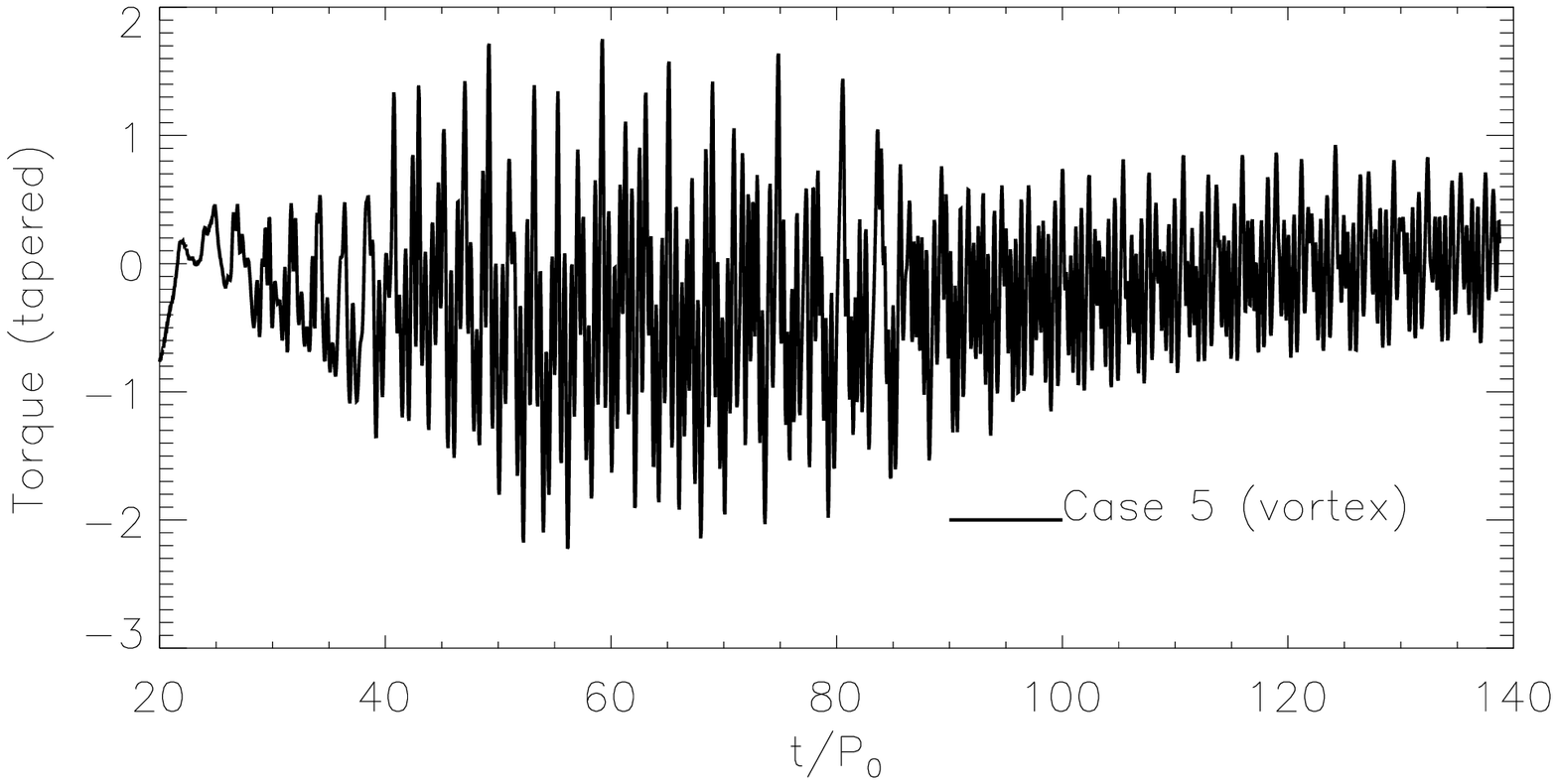} 
  \caption{Instantaneous disc-on-planet torques in simulations where the vortex mode develops. 
    Top: Case 3. Bottom: Case 5. Note that Case 5 has been extended to $t=135P_0$. \label{torque3}}
\end{figure}

Fig. \ref{torque3} shows the disc-on-planet torques in Case 3 and Case 5, which develop
the $m=5$ and $m=6$ vortex modes, respectively. These plots are qualitatively similar to
2D simulations \citep[e.g.][]{li05}. The torques oscillate on orbital 
time-scales and its instantaneous values can be of either sign. However, 
upon averaging over the simulation we find the total torques are negative in both cases. This
means inwards migration is still favoured. 

We extended Case 5 to $t=135P_0$ and find the vortices 
have similar over-densities as at $t=50P_0$. However, 
Fig. \ref{torque3} show the torque oscillation amplitudes decrease
towards the end of Case 5 compared to $t\in[40,80]P_0$.  
At $t=50P_0$ the vortices are located in $r-r_p\in[3.5,5.5]r_h$ but by
$t=135P_0$ they are located in $r-r_p\in[4,6]r_h$. 
Given that $t\in[40,80]P_0$ is only $20P_0$ to $50P_0$ after the planet  
potential has been fully introduced, gap-formation is probably 
ongoing during this time. We expect torque amplitudes to be larger
during  gap-formation since the vortices lie closer to the planet. 


\begin{figure}
  \centering
  \includegraphics[scale=.425,clip=true,trim=0.2cm 1.825cm 0cm
    0cm]{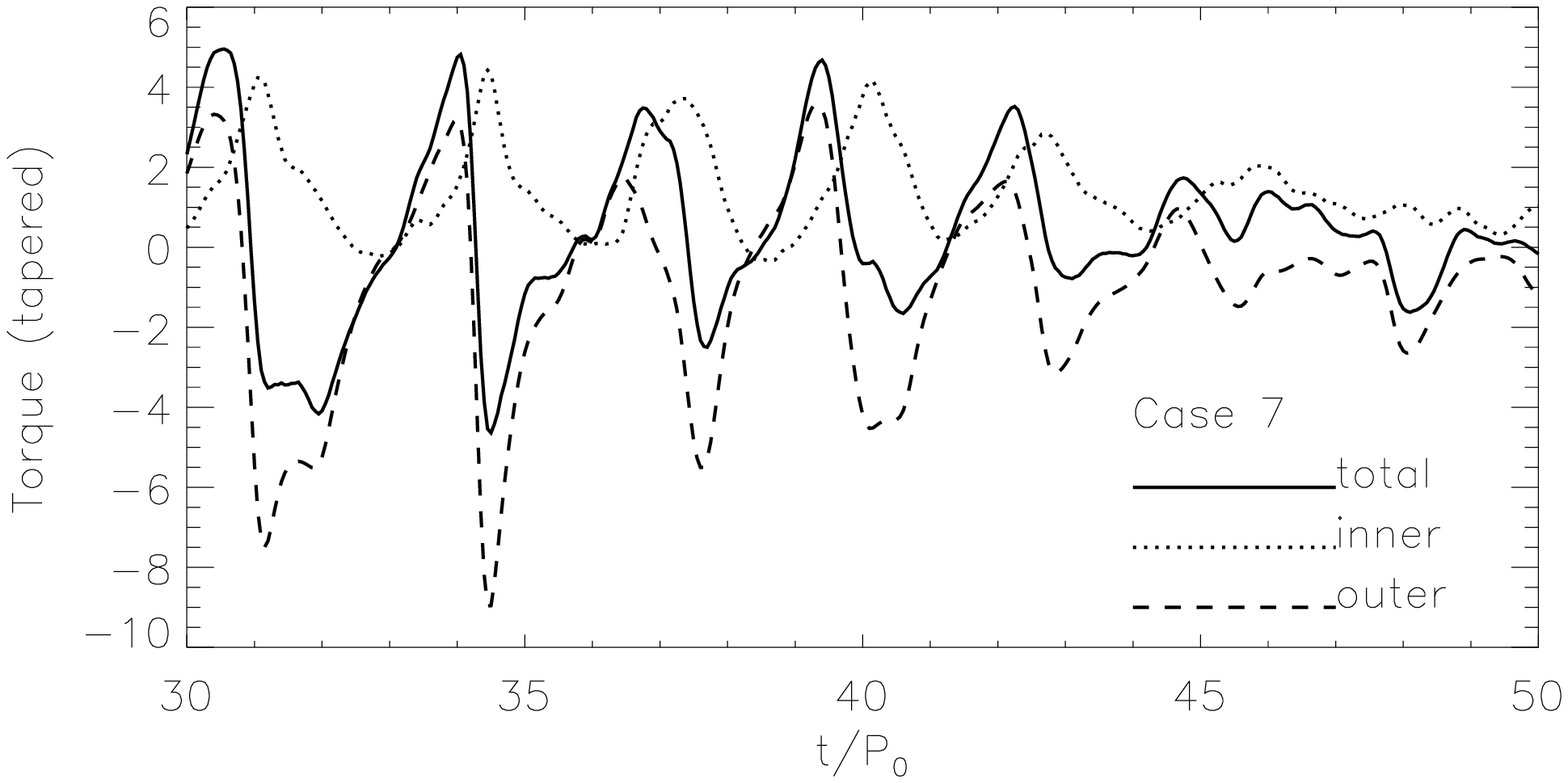}  
  \includegraphics[scale=.425,clip=true,trim=0.2cm 0cm 0cm 0.2cm]{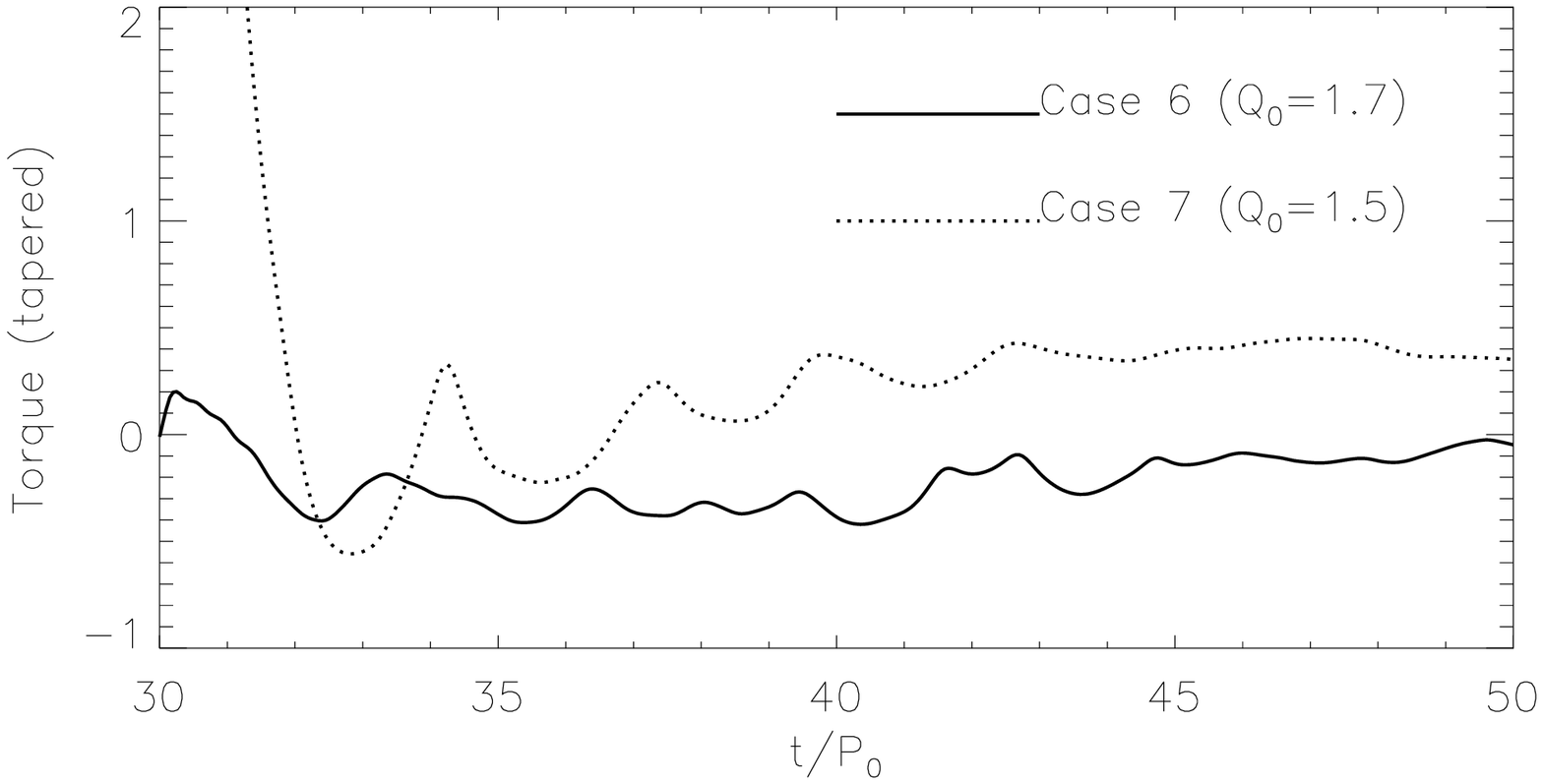} 
  \caption{Disc-on-planet torques in the presence of spiral modes
    associated with the outer gap edge. Top: total torque (solid),
    torque from the inner disc (dotted) and from the outer disc
    (dashed) in Case 7. Bottom: time-averaged torques in Case 6
    (solid) and Case 7 (dotted).
\label{torque3_spiral}}
\end{figure}

Next we examine disc-planet torques in the presence of the spiral
modes. The top panel of Fig. \ref{torque3_spiral} shows the
instantaneous disc-on-planet torques. Contributions from the
inner disc ($r < r_p$) and outer disc ($r > r_p$) are plotted
separately for comparison with Fig. 18b in \cite{lin11b}, which is 
similar to the present plot. Large oscillations in the outer torque
due to edge mode spirals cause the total torque to be positive or
negative at a given instant. 

Unlike the vortex modes, Fig. \ref{torque3_spiral} shows that spiral
modes can lead to a positive running-time averaged torques (bottom panel). 
The average torques become more positive with time after spiral modes develop, 
and with increasing self-gravity (which increases the instability strength). 
This was also observed in high-resolution 2D simulations in
\cite{lin11b}. There it was suggested that the creation of large   
`voids' in between spiral arms decreases the time-averaged density in  
the planet-induced wakes, thereby reducing associated torque
magnitudes. Since the outer planetary wake normally provide a negative
torque, the spiral modes make the total torque more positive. 
The similarity between 2D and 3D results indicate that
outwards migration induced by spiral modes, which was seen in 2D by
\cite{lin11b,lin12b}, will also operate in 3D.      




\section{Summary and discussion}\label{summary}
We have performed customised numerical simulations of  
three-dimensional self-gravitating discs, in which an embedded 
satellite or planet has opened a gap. We explicitly verified in 
3D the main results on gap stability previously obtained from 
2D calculations \citep{koller03,li05,valborro07,meschiari08}, in
particular those by \cite{lin11a,lin11b}.  

Planetary gaps are potentially unstable because of the existence of 
vortensity extrema generated by planet-induced shocks. Disc-satellite
interaction also occur in other systems such as stars in black hole
accretion discs \citep{kocsis11,mckernan11,baruteau11b}.  Furthermore, 
gaps opened by satellites are just one type of structured disc. 
Other examples 
include dead zone boundaries mentioned in \S\ref{intro} and
transition discs \citep{regaly12}.  

Thus, although we were motivated by previous works on planetary gap 
stability, and hence considered disc-planet systems, we expect our
results to be applicable to discs with radial structure of other    
origin, provided the vortensity profiles involve stationary points and
therefore prone to the same instabilities.    

\subsection{Confirmed 2D results}
We demonstrated the development of the vortex instability
at the outer gap edge opened by a giant planet in 3D 
discs. We began with a non-self-gravitating disc, in which a few vortices
develop then quickly merge. The quasi-steady state is a single
azimuthally extended vortex at the gap edge. This evolution is similar
to the 2D simulation in \cite{valborro07}.  

We also showed the effect of self-gravity on the vortex
instability observed in 2D \citep{lyra09,lin11a} persists in 3D.
We observe more vortices as the strength of self-gravity is increased
by increasing the density scale. These vortices resist merging
on dynamical timescales. In our disc models with $Q_0=3$, the 
multi-vortex configuration lasts until the end of the
simulations.        

As expected from 2D linear theory \citep{lin11b}, 
vortex modes are suppressed in our massive disc models. Instead, a 
global spiral instability develops which is associated with the local
vortensity maximum just inside the outer gap edge. These are distinct
from vortex modes since self-gravity is essential. 

Our limited numerical resolution does not permit accurate disc-planet
torque measurements, but 
the qualitative effect of the vortex and spiral instabilities,
previously studied in 2D, have been reproduced in 3D --- oscillatory
torques of either sign and the tendency for spiral modes to provide on
average a positive torque. 
The similarity to 2D results is not surprising since for  
giant planets $r_h\gtrsim H$, so the razor-thin disc approximation is
expected to be valid as far as disc-planet interaction is
concerned. Furthermore, vertical self-gravity increases the midplane
density while reducing that in the atmosphere (Appendix
\ref{vertsg_mod}), so that given a fixed temperature profile the 2D
approximation is even better for self-gravitating discs.

 It is worth mentioning here that in shearing sheet
  simulations,  \cite{mamatsashvili09} found 
  self-gravity to favour vortices of smaller scale. They initialized
  a local patch of an unstructured disc with random velocity perturbations.
  Their gravito-turbulent discs are dominated by small vortices limited by
  the local Jeans length (which is smaller than the scale-height). Without self-gravity, 
  they merge to form larger vortices. 
  These observations are similar to the above results for the vortex mode.  
  However, the setups are quite different 
  as we consider radially structured, laminar global discs. Our
  large-scale vortices develop from a linear instability and have horizontal sizes
  of a few scale-heights. Thus, confirmation of the above results in 3D is only valid
  for the edge instabilities considered in this paper.


\subsection{Effects of 3D}
In our simulations the dominant three-dimensional effect is
vertical self-gravity on the density field. In the
non-self-gravitating limit, the relative density perturbation
associated with a vortex is columnar with weak vertical dependence. This is
consistent with with 3D linear and nonlinear simulations
\citep{meheut10, meheut12a,  meheut12b, lin12}.       

As the strength of self-gravity is increased, vortices become 
more vertically stratified --- they are condensed towards the
midplane. For moderately self-gravitating discs, the vortex
midplane density enhancement can be twice that near the upper
boundary.  The spiral modes display significant vertical structure
near the gap edge, while the density waves they launch in the outer
disc is columnar.  The latter is probably due to the chosen equation of state (see below).

The effect of vertical self-gravity on the vortex mode is seen even in
our least massive disc model with $Q_0=8$. One can consider an
initially smooth disc which is justified to be
non-self-gravitating. However, this approximation may become less good
with the creation of a vortensity minimum, because it
can also be a local minimum in the Toomre $Q$
(Eq. \ref{toomreq_vortensity}).  That is, the Toomre parameter is
decreased with the development of local radial structure (such as a density bump). 
The non-self-gravitating approximation worsens further when the vortex
instability associated with $\mathrm{min}(\eta)$ develops, because the   
vortices are regions of enhanced density, especially if they merge
into a single large vortex. Thus, the
non-self-gravitating approximation is not guaranteed to hold 
in the perturbed state even if it does in the initial disc.

It is interesting to note that linear calculations of the vortex
instability in vertically isothermal , non-self-gravitating discs 
show that the vertical velocity vanishes near the vortex
centroid  \citep{meheut12a,lin12}, but this is not observed in
nonlinear calculations \citep[][ and in the present
  simulations]{meheut12a}.  This contrasts to their anti-cyclonic
horizontal flow, which can be computed in linear theory and seen in
hydrodynamic simulations \citep{li00,li01}.  This suggests that vertical 
motions in this case may be associated with secondary processes
\citep[e.g][]{goodman87}. 

Although we observe somewhat complicated vertical flow for both
vortex and spiral instabilities, the vertical Mach number is
at most a few per cent in magnitude. Also, the magnitude of
vertical motion is at most $\sim 20\%$  of the radial motion on
average. This suggests that the disturbances at the gap edge is 
roughly two-dimensional in the present disc models. This would be 
consistent with early studies which find instabilities associated with
co-rotation singularities are two-dimensional
\citep{papaloizou85,goldreich86}.  Recent 3D linear 
  calculations also find that, even with a vertical temperature gradient, 
  the vortex mode (without self-gravity) is
  largely 2D near the vortensity minimum \citep{lin12}.

\subsection{Caveats and outlooks}\label{caveats}

In order to provide 3D examples of previous 2D results, a range of 
disc models had to be simulated, each for many dynamical 
timescales with full self-gravity. To maintain reasonable
computational cost, numerical resolution in the $r\phi$ plane is much
reduced compared to razor-thin disc simulations ($\sim  4$---6 cells
per $H$ compared to $\sim 16$ in 2D).  Despite this, our
plots in the $r\phi$ plane closely resemble those obtained from 2D
simulations.  

On the other hand, the low resolutions adopted here are unlikely to
capture elliptic instabilities which may destroy 
vortices in 3D \citep{lesur09b,lesur10}. 
This might not be a serious
issue when the condition for the vortex instability is maintained 
(by a planet in the present context). Also, the vortex 
grows on dynamical timescales whereas the elliptic instability takes
much longer \citep{lesur09b}. The initial development of vortices is not expected to be 
suppressed by the elliptic instability \citep[as found
  by][]{meheut12a}.   

The vortex instability produces smaller vortices with increasing
self-gravity. According to \cite{lesur09b}, vortices with small
aspect-ratios ($\lesssim 4$) are strongly unstable in 3D, but note that their  
model is a local patch of a smooth disc without self-gravity, 
whereas we considered a gap edge in a global self-gravitating disc. 

Inclusion of self-gravity may change the stability properties of
vortices. In particular, we found that a vortex can flatten somewhat under its own
weight. \cite{lithwick09} suggested vertical gravity helps to 
stabilise vortices in 3D. Vertical \emph{self}-gravity 
can enhance this effect.  \citeauthor{lithwick09} 
found in local 3D simulations that `tall' vortices are unstable 
whereas `short' vortices survive as in 2D \citep{godon99}. 
We may expect the more stratified vortices formed in self-gravitating 
discs to be more stable than those in non-self-gravitating discs \citep{barranco05}. 
The elliptic instability is also weakened by stratification \citep{lesur09b}, 
so vertical self-gravity may also be stabilising in this respect. 

%

We have adopted the locally isothermal EOS for simplicity and direct
comparison with previous 2D results. This EOS limits the 3D 
structure of density waves compared to thermally 
stratified discs \citep[e.g][]{lin90,lubow98,ogilvie99}, which can 
cause refraction of waves out of the midplane.  
The locally isothermal EOS represents the limit of efficient cooling
\citep{boss98}. This might apply in optically thin
regions of a disk\footnote{For example, \cite{cossins10} find optical
depths $\tau < 0.2$ beyond $\sim100\mathrm{AU}$ in their models of protoplanetary
discs.}, but is violated if high densities develop, such as 
self-gravitating clumps \citep{pickett00}. Our edge instabilities only 
reach moderate over-densities. Nevertheless, enhanced
vertical stratification of the edge disturbances observed in our
simulations are likely exaggerated by the EOS. Clearly, it is necessary to
extend models of edge instabilities in self-gravitating discs to
include an appropriate energy equation.

Another important issue is vertical boundary conditions. We simulated
a thin disc and imposed a reflecting upper boundary to prevent mass
loss from above. This setup  may enhance the two-dimensionality
of the problem. The vortex instability tends to involve the 
entire column of fluid, especially in the weak self-gravity regime. It
is therefore a global instability in $z$. We suspect the spiral mode
is less affected by vertical boundary conditions because the
instability tends to concentrate material at the midplane. Future
simulations will consider varying the vertical domain size and upper
disc boundary conditions. 

Our torque measurements indicate migration will be significantly
affected by the vortex and spiral instabilities. Preliminary 3D
simulations with a freely migrating planet have been performed. We
recover the vortex-planet scattering and the spiral-induced outward
migration described by \cite{lin10,lin12b}.   
The disc-planet torque is determined by the density field, and
the above instabilities have density perturbations that either
have weak vertical dependence or concentrated at the midplane. Thus, we
believe that at present, 2D simulations are more advantageous for studies
focusing on migration, because high resolution is feasible and needed.

\section*{Acknowledgments}
I thank C. Baruteau and G. Mamatsashvili for comments on the first version of this paper. 
Computations were performed on the GPC supercomputer at the SciNet HPC Consortium. 
SciNet is funded by: the Canada Foundation for Innovation under the auspices of Compute Canada; 
the Government of Ontario; Ontario Research Fund - Research Excellence; and 
the University of Toronto. 


\appendix
\section{Modification to vertical structure by self-gravity}\label{vertsg_mod}
We describe a simple procedure to set up the vertical structure of a locally
isothermal, self-gravitating disc. We imagine setting up a non-self-gravitating disc,
then slowly switch on the vertical force due to self-gravity. We expect the midplane
density to increase at the expense of gas density higher in the atmosphere. It is assumed
that the temperature profile remains unchanged. 

Vertical hydrostatic equilibrium between
gas pressure, stellar gravity and self-gravity reads
\begin{align}
c_\mathrm{iso}^2(R)\frac{\p\ln{\rho}}{\p z} = -\frac{\p\Phi_*}{\p z} - \frac{\p\Phi}{\p z}. 
\end{align}
Assuming a smooth radial density profile, we use the plane-parallel atmosphere
approximation for the disc potential, i.e.
\begin{align}
\frac{\p^2\Phi}{\p z^2} = 4\pi G \rho. 
\end{align}
Next, we write the density field as
\begin{align}
\rho(R,z) = \rho_N(R, z) \times \beta(z; R)
\end{align}
where $\rho_N$ is the density field corresponding to the non-self-gravitating disc:
\begin{align}
&c_\mathrm{iso}^2(R)\frac{\p\ln{\rho_N}}{\p z} = -\frac{\p\Phi_*}{\p z},\\
&\rho_N = \frac{\Sigma(R)}{\sqrt{2\pi}H(R)}\exp{(-z^2/2H^2)} \notag\\
&\phantom{\rho_N}\equiv \rho_{N0}(R)\exp{(-z^2/2H^2)},
\end{align}
where $\rho_{N0}=\rho_N(R,z=0)$ is the midplane density. The explicit
expression for $\rho_N$ above assumes a thin disc. The function
$\beta$ describes the modification to the local density in order to be
consistent with self-gravity. By construction, its governing equation
is 
\begin{align}
c_\mathrm{iso}^2(R)\frac{\p^2\ln{\beta}}{\p z^2} = -4\pi G\rho_N\beta. 
\end{align}
Let 
\begin{align}
& \chi  = \ln\beta - z^2/2H^2 ,\\
& \xi= \left(\frac{4\pi G \rho_{N0} }{c_\mathrm{iso}^2}\right)^{1/2} z,
\end{align}
then the governing equation can be written in dimensionless form
\begin{align}
&\frac{\p^2 \chi}{\p\xi^2} = -K -\exp{\chi}\label{vertsg_eqn},\\
& K \equiv \frac{c_\mathrm{iso}^2}{4\pi G \rho_{N0} H^2}. \notag
\end{align}
Note that $K(R)$ is proportional to the local Keplerian Toomre
parameter. Eq. \ref{vertsg_eqn} can be further reduced to a first
order differential equation, but this is unnecessary because we
pursue a numerical solution at the end. Appropriate 
boundary conditions are
\begin{align}
&\chi(z=0)  =  \ln\beta_0, \\
&\left.\frac{\p \chi}{\p \xi} \right|_{z=0} = 0.
\end{align}
$\beta_0(R)$ is the midplane density enhancement. To determine its value, we impose
the surface density before and after modification by self-gravity to
remain the same. Then we require 
\begin{align}\label{beta0_eqn}
&F(\beta_0) \equiv \sqrt{\frac{2K}{\pi}}\int_0^{ n /\sqrt{K} } 
\exp{\chi(\xi;\beta_0)}d\xi - \erf{\left(\frac{n}{\sqrt{2}}\right)} \notag\\
&\phantom{F(\beta_0)} = 0,  
\end{align}
where $n$ is the number of scale-heights of the non-self-gravitating
disc we originally considered. At a given cylindrical radius $R$, we
solve Eq. \ref{beta0_eqn} using Newton-Raphson iteration. Each
iteration involves integrating the governing ODE for $\chi$
(Eq. \ref{vertsg_eqn}). At the end of the iteration, we have $\beta(z;
R)$ and the midplane enhancement $\beta_0(R)$.  

We comment that the procedure outlined above can be extended to
polytropic discs. In this case, there is an additional unknown --- 
the new disc thickness after adjustment by self-gravity and an
additional constraint --- the density should vanish at the new disc
surface.

\end{document}